\documentclass[11pt]{article}
\usepackage{jheppub}
\usepackage{amsmath,amssymb,amsfonts,graphicx,slashed,amsthm,mathtools,upgreek, enumerate, tensor, subfig}
\usepackage[dvipsnames]{xcolor}
\usepackage{arydshln}

\usepackage{comment}
\usepackage{hyperref}
\usepackage[utf8]{inputenc}
\usepackage[titletoc]{appendix}

\numberwithin{equation}{section} 
\allowdisplaybreaks

\allowdisplaybreaks

\newcommand{\beq}{\begin{equation}}
\newcommand{\eeq}{\end{equation}}
\newcommand{\bea}{\begin{equation}\begin{aligned}}
\newcommand{\eea}{\end{aligned}\end{equation}}

\subheader{\small CERN-TH-2020-039}

\title{{\Large Geometric secret sharing in a model of Hawking radiation}}

\author[a, b]{Vijay Balasubramanian}
\author[a]{\!, Arjun Kar}
\author[c]{\!, Onkar Parrikar}
\author[d]{\!, G\'{a}bor S\'{a}rosi}
\author[a]{\!, Tomonori Ugajin}

\affiliation[\,a]{David Rittenhouse Laboratory, University of Pennsylvania,\\
209 S.33rd Street, Philadelphia, PA 19104, USA}
\affiliation[\,b]{Theoretische Natuurkunde, Vrije Universiteit Brussel (VUB), and \\ International Solvay Institutes, Pleinlaan 2, B-1050 Brussels, Belgium}
\affiliation[\,c]{Stanford Institute for Theoretical Physics, 
Stanford University, Stanford, CA 94305, USA}
\affiliation[\,d]{CERN, Theoretical Physics Department, 1211 Geneva 23, Switzerland}

\emailAdd{vijay@physics.upenn.edu}
\emailAdd{arjunkar@sas.upenn.edu}
\emailAdd{parrikar@stanford.edu}
\emailAdd{gabor.sarosi@cern.ch}
\emailAdd{loop.diagram@gmail.com}

\abstract{
We consider a black hole in three dimensional AdS space entangled with an auxiliary radiation system.   
We model the microstates of the black hole in terms of a field theory living on an end of the world brane behind the horizon,  and allow this field theory to itself have a holographic dual geometry.  This geometry is also a black hole since entanglement of the microstates with the radiation leaves them in a mixed state.  This ``inception black hole'' can be purified by entanglement through a wormhole with an auxiliary system which is naturally identified with the external radiation, giving a realization of the ER=EPR scenario.  In this context, we propose an  extension of the Ryu-Takayanagi (RT) formula, in which extremal surfaces computing entanglement entropy are allowed to pass through the brane into its dual geometry.  This new rule reproduces the Page curve for evaporating black holes, consistently with the recently proposed ``island formula''.  We then separate the radiation system into pieces.  Our extended RT rule  shows that the entanglement wedge of the union of radiation subsystems covers the black hole interior at late times, but the union of entanglement wedges of the subsystems may not.  This result points to a secret sharing scheme in Hawking radiation wherein reconstruction of certain regions in the interior is impossible with any subsystem of the radiation, but possible with all of it.
}


\begin{document}

\maketitle

\parskip=10pt

\section{Introduction}

The semiclassical calculation of Hawking \cite{Hawking:1974sw} predicts that pure states of quantum theories can collapse to make black holes and then evaporate into mixed states, thus destroying information, a scenario which is forbidden if quantum mechanics is correct.  By contrast, the AdS/CFT correspondence suggests that information can be recovered from black holes because the whole spacetime is dual to a unitary CFT.   Many authors have suggested that the mechanism for information recovery from black holes is entanglement of Hawking quanta with the interior microstate. Naively quantifying this entanglement as the von Neumann entropy of effective field theory on the black hole background leads to a contradiction --  the entanglement seems to grow continuously with time, which is impossible if there are a finite number of black hole microstates in the first place with which the radiation may be entangled.   In flat space, where black holes evaporate completely, the entropy of the radiation must decline back to zero eventually, while in AdS space, where large black holes come into equilibrium with the radiation, the entanglement entropy should level off at a plateau.  The time at which the entropy growth stops is called the Page time \cite{Page:1993wv}.

\begin{figure}
    \centering
    \includegraphics[scale=.4]{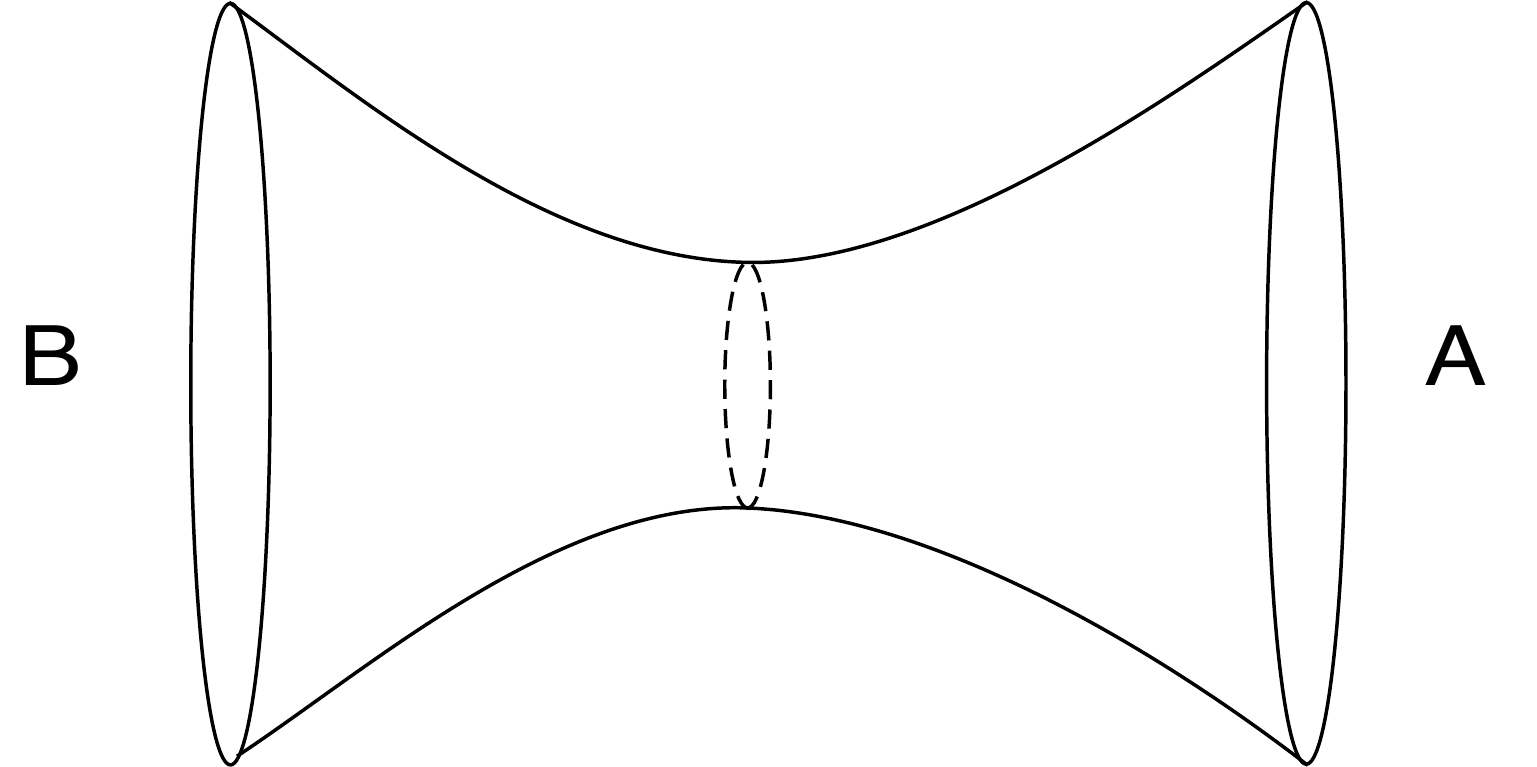} \hfil
    \includegraphics[scale=.4]{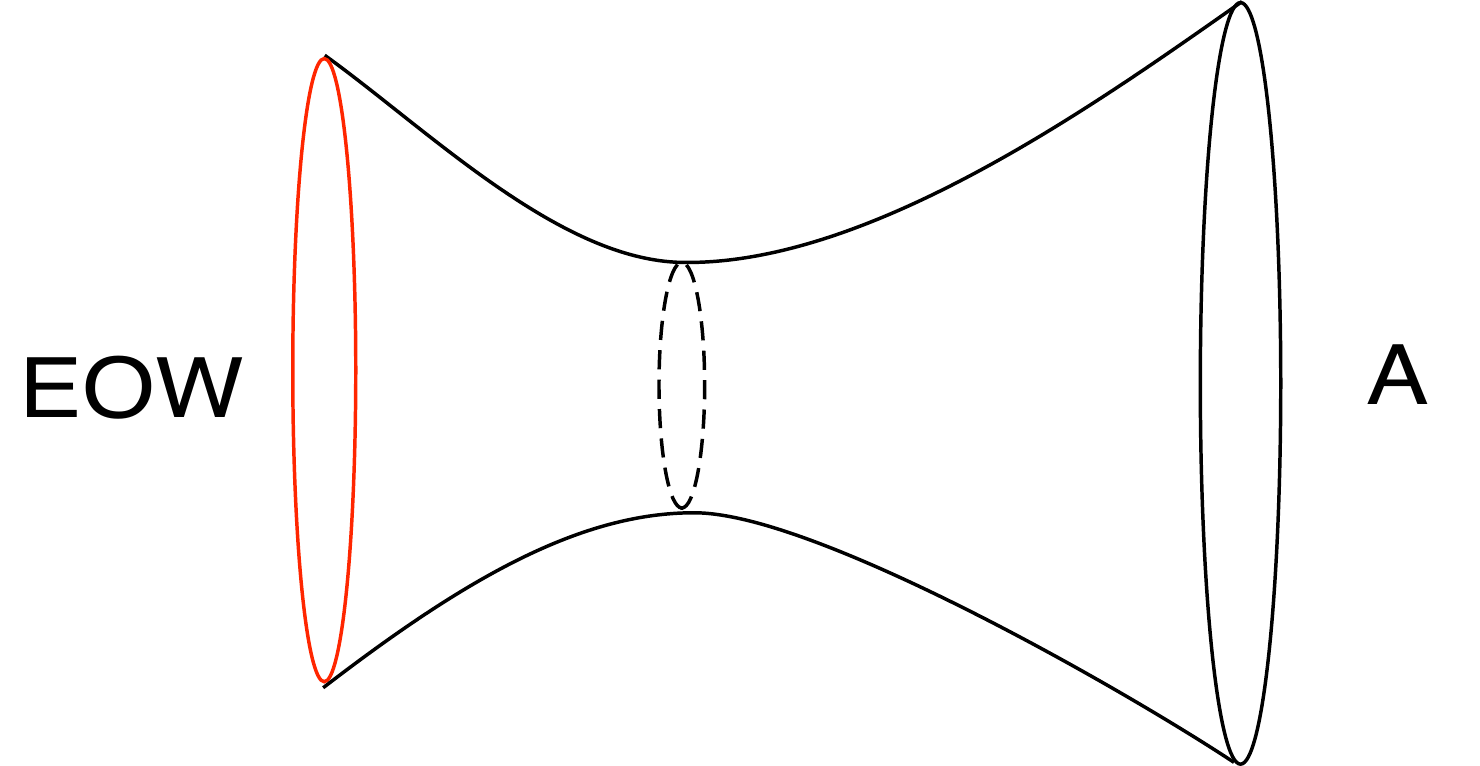} \hfil \\
       ({\bf a}) \hspace{2.5in} ({\bf b}) \\
     \includegraphics[scale=.25]{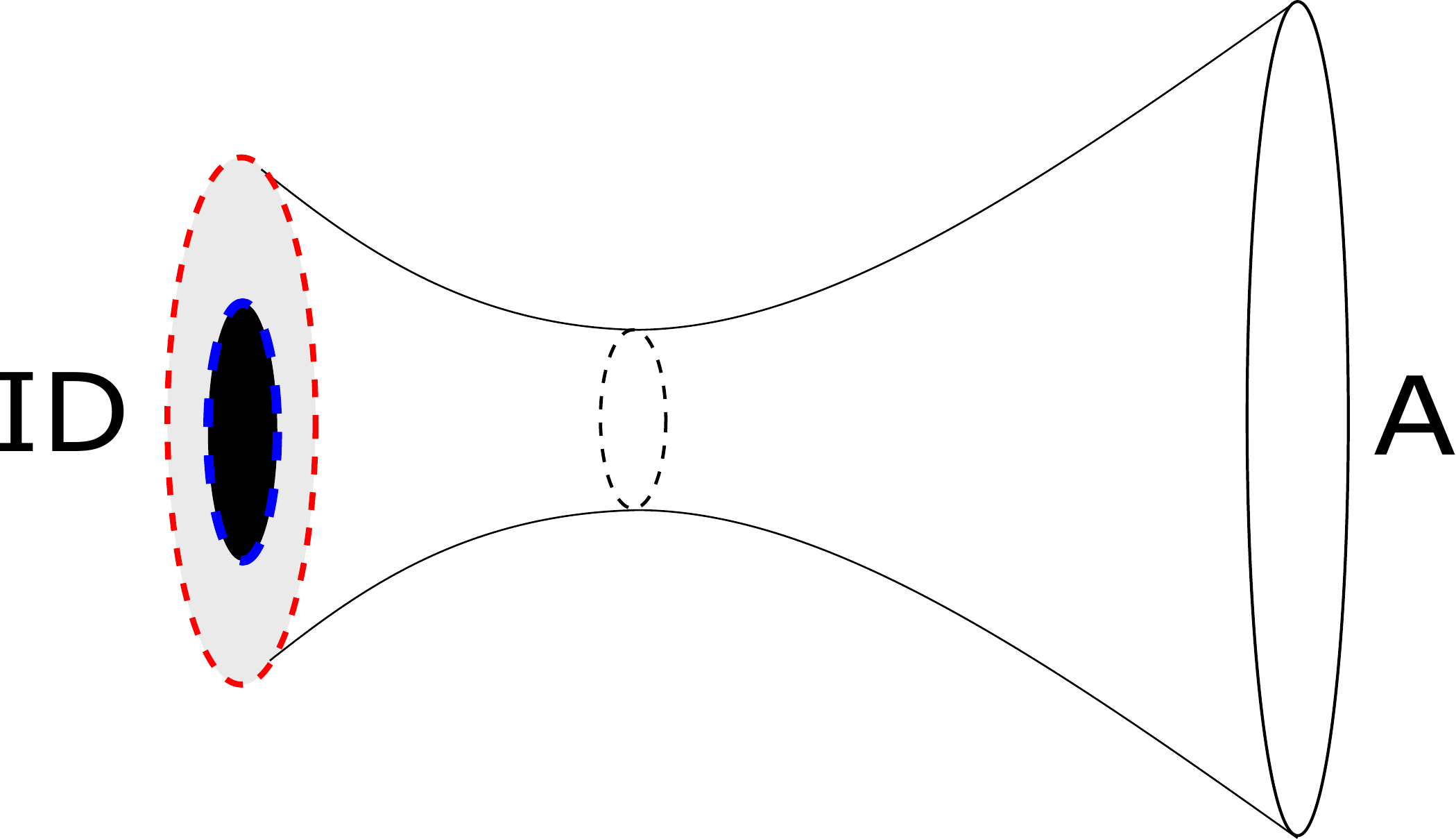}
    \\
        ({\bf c}) 
    \caption{\small ({\bf a}) The time-reflection-symmetric slice of the eternal BTZ black hole is a wormhole between two asymptotic regions, $A$ and $B$ .   The dashed circle is the bifurcate horizon.  This geometry is dual to the thermofield double state. ({\bf b}) The two-sided wormhole with one side truncated and replaced with an EOW brane (red circle) at a finite distance from the bifurcate horizon.  In our model the EOW brane has an internal state structure which matches that of a 2d CFT. ({\bf c}) The EOW brane can be replaced with its holographic dual geometry.    When the brane CFT is  in a thermal state above the Hawking-Page transition, this geometry is a black hole within the inception disk (ID, gray) which is glued to the original geometry at the location of the brane (dashed red circle).  The  entropy associated with the brane is proportional to the area of the black hole horizon in the inception disk.  This ``inception horizon''(dashed blue circle) is an extremal surface  homologous to the asymptotic region $A$.  As such, it competes with the usual bifurcate horizon (dashed black circle) when we use the RT formula to search for the minimal surface which computes the entropy of $A$.     In other words, the homology constraint in the RT formula can be satisfied by pulling curves through the circle where the inception disk meets the real geometry.}
    \label{fig:tfd}
\end{figure}

Recent work \cite{Almheiri:2019hni} has clarified the problem in the simplified setting of two-dimensional Jackiw-Teitelboim gravity coupled to a non-gravitating 2d CFT representing radiation degrees of freedom distant from the black hole.  This work suggests that the rectification of the Page curve is already visible in the semiclassical theory if the entanglement entropy $S_{A}$ of  Hawking radiation collected in a weakly- or non- gravitating region $A$  is actually computed by a new ``island formula":
\begin{equation}
  S_{A} = \underset{B}{\text{min ext}} \left[ \frac{{\rm Area}(\partial B)}{4G_{N}} +S^{{\rm eff}}_{AB} \right]
  \label{eq:island}
    \end{equation}
where $B$ is an ``island" in the gravitating region,  $ \text{Area}(\partial B)$ is the area of the boundary of the island, and $S^{{\rm eff}}_{AB}$ is the effective field theory entanglement entropy of  quanta in the union of the regions $A$ and $B$.  This formula was inspired by the quantum extremal surface formula for holographic entanglement entropy \cite{Engelhardt:2014gca}, and the papers \cite{Almheiri:2019psf,Penington:2019npb} showing new quantum extremal surfaces inside evaporating black holes. Technically, the authors of \cite{Almheiri:2019hni} demonstrated their ideas by taking the radiation CFT to be holographically dual to a three dimensional classical gravity in which entropies of subregions could be computed by the Ryu-Takayanagi (RT) formula \cite{Ryu:2006bv} in one higher dimension. 
In this context, the island formula ensures that entropy growth in  Hawking radiation terminates at the Page time.  Further work has demonstrated that the island formula follows when the radiation entropy is computed via the replica trick, provided certain novel Euclidean wormholes between replicas are included \cite{Almheiri:2019qdq,Penington:2019kki}.\footnote{See \cite{Rozali:2019day,Almheiri:2019yqk,Almheiri:2019psy,Chen:2019iro,Liu:2019svk,Liu:2020gnp} for further work on quantum extremal islands in evaporating black holes.}

In this paper, we examine these ideas in three dimensions with a negative cosmological constant, while following  \cite{Kourkoulou:2017zaj,Almheiri:2018ijj,Cooper:2018cmb,Penington:2019kki} to model black hole microstates as excitations of an End-of-the-World (EOW) brane that truncates the geometry behind the horizon of an eternal black hole (Figs.~\ref{fig:tfd}a,b).  Because we are working with 3d gravity, the EOW brane is 2-dimensional.  The EOW brane has a Hilbert space containing the black hole microstates, and we imagine this Hilbert space comes from a quantum theory living on the brane. 
We take this brane theory to itself be a (deformation of a) 2d conformal field theory (the Brane CFT, which is distinct from the CFT living on the asymptotic  boundary of the spacetime).  We then consider the holographic dual to the Brane CFT, ``filling in'' the EOW brane to give a complete 3d geometry. In the region behind the brane there is a different cosmological constant  related to the central charge of the Brane CFT, which is in turn associated to the number of black hole microstates.   We call the 3d geometry behind the brane the \emph{Inception Geometry}.\footnote{This terminology is inspired by the 2010 film \textit{Inception} (spoilers ahead).  In the context of the film, ``inception" refers to planting an idea in someone's mind.  Here we use it instead to refer to the dream within a dream, as we have constructed a geometry within a geometry.}  As an example, consider a situation where $k$ microstates (in some basis) are maximally entangled with  external radiation quanta. In this case, the state of the Brane CFT will be thermal; thus the Inception Geometry will itself contain a black hole. Then the total geometry has two horizons, one in the real space and one in the inception space (Fig. \ref{fig:tfd}c). We will realize this scenario by imposing a modified version of the Israel junction conditions \cite{Israel:1966rt} to glue the inception geometry to the real one.  This procedure will lead us to the most general way to glue two BTZ black holes with different temperatures, curvature radii, and Newton's constant, and no additional non-holographic stress energy on the gluing surface.

In this context, we propose a new form of the Ryu-Takayanagi (RT) formula, whereby entanglement entropy in the CFT dual to AdS space is computed by the area of extremal surfaces that are allowed to pass through the EOW brane into the Inception Geometry, and motivate the new rule by setting up a replica derivation. 
The surfaces computed with this form of the RT formula are models of Quantum Extremal Surfaces \cite{Engelhardt:2014gca,Penington:2019npb,Almheiri:2019psf}, where the bulk effective field theory contribution to the quantum gravity entropy is modeled by a contribution from the inception geometry.
We  apply our prescription to a setting where EOW brane microstates are maximally entangled with radiation that has escaped through a transparent AdS$_3$ boundary (marked $A$ in Fig.~\ref{fig:tfd}c) into an external reservoir. The CFT dual to the AdS geometry captures the physics of the black hole microstates but not the emitted radiation, and so it must display an entanglement entropy computed by the area of extremal surfaces homologous to the AdS boundary.  In our extended RT prescription  there are two competing extremal surfaces satisfying this homology condition: the horizon of the real black hole, and the horizon of the inception black hole, shown in Fig.~\ref{fig:tfd}c. The smaller area gives the entanglement entropy. We will see that this prescription recovers the predicted Page behavior of Hawking radiation. 
 
Our proposal helps to uncover new aspects of information recovery from  Hawking radiation. First, we can purify the inception black hole by entanglement of the microstates through a wormhole with an auxiliary system, which is naturally identified with the external radiation.  This construction gives a realization of the ER=EPR idea \cite{Maldacena:2013xja}.   Furthermore, we can split the auxiliary system (or, equivalently, the radiation) into multiple distinct parts.  Such a split can be modeled by purifying the inception black hole with a multiboundary wormhole \cite{Aminneborg:1997pz,Skenderis:2009ju,Balasubramanian:2014hda}.  Each leg of the wormhole corresponds to a different part of the Hawking radiation.  We focus in particular on a case where the radiation is split into two parts  corresponding to early and late time Hawking radiation which will be naturally separated by great distances on any equal time slice, and thus will not directly interact.  By studying our extended RT prescription in the total geometry (real + inception) we find a new class of extremal surfaces homologous to the spacetime boundaries (the ``infalling geodesics'') that do not coincide with horizons (see Fig.~\ref{fig:octopus_1}).  The existence of these new extremal surfaces has an interesting consequence:  the  entanglement wedge of any part of the radiation  contains a part of the interior of real black hole.  However, there can be a region inside the black hole which is not in the union of the entanglement wedges of any set of subsystems of the radiation, even though it is in the entanglement wedge of the union of subsystems.  This missing region corresponds to a shared secret that is embedded in the entanglement between radiation subsystems, and is only recoverable if we have access to all of them at the same time.  In this sense, Hawking radiation implements a quantum secret sharing scheme.


\section{Holographic inception for black hole microstates}
\label{section:ads3}

Consider the eternal BTZ black hole. On the time-reflection-symmetric slice this geometry has a spatial section with two asymptotic AdS$_3$ regions separated by a horizon (Fig.~\ref{fig:tfd}a) -- i.e., it is a two-boundary wormhole which acts like a black hole when observed from outside the horizon.  To set up the Hawking paradox we need two ingredients: (a) a model of the microstates, and (b) a model of the radiation system.  We  follow the trick of \cite{Kourkoulou:2017zaj,Cooper:2018cmb,Penington:2019kki} to model the microstates by an End-of-the-World (EOW) brane placed behind the horizon.  Schematically, the EOW brane cuts off the eternal black hole geometry, removing the second asymptotic region (Fig.~\ref{fig:tfd}b).  Modeling the radiation is tricky because the AdS geometry acts as a box confining
finite energy particles, so that the Hawking quanta will not escape, and rather come to equilibrium with the black hole.  This is awkward because if they remain in the gravitating geometry, their entanglement entropy should include a quantum gravity component that is difficult to compute.  We will follow \cite{Rocha:2008fe,Almheiri:2019psf,Penington:2019npb} to avoid this difficulty by imagining  transparent boundary conditions at the AdS boundary that allow  radiation to be collected in a non-gravitating reservoir just outside the AdS boundary.

Suppose that the black hole has formed from collapse of a shell of matter dropped in from the AdS boundary.  Then, at early times no Hawking quanta have been collected in the reservoir.  After the black hole forms, it radiates and the number of quanta collected in the reservoir increases.  Eventually we have a black hole of a certain horizon area entangled with radiation.  
We will model the Hawking quanta at a given time as occupying a $k$-dimensional subspace of the reservoir  Hilbert space ($\mathcal{H}_{R}$) which is maximally entangled with a $k$-dimensional subspace of the black hole microstates.  A unitary transformation can distill the entangled part of the microstate Hilbert space into a separate factor $\mathcal{H}_{B}$.  We can then write the total state in terms of  $|i \rangle_{R} \in \mathcal{H}_{R}$  entangled with   $|\psi_{i} \rangle_{B} \in \mathcal{H}_{B}$ as\footnote{Such states were previously considered in the context of black hole evaporation in \cite{Verlinde:2012cy,Verlinde:2013qya}.}
\beq
|\psi \rangle  = \sum_{i=1}^{k} \; |i \rangle_{R}  \otimes  |\psi_{i} \rangle_{B} \label{eq:HHstate} \, .
\eeq
When $k > e^{S_{\text{BH}}}$ the states $|\psi_{i} \rangle_{B}$ cannot be orthogonal because the microstate Hilbert space has dimension $e^{S_{\text{BH}}}$, where $S_{\text{BH}}$ is the coarse-grained black hole entropy. But when $k < e^{S_{\text{BH}}}$, we can take the $|\psi_{i} \rangle_{B}$ to be approximately orthogonal.    We can understand this orthogonality conceptually as follows.  Imagine a black hole in a particular microstate, and let it radiate some quanta. Each possible radiated configuration $|i \rangle$ should be in a product with a different  underlying microstate $|\psi_i\rangle$.  
Since the final state will be a superposition of such tensor products, it has the general form \eqref{eq:HHstate}.
Given the chaotic dynamics expected for black holes, it is reasonable to assume that the $|\psi_i\rangle$ will be random vectors in $\mathcal{H}_{B}$.   When $k$ is small, the required number of such random $|\psi_i\rangle$  is much smaller than the dimension of the microstate Hilbert space, and so they should be orthogonal with high probability. 

\subsection{Inception for microstates}
\label{sec:inceptionformicristate}
To gain further insight, the authors of \cite{Almheiri:2019hni}  used a trick: they imagined that the matter fields traveling through the black hole background and in the radiation reservoir formed a conformal field theory with a holographic dual.
We will instead take the theory on the EOW brane to be (a deformation of) a conformal field theory, the brane CFT.
This brane CFT is {\it different} from the CFT living on the asymptotic boundary of AdS, and in particular has a different central charge (see details in Sec.~\ref{sec:quantincep}).  In fact, there is reason to expect in string theory that the microscopic description of black hole microstates may generally occur through such  CFTs associated to D-brane sources (e.g., besides \cite{Strominger:1996sh} and \cite{Mathur:2005zp} see \cite{Balasubramanian:2007bs} for examples involving charged black holes in AdS$_5$). We will then assume that this brane CFT admits its own holographic dual.  The equal time slice of this dual theory will be a two-dimensional disk ``filling in" the EOW brane which lives on its boundary.  We will call this the ``Inception Disk".  Since the states of the EOW brane are entangled with the radiation, tracing out the radiation will leave the brane CFT in an approximately thermal state, which is dual to a black hole in the Inception Disk (Fig.~\ref{fig:tfd}c).   If the EOW microstates are maximally entangled with $k$ radiation quanta, the entropy of the inception black hole will be $\log k$. 

Thus, the total system is now described by a single three-dimensional geometry in which the cosmological constant changes across the EOW surface, since the central charge of the EOW brane CFT is different from the central charge of the CFT on the AdS boundary.   In addition, there are {\it two} horizons.  The first, associated to the asymptotic observers, is the original one of the black hole.  The second, associated to the microstates and any observer who directly interacts with them in the black hole interior, is the inception horizon.   \emph{We propose that the standard Ryu-Takayanagi prescription for entanglement entropies can be applied in the total geometry with extremal surfaces traveling into the inception region as necessary, subject to a refraction condition because of the changing cosmological constant, and to a condition that they are homologous in the {\it complete} geometry to the region whose entanglement entropy we are trying to compute.} Such surfaces can be thought of as models of Quantum Extremal Surfaces if we think about the bulk effective field theory contribution to the generalized entropy as the entropy of the brane segment that is captured by the part of the RT surface in the real side of the geometry. This entropy of the brane segment is geometrized by holographic inception and captured by the piece of the RT surface living in the inception side of the geometry. We will argue for this rule in such glued geometries using the replica trick in Sec.~\ref{sec:replica}.

\subsection{Quantifying the inception theory}
\label{sec:quantincep}

Strictly speaking, the EOW brane theory does not have to be a CFT all the way into the ultraviolet, and can be an irrelevant deformation of a CFT that introduces a cutoff on the spectrum, provided that the complete brane Hilbert space has at least $e^{S_{\text{BH}}}$ dimensions.  In practical terms, this means that the holographic dual of the EOW brane in the inception disk can have a finite cutoff near its boundary, and can perhaps be understood as a ${\rm T} \bar{{\rm T}}$ deformation of a CFT \cite{McGough:2016lol}.   This freedom will be important for physically gluing the original black hole to the inception geometry.



In a top-down approach we would derive the theory of the microstates and its holographic dual from the underlying quantum gravity. For example, if the black hole were created by a collection of intersecting D-branes, we would have to work out their effective theory.   We will instead consider consistency conditions that the inception geometry must satisfy: (1) it must produce a ``long wormhole''\footnote{By long wormhole we mean a wormhole which is lengthier than a standard wormhole and can have two interior horizons; in our context, such wormholes appear when we perform inception on an EOW brane.} region in the black hole interior leading to a second horizon, (2) it must have two adjustable parameters for the masses of the real and inception black holes in order to allow a tunable amount of entanglement with the radiation reservoir, and (3) the total geometry must solve the equations of motion so that we can define a generalization of the Ryu-Takayanagi formula.


We could follow several strategies to achieve these consistency conditions while gluing together black hole geometries with different temperatures.  We will take an approach in which the Newton constant in the real and inception geometries are different.   This is natural because the theory dual to the EOW brane can certainly have parameters that are different from those of the real spacetime.  With this approach we will find that the entire stress energy on the EOW brane can be taken to be the  holographic stress tensor coming from the inception disc.  Then an ``evaporation protocol'' that changes the horizon area of the inception black hole to reflect varying entanglement of the EOW brane  with radiation will also have to vary the inception curvature scale and Newton constant.  An alternative approach is to allow the EOW brane to contain a non-holographic component in its stress-energy, or non-trivial topology behind the horizon.  In these settings, which have more parameters, it is possible to define an evaporation protocol in which the curvature scale and Newton constant in the inception geometry remain fixed as we increase the entanglement with radiation.  We will analyze the simplest scenario without such  additional parameters, and will argue that all these models give similar physical results for the Page transition and secret sharing in Hawking radiation, essentially because the inception geometry is a robust representation of the overall entanglement structure of the states, and does not seek to capture the detailed properties.

\subsubsection*{Details of the inception geometry}

The physical and EOW brane CFTs are allowed to have different parameters, such as their central charges. Thus their holographic duals will have different curvature radii and Newton constants. We will denote quantities associated to the inception geometry with a prime: e.g., the original AdS radius will be $\ell$ and the inception AdS radius will be $\ell'$.

\begin{figure}
    \centering
    \includegraphics[width=0.5\textwidth]{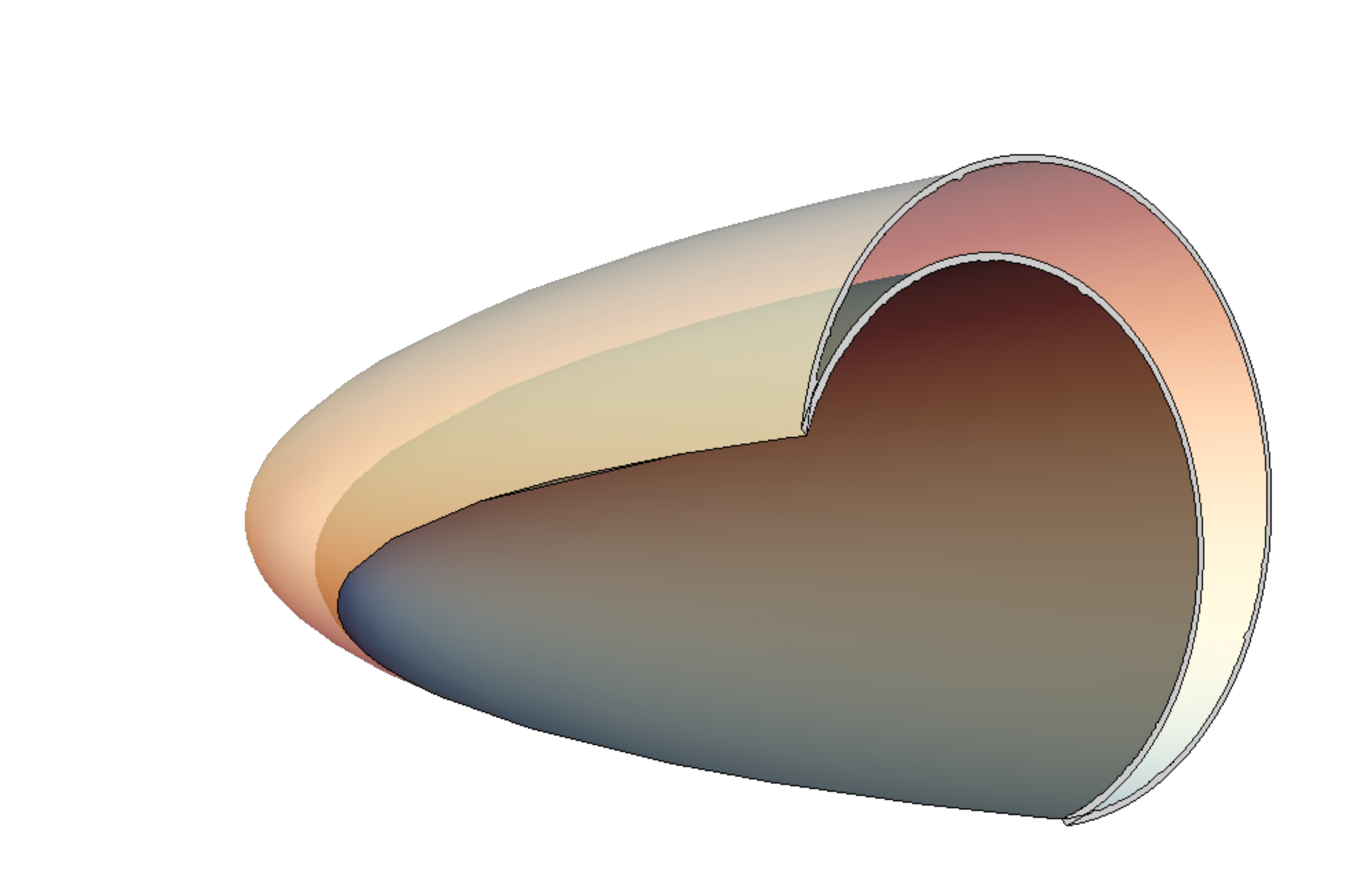}
    \caption{\small Visualization of the glued Euclidean geometry. The $\varphi$ direction is suppressed and the partial circles are the Euclidean time circles $\tau$ and $\tau'$.}    
    \label{fig:cigars}
\end{figure}

\begin{figure}
    \centering
    \includegraphics[width=0.35\textwidth]{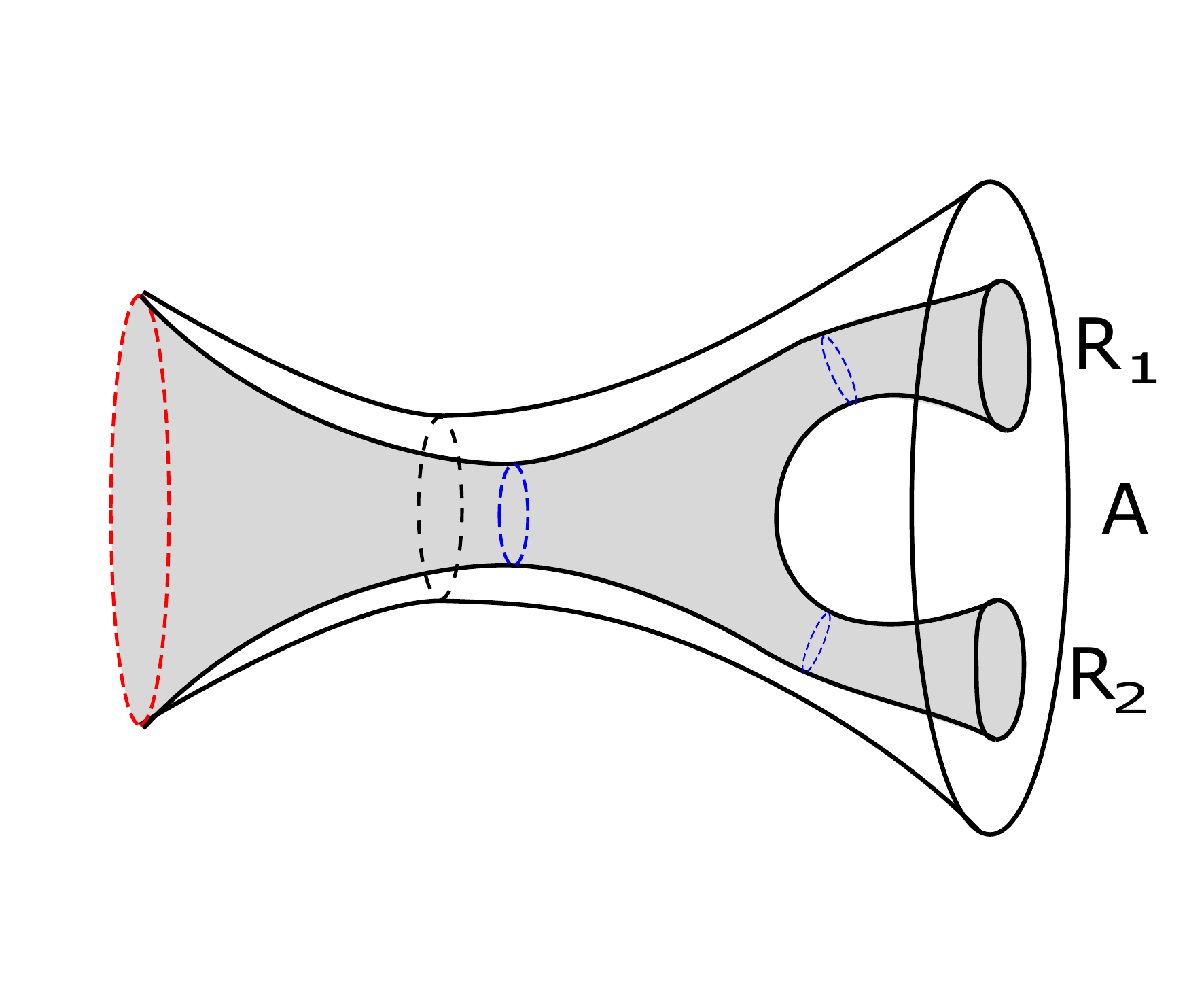} \hspace{0.5cm}
    \includegraphics[width=0.5\textwidth]{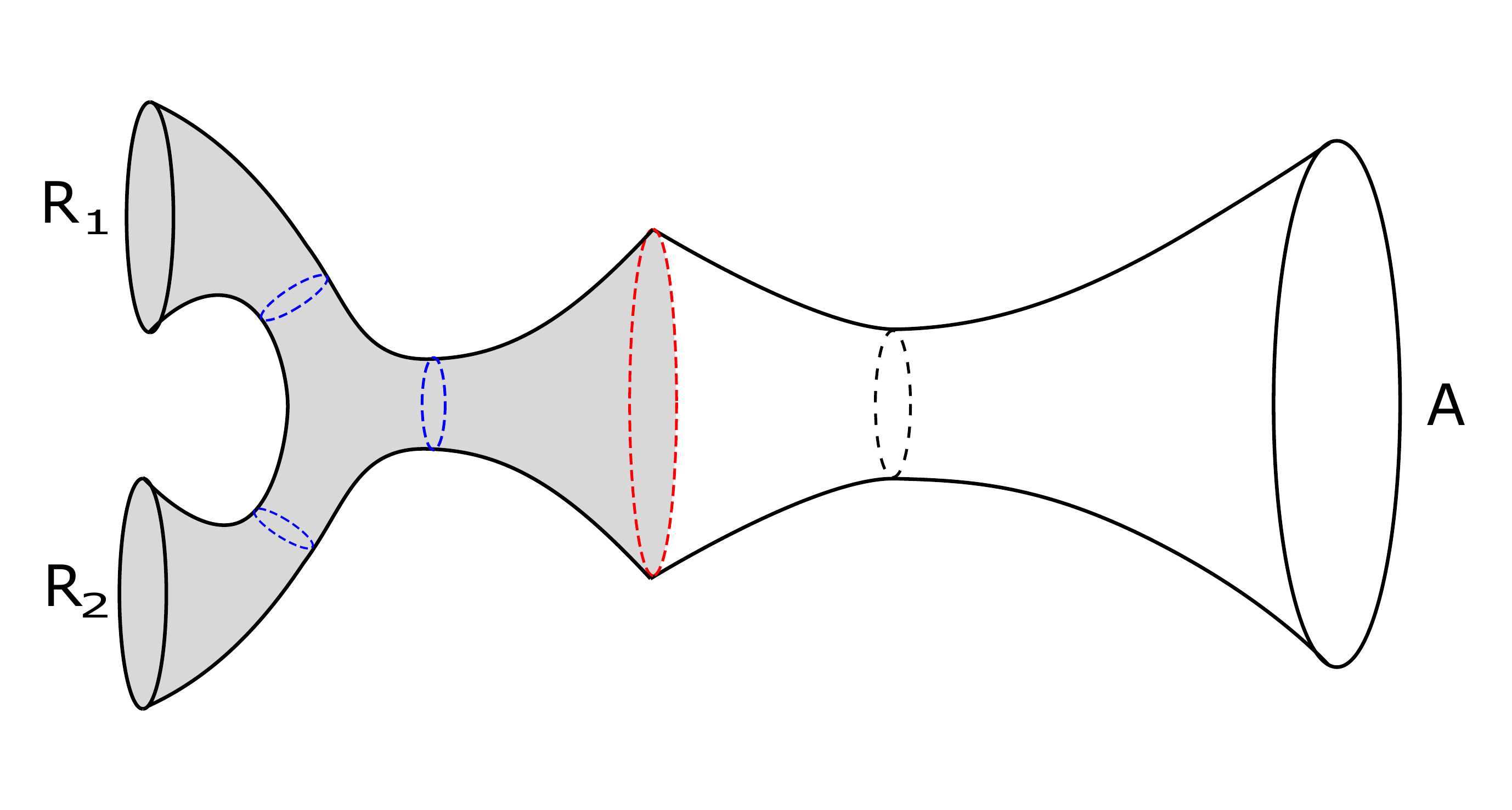}
    \caption{\small The spatial slice of the glued geometry when we purify the inception black hole with a two boundary wormhole. Left: since we glue convex surfaces to convex surfaces, the original boundary and the purifying systems are naturally on the same side. Right: in order to increase clarity, we ``unfold" the diagram on the left when we depict its various features.
    }
    \label{fig:octopus_0}
\end{figure}

The original (Euclidean) black hole geometry  is just described by the BTZ metric, 
\beq
\label{eq:btz}
ds^2=f(r)d\tau^2 + \frac{dr^2}{f(r)}+r^2d\varphi^2,
\eeq
\beq
f(r)=\frac{r^2-r_h^2}{\ell^2}, \quad \quad \beta= \frac{2\pi \ell^2}{r_h},
\eeq
where $r_h$ is the horizon radius and $\beta$ is the inverse temperature. 
We model microstates following \cite{Cooper:2018cmb,Kourkoulou:2017zaj,Hartman:2013qma}, by putting an end-of-the-world (EOW) brane behind the horizon, and we associate a state of the inception theory with the brane. In addition to just being a label on the brane, the choice of inception state will affect the brane trajectory.  We require the Brown-York stress tensor on the brane induced from the real side to be equal to the stress tensor of the state we pick in the inception theory. Since we work with Neumann boundary conditions, this condition will give an equation of motion for the brane trajectory. When the inception theory is holographic, it is natural to impose the condition that this stress tensor  agree with the Brown-York stress tensor induced from the inception side.\footnote{We can choose to include holographic counterterms \cite{Balasubramanian:1999re}, but then we need to include them on both sides so that we do not compare a renormalized stress tensor with a bare one. In this case they will not change the discussion below.} 
This motivates us to glue the real and inception geometries using the following junction conditions
\beq
\label{eq:metricjunction}
h_{ab}=h_{ab}',
\eeq
\beq
\label{eq:extrinsicjunction}
\frac{1}{G_N}K_{ab}=\frac{1}{G_N'}K_{ab}'.
\eeq
Here $h_{ab}$ is the induced metric and $K_{ab}$ is the extrinsic curvature on the gluing surface.
These conditions differ from the usual Israel junction conditions \cite{Israel:1966rt} in two important way. First, we can have $G_N\neq G_N'$ in which case the extrinsic curvatures must differ across the surface. Second, we will glue the geometries so that the orientations of normal vectors in \eqref{eq:extrinsicjunction} is the \textit{same}, that is, we glue convex surfaces to convex surfaces, while the Israel conditions glue convex surfaces to concave surfaces. In Euclidean signature, this leads to two cigar geometries that are glued in the way depicted on Fig.~\ref{fig:cigars}.\footnote{A similar Euclidean geometry was considered in \cite{Bernamonti:2018vmw} (see also \cite{Goel:2018ubv} for a JT gravity version leading to long wormholes), with a conventional convex to concave gluing. The main difference is that in our case there is no stress tensor localized on the shell, which is possible only because the two sides of the gluing have different parameters in their actions.}
In Lorentzian signature, we need to imagine the glued geometry as a folded piece of paper when embedded in higher dimensional space (see left of Fig.~\ref{fig:penrose}), so that the real and inception bulks are living ``on top of each other". While this choice of orientation is required to find interesting solutions for our purposes, it is also appealing as a model of both black hole complementarity and ER=EPR. Indeed, we can imagine that the radiation degrees of freedom in the real CFT $A$ are connected via ER=EPR bridges to the EOW brane behind the horizon. When we remove radiation from $A$ into the reservoir $R$, we also ``split" the bulk dual into two, and the removed ER=EPR bridges make up the inception part of the geometry, that lives on top of the real geometry, but after the splitting, only connects to it at the EOW brane (Fig.~\ref{fig:octopus_0}).

We will be interested in the situation when the microstates are entangled with the radiation, and so the inception geometry is itself a black hole.  The inception geometry is then also given by the BTZ metric \eqref{eq:btz}, with parameters $r_h'$ and $\ell'$. We will now find the most general rotationally invariant surfaces that solve the junction conditions \eqref{eq:metricjunction},\eqref{eq:extrinsicjunction} in this case.

We will work in Euclidean signature. The most general such surface is given by some trajectory $r(\tau)$. Pulling back the metric \eqref{eq:btz} we find that $h_{\varphi \varphi}=r^2(\tau)$. Then the equation $h_{\varphi \varphi}=h_{\varphi \varphi}'$ fixes the change of coordinate between the two sides to be $r(\tau)=r(\tau')$. It is therefore useful to change the coordinate $\tau$ to $r$ on the surface, which is a common coordinate. After the change of coordinates we have
\beq
h_{rr}=\frac{\ell^4 + (\frac{d\tau}{dr})^2(r^2-r_h^2)^2}{\ell^2(r^2-r_h^2)}, \quad \quad K_{\varphi \varphi}=-\frac{r(r^2-r_h^2)^{3/2}}{\ell \sqrt{\frac{\ell^4}{(\frac{d\tau}{dr})^2}+(r^2-r_h^2)^2}}.
\eeq
We can now solve the equations $h_{rr}=h'_{rr}$ and $K_{\varphi \varphi}/G_N=K_{\varphi \varphi}'/G_N'$ for the derivatives of the real and inception gluing surfaces. After integrating, the solution is
\beq
\label{eq:genbranetrajectory}
\tau(r) = \ell^2 \sqrt{\frac{r_t^2-r_h^2}{r_h^2-r_b^2} }\int_{r_t}^r d\tilde r\frac{\sqrt{{\tilde r}^2-r_b^2}}{({\tilde r}^2-r_h^2)\sqrt{{\tilde r}^2-r_t^2}}
\eeq
\beq
\tau'(r) = \ell'^2 \sqrt{\frac{r_t^2-r_h'^2}{r_h'^2-r_b^2} }\int_{r_t}^r d\tilde r\frac{\sqrt{{\tilde r}^2-r_b^2}}{({\tilde r}^2-r_h'^2)\sqrt{{\tilde r}^2-r_t^2}},
\eeq
(the prime on $\tau'$ denotes inception quantity and not derivative). We have defined in the above expressions
\beq
\label{eq:braneparameters}
r_t= \sqrt{\frac{\ell^2 G_N^2 r_h'^2-\ell'^2 G_N'^2 r_h^2}{\ell^2 G_N^2-\ell'^2 G_N'^2}}, \quad \quad r_b = \sqrt{\frac{\ell^2 r_h'^2-\ell'^2 r_h^2}{\ell^2-\ell'^2}},
\eeq
where $r_t$ corresponds to the turning point where $\frac{dr}{d\tau}=0$, while at $r_b$, $\frac{dr}{d\tau}=\infty$.  To get a real solution for the brane trajectory we need $r_b < r_h, r_h'$ and $r_h,r_h'<r_t$. The integrals can be given in terms of elliptic functions, but we will not need their explicit forms. To write the EOW brane trajectory  we will need the turning point $r_t$, which is the location of the brane on the $\tau=0$ slice where we continue to Lorentzian signature.

The $rr$ component of the extrinsic curvature reads as
\beq
K_{rr}=-\frac{(\frac{d\tau}{dr})^2\left( \frac{3 \ell^4 r}{(\frac{d\tau}{dr})^2}+(r^2-r_h^2)\left[\frac{\frac{d^2\tau}{dr^2} \ell^4}{(\frac{d\tau}{dr})^3}+r^3-r r_h^2\right]\right)}{\ell^3 \sqrt{r^2-r_h^2}\sqrt{\frac{\ell^4}{(\frac{d\tau}{dr})^2}+(r^2-r_h^2)^2}},
\eeq
and one can check that the solution \eqref{eq:genbranetrajectory}-\eqref{eq:braneparameters} automatically satisfies the last junction condition $K_{rr}/G_N=K'_{rr}/G_N'$.

\begin{figure}
    \centering
    \includegraphics[width=0.7\textwidth]{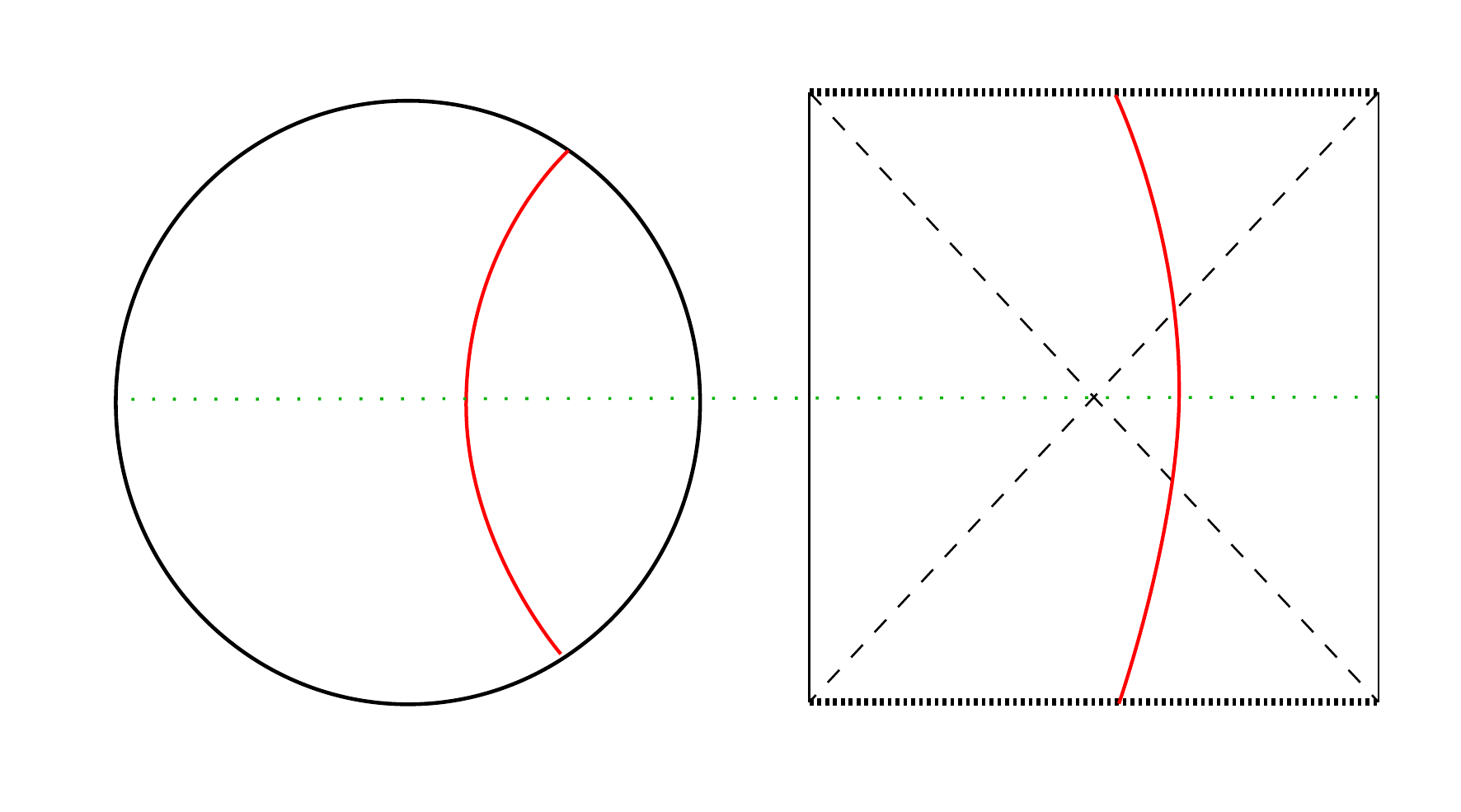}
    \caption{\small Illustration of a healthy brane trajectory in Euclidean (left) and Lorentzian (right). The two figures share the time reflection symmetric slice (green dotted).}    
    \label{fig:branetrajectory}
\end{figure}

It is important that a non-singular (without cusps or self-intersections) Euclidean brane trajectory should start from the boundary and return to the boundary after reaching the turning point $r_t$ (Fig.~\ref{fig:branetrajectory}). For this to happen, we need that $r_t>r_h,r_h'$, which are the origins of the Euclidean real/inception geometries, and that $r_t>r_b$ (to avoid singularities). A real trajectory in \eqref{eq:genbranetrajectory} then requires $r_b<r_h,r_h'<r_t$ (or just $r_h,r_h'<r_t$ when $r_b$ is purely imaginary). These inequalities give constraints on the possible choice of parameters.\footnote{For future reference we note that we can trade the parameters $G_N'$, $\ell'$ of the inception theory for $r_t$ and $r_b$. They are re-expressed as
\beq
\label{eq:primeswithradius}
\ell'=\ell \sqrt{\frac{r_h'^2-r_b^2}{r_h^2-r_b^2}}, \quad \quad G_N'=G_N \sqrt{\frac{(r_h^2-r_b^2)(r_t^2-r_h'^2)}{(r_h'^2-r_b^2)(r_t^2-r_h^2)}},
\eeq
and are real when $r_b<r_h,r_h'<r_t$.
} The spatial slice of the glued geometry is given in Fig.~\ref{fig:tfd}c, and the Penrose diagram after continuation to Lorentzian looks like Fig.~\ref{fig:penrose}. Note that later on, we will only use the time reflection symmetric slice of this geometry.

The solutions \eqref{eq:genbranetrajectory} include the constant tension brane ($K_{ab}=T h_{ab}$) as a special case when $r_b=0$ or $r_h'=\frac{\ell'}{\ell}r_h$. In this case, the brane trajectory has an elementary form \cite{Cooper:2018cmb}
\beq
\label{eq:branetrajectory}
r(\tau)=r_h\sqrt{\frac{1+\ell^2 T^2 \tan^2(\frac{r_h}{\ell^2} \tau)}{1-\ell^2 T^2}} \, ,
\eeq
where
\beq
T=\frac{G}{\ell \ell'}\sqrt{\frac{\ell'^2-\ell^2}{G^2-G'^2}},
\eeq
and the trajectory on the inception side is obtained by swapping primed and unprimed parameters.
 One can calculate the induced metric on the brane, and it describes a Big Bang-Big Crunch cosmology. It has a simple form if we introduce the new time coordinate $\hat \tau=\frac{\ell}{r_h}\arctan \left(\ell T \tan \frac{r_h \tau}{\ell^2} \right)$, in which it reads as
 \beq
 \label{eq:induced}
 ds^2 =\frac{r_h^2}{(1-\ell^2 T^2)\cos^2(\frac{r_h \hat \tau}{\ell})}(d\hat \tau^2 + d\varphi^2).
 \eeq
 This is Euclidean AdS$_2$ in global coordinates, with $\hat \tau \in (-\frac{\pi \ell}{2r_h},\frac{\pi \ell}{2r_h})$ playing the role of the global spatial coordinate of AdS$_2$, and $\varphi$ playing the role of the time of AdS$_2$.

In the above construction, we will allow the black hole in the inception geometry to have an entropy bigger than that of the original black hole in the real geometry.
From the microscopic perspective the logic of this scenario may seen puzzling.  How is it possible to conceive of an inception black hole with entropy bigger than the coarse-grained entropy of the original black hole if the EOW brane theory is supposed to be modeling the black hole microstates?  Indeed, from the microscopic point of view, the brane theory modeling the black hole microstates must have a Hilbert space dimension set by the Bekenstein-Hawking entropy, preventing any information paradoxes.  However, here we are considering an effective field theory situation in which the entropy of the Hawking quanta seems to rise indefinitely because subtle correlations are not included in the low-energy calculation.  From the black hole interior point of view this means that we must correspondingly imagine that a very large Hilbert space can hide behind the black hole horizon, and that these states are entangled with the Hawking radiation giving rise to its thermal character.   In effect, this means that in (\ref{eq:HHstate}) we continue to take the microstates $|\psi_i\rangle$ to be orthogonal in the effective description even when their number exceeds $e^{S_{BH}}$.  Below, we will show that if we use this ruse to try to generate a paradox in the effective theory, the Ryu-Takayanagi formula extended to include the presence of the EOW brane and its dual geometry will contrive to rescue the consistency of the theory \textit{without} the need for a microscopic description, an effect which we may refer to as \emph{entropic censorship}.

\subsection{Reproducing the Page curve: microstates, islands, and inception}
\label{sec:reproduce}

\begin{figure}
    \centering
    \includegraphics[width=0.32\textwidth]{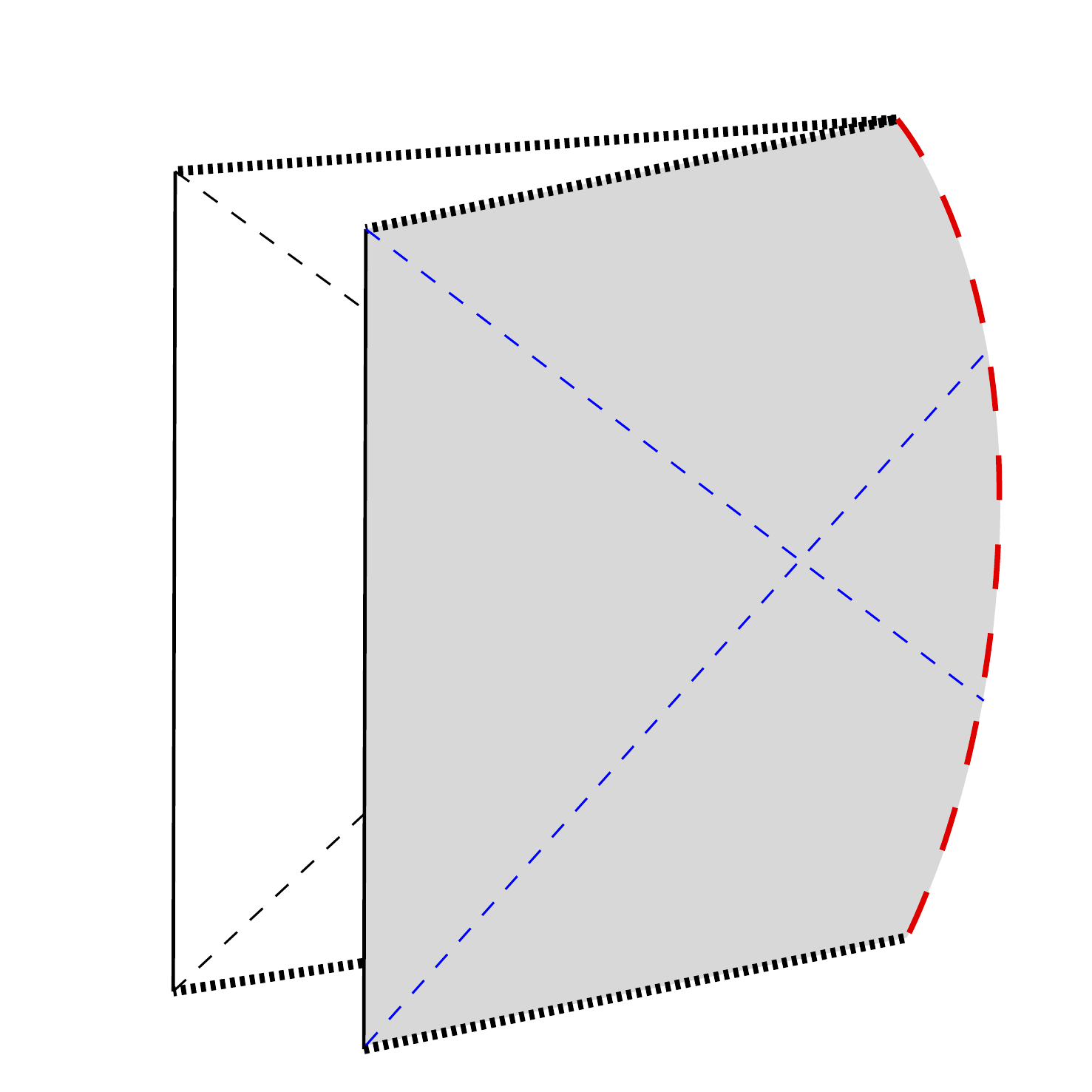}
    \includegraphics[width=0.52\textwidth]{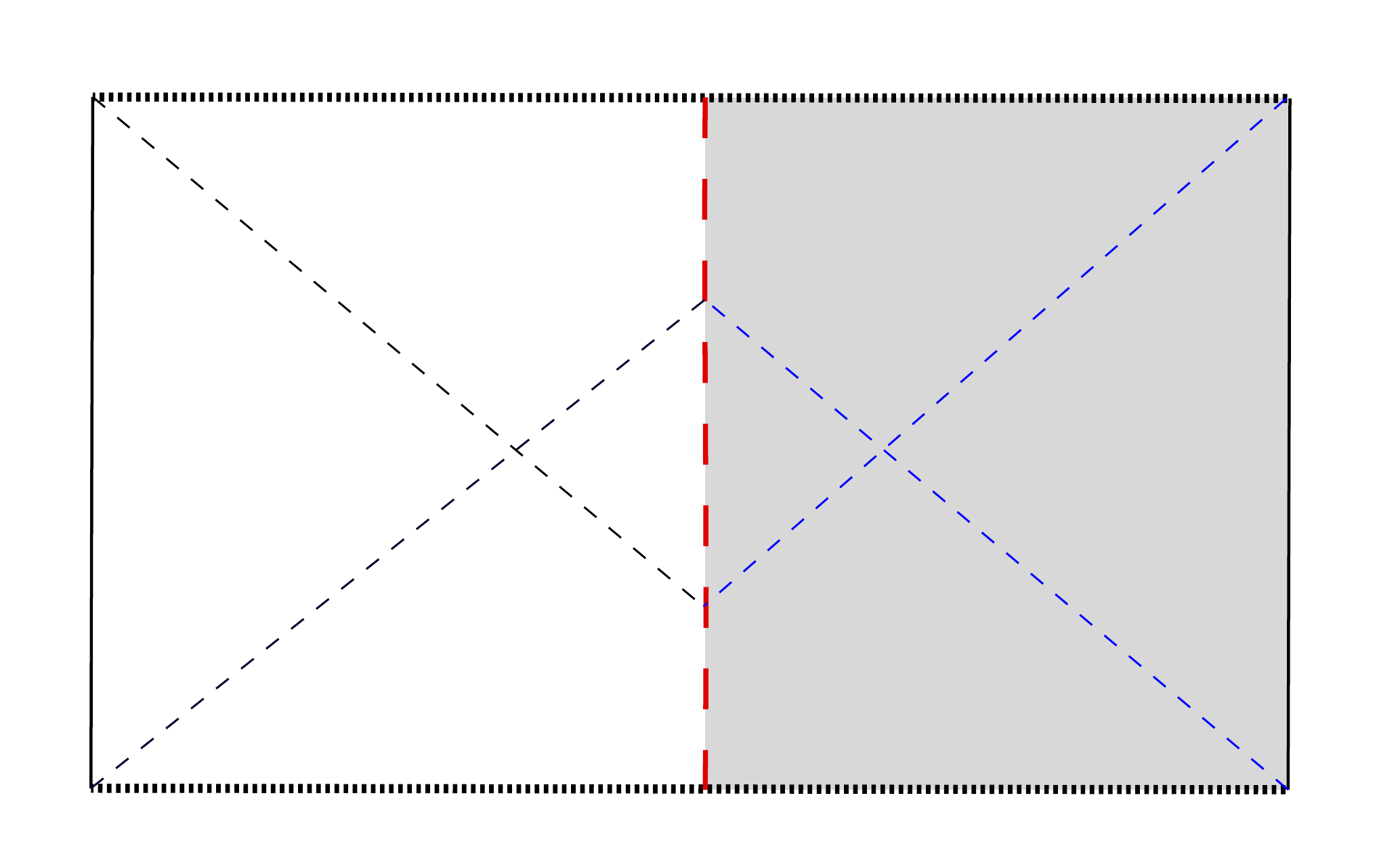}
    \caption{\small The maximally extended Penrose diagram of the spacetime after  inception. The real black hole spacetime (white) terminates at the 
EOW brane (red dashed line), then holographic inception creates the region behind (gray).
 The inception region also contains a black hole, due to the entanglement in the state \eqref{eq:HHstate}. Left: since we glue convex-to-convex, we need to imagine the diagram as a folded piece of paper. Right: the causal structure is better visualized when we unfold the diagram. Note that a conventional convex-to-concave gluing would not lead to a long wormhole: it would require us to delete the other side of the brane in the inception geometry.}    
    \label{fig:penrose}
\end{figure}

In our geometric description of the state \eqref{eq:HHstate} (Fig.~\ref{fig:tfd}c), we can simply compute the entanglement entropy of the theory on the AdS boundary by computing the Ryu-Takayanagi (RT) formula applied to extremal surfaces homologous to the entire AdS boundary.   
There are two such extremal surfaces in Fig.~\ref{fig:tfd}c -- the original horizon which gives entropy $S_{\text{BH}}$, and the inception horizon which gives entropy $\log k$.  This is related to the horizon size of the inception black hole $r_h'$ and the Newton constant  $G_{N}'$ in this region through 
\beq
\log k=\frac{\pi}{2}\frac{r_h'}{G_N'}  =\frac{2\pi c'}{3\beta'}, 
\eeq
where  $c'$ is the central charge of the brane CFT, and $\beta'$ is the temperature of the inception black hole.

Clearly when $k < e^{S_{\text{BH}}}$ the RT prescription selects the inception horizon as a measure of the entanglement entropy, but when $k> e^{S_{\text{BH}}}$ the original horizon is selected, giving $S_{\text{BH}}$ as the entanglement entropy.  In the dynamical scenario discussed above, this implies that the entanglement entropy of the radiation in this holographic computation will increase until it equals $S_{\text{BH}}$ and will plateau there.  Our prescription thus reproduces the result of the island formula \cite{Almheiri:2019hni}, in which the island would have been the region between the original horizon and the EOW brane.  However, in our setup there was no need to invoke an island.  The standard prescription for holographic entanglement entropy reproduces the result, perhaps giving an alternative justification for it in our three-dimensional setting.   From a slightly different perspective,  we can regard the ``islands'' of \cite{Almheiri:2019hni} as regions of space disconnected from the AdS boundary that can be reconstructed due to their entanglement with the radiation.  We will see that when enough radiation has been collected, an ``island" in this sense appears because the entire region behind the black hole horizon becomes part of the entanglement wedge of the radiation.

\begin{figure}
    \centering
    \includegraphics[scale=.3]{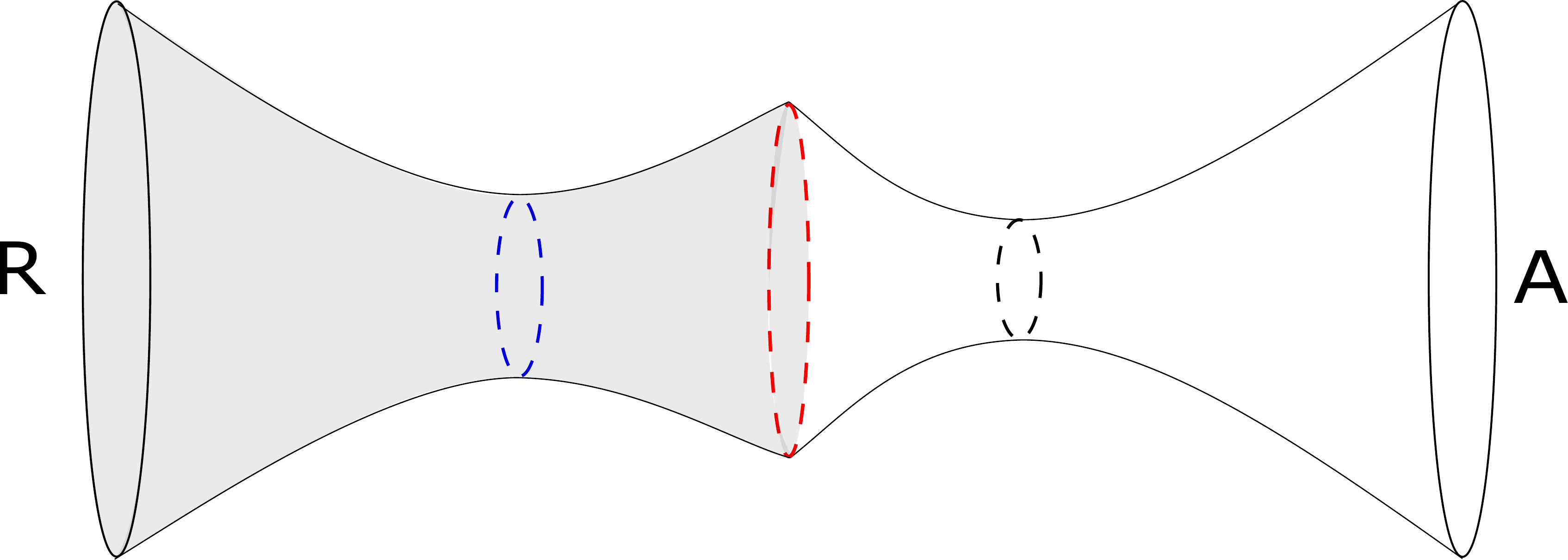}
    \includegraphics[scale=.3]{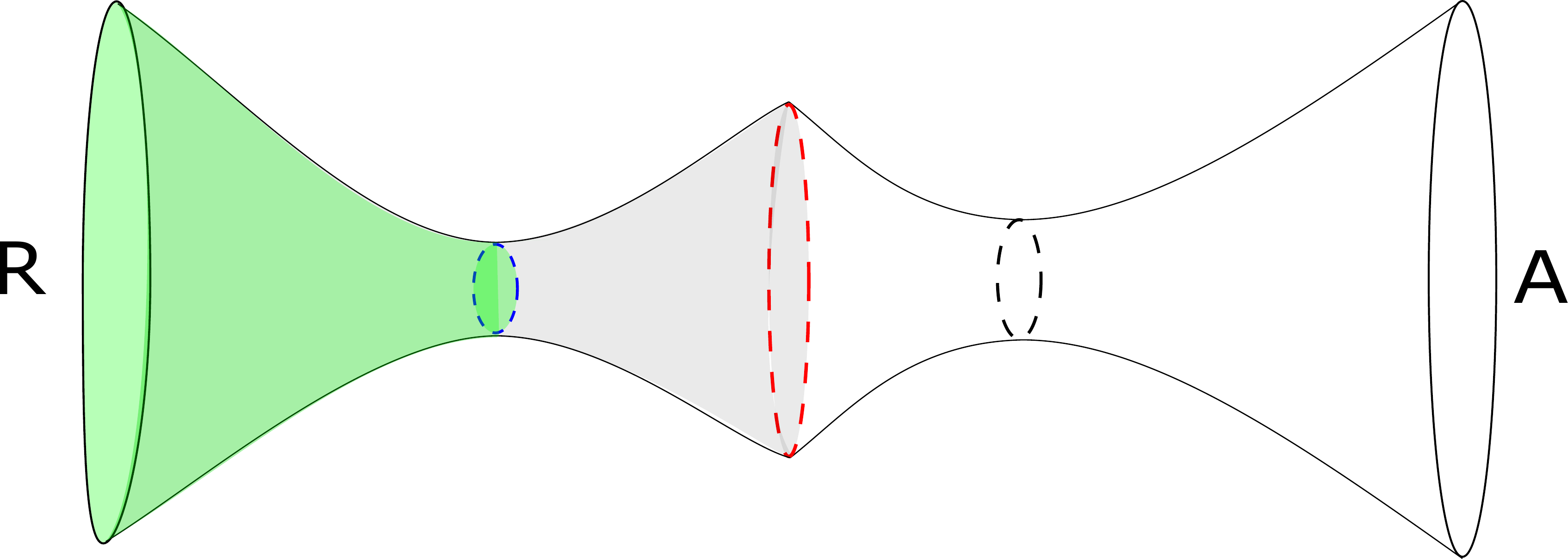}
    \includegraphics[scale=.3]{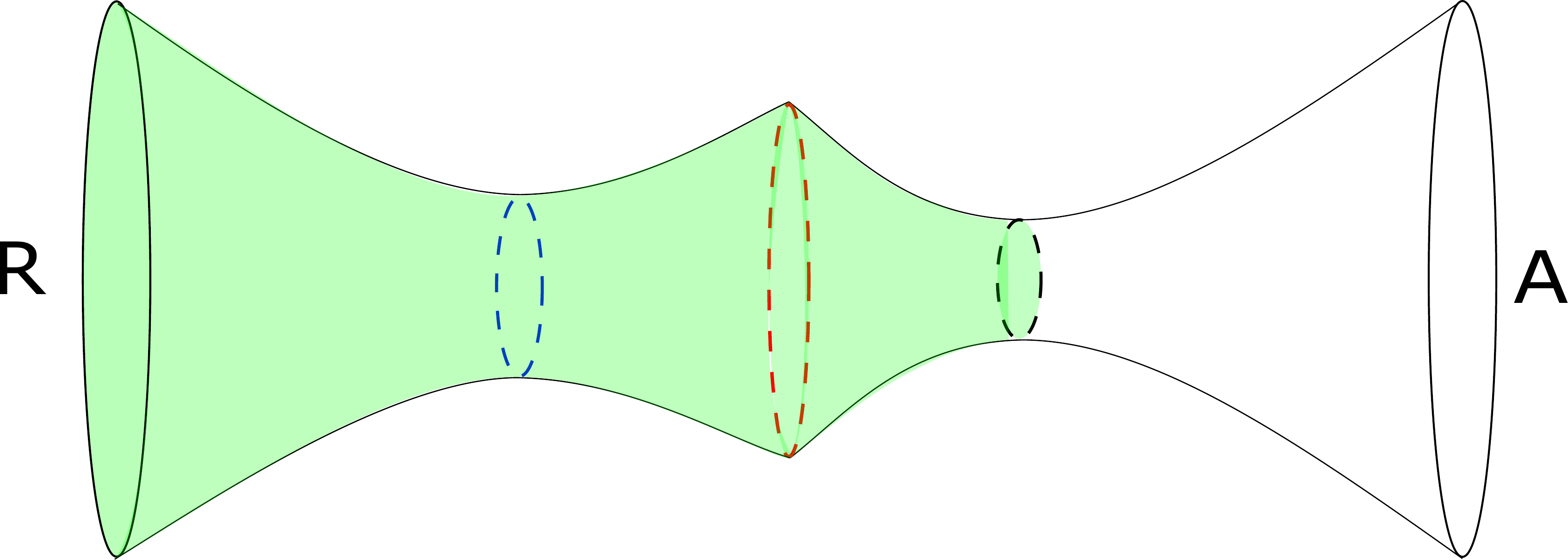}
    \caption{\small Top: the purification of Fig.~\ref{fig:tfd}c.
    The auxiliary radiation system (the reservoir) is identified with a new asymptotic boundary $R$ in the inception geometry.
    Middle: the entanglement wedge (green) of the radiation before the Page time when the entropy of the inception black hole is smaller than the entropy of the real black hole.
    Bottom: the entanglement wedge (green) of the radiation after the Page time when the entropy of the inception black hole is greater than the entropy of the real black hole.
    }
    \label{fig:purified-tfd}
\end{figure}
As discussed above, the inception geometry contains a black hole, and is dual to a thermal state in the brane CFT.   We can purify this state by introducing a thermofield double auxiliary  system.  It is natural to identify this auxiliary system with the radiation that purified the microstates in the first place in \eqref{eq:HHstate}.  Pictorially, this identifies the new asymptotic boundary of the inception wormhole with the reservoir where the original radiation was captured  (top of Fig.~\ref{fig:purified-tfd}) directly realizing the ER=EPR idea of \cite{Maldacena:2013xja}  (see the related discussion \cite{Verlindetalk}). 
Geometrically, this procedure corresponds to maximally extending the black hole in the inception disk beyond its horizon and through a wormhole to a second boundary (Figs.~\ref{fig:penrose}, \ref{fig:purified-tfd}). 
This construction effectively produces a two boundary ``long wormhole" which, following Sec.~\ref{sec:quantincep}, glues together two regions with different curvature scales and Newton constants.  The long wormhole has {\it two} extremal cross-sections -- one is the horizon of the original black hole, and the other is the horizon in the inception geometry. The causal horizon of the original black hole measures the coarse-grained entropy of the microstate after tracing out the radiation, and the causal horizon of the inception geometry measures the coarse-grained entropy of the radiation after tracing out the microstates.
It is the minimum of these two coarse-grained entropies which yields the entanglement entropy of the overall pure state after tracing out either factor.

Note that the brane CFT is related to the black hole interior and the infalling observer, while the radiation is measured by the asymptotic observer.  The identification between the inception wormhole and the reservoir implies that  dynamics on a part of the Hilbert space of the brane CFT is equivalent to the dynamics on a part of the Hilbert space of the radiation.  This could be regarded as a concrete manifestation of black hole complementarity \cite{Susskind:1993if}.  In particular, a measurement in the brane CFT would result in a projection in the radiation also, and vice versa, because the systems are maximally entangled.

When $k> e^{S_{\text{BH}}}$, an island forms in the sense of \cite{Almheiri:2019hni,Penington:2019kki,Almheiri:2019qdq}.  In this regime, the region between the real horizon and the inception horizon is no longer in the entanglement wedge of the physical boundary (region $A$ in Fig.~\ref{fig:purified-tfd}), because the RT surface is at the real black hole horizon.  The purity of the full quantum state then implies that the interior region must be reconstructible from the radiation.  The part of space that can be reconstructed from the radiation can be computed in our construction by looking at the entanglement wedge of the boundary of the inception wormhole (region $R$ in Fig.~\ref{fig:purified-tfd}) after using our extended version of the RT formula.   Below the Page transition ($k < e^{S_{\text{BH}}}$) the entanglement wedge of $R$ stops at the inception horizon since this has smaller area than the real horizon.  But, as the Page transition occurs (i.e. as $k$ begins to exceed $e^{S_{\text{BH}}}$), the entanglement wedge of $R$ extends through to the real horizon which now has a smaller area.  We can interpret this as saying that  a region in the real spacetime that is reconstructible from the radiation and which is disconnected from the boundary of space, i.e., an ``island'' in the sense of \cite{Almheiri:2019hni,Penington:2019kki,Almheiri:2019qdq},  forms between the real black hole horizon and the EOW brane.

This information recovery is sudden: when $k$ is smaller than $e^{S_{\text{BH}}}$, none of the interior can be reconstructed from the radiation, and when $k$ becomes larger than $e^{S_{\text{BH}}}$ all of the interior can be reconstructed, where by ``interior'' we  mean the region between the EOW brane and real black hole horizon (Fig.~\ref{fig:purified-tfd}).  

\subsection{Details of the evaporation protocol}\label{sec:protocol}

In order for the above-described picture of the Page transition to be compatible with the gluing conditions that we described in Sec.~\ref{sec:quantincep}, we need to make sure that we can change the entropy associated to the inception horizon in a way that the gluing surface remains real and non-singular. In other words, we need to maintain $r_b<r_h'<r_t$ during the process.  To mimic the evaporation process, we  increase $r_h'$ starting from some small value in order to increase the entanglement of the brane microstates with the radiation reservoir.  At $r_h'=r_h$, the gluing surface \eqref{eq:genbranetrajectory} becomes the horizon, which means that the EOW brane does not ``fit inside" the black hole anymore. Since the Page transition happens when $r_h/G_N=r_h'/G_N'$, we see that we need $G_N'<G_N$ in order to see the transition before $r_h'=r_h$.   While performing this protocol, we fix $\ell$ and $G_N$ on the real side, and hence the central change $c=3\ell/2G_N$ of the holographic dual, because the real space theory should be fixed through the evaporation.  In addition, we fix the ratio of central charges $\hat c=c/c'$, because $c'$ becomes the central charge of the radiation CFT after inception through a long wormhole, and we expect the radiation theory to also be fixed through the protocol.  However, we can let the inception parameters $\ell'$ and $G_N'$ vary during the process as long as their ratio, which determines $c'$, is fixed.    Equivalently, we can vary the position of the EOW brane $r_t$, which is a function of  $\ell'$ and $G_N'$ through \eqref{eq:braneparameters}. 

\begin{figure}
    \centering
    \includegraphics[scale=.45]{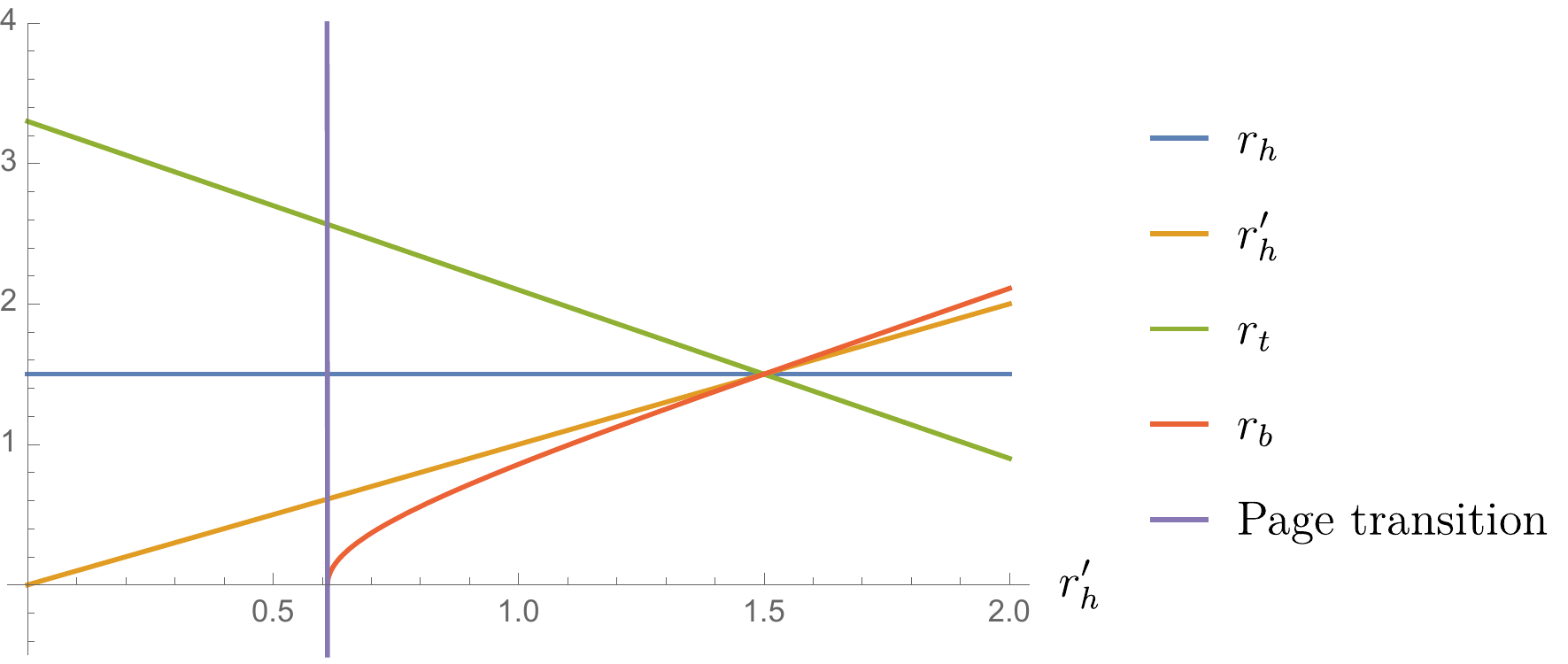}  \includegraphics[scale=.45]{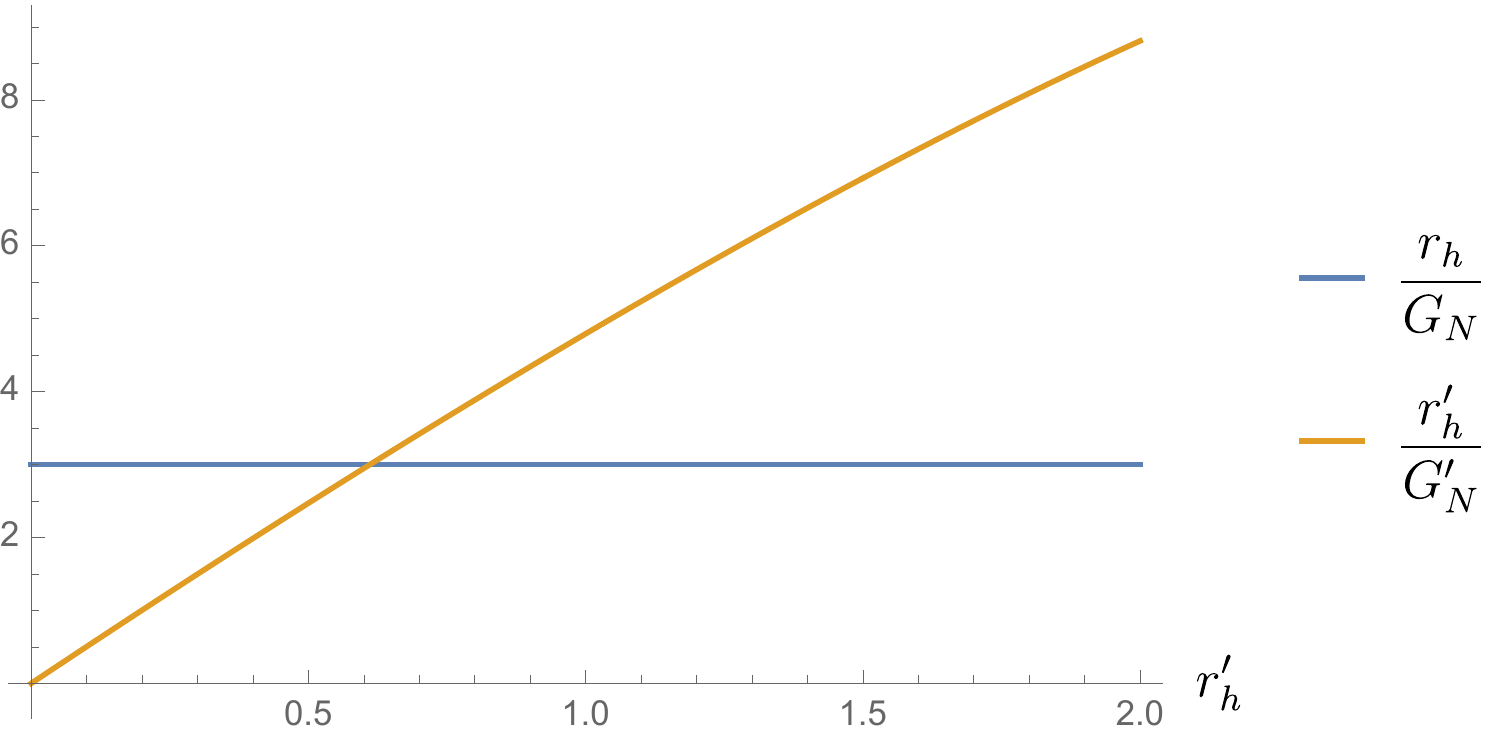}
    \caption{\small Left: horizon radii, the brane location $r_t$ and the scale $r_b$ as a function of $r_h'$ in the protocol described in the main text. We have $r_b<r_h'<r_h<r_t$, i.e. a healthy gluing surface in the neighborhood of the Page transition. Right: the Bekenstein-Hawking entropies associated to the causal horizons during the same protocol. The orange curve is not a straight line because $G_N'$ slightly changes during the protocol.}    
    \label{fig:protocol1}
\end{figure}

We will fix this one-parameter ambiguity, to define an evaporation protocol with non-singular brane trajectories, by requiring that brane position $r_t$ changes from some fixed value to the horizon size $r_h$ as the inception horizon increases from $r'_{h}$  to $r_{h}$.  In fact, the equation of motion of the brane \eqref{eq:braneparameters} enforces that $r_t = r_{h}$ when  $r'_{h} =r_{h}$, so that the choice here is the initial value of the brane location and the subsequent trajectory during the protocol.  As discussed above, the variation of $r_t$ is equivalent to a variation of $\ell'$ and $G_N'$  with $c'$ fixed through \eqref{eq:braneparameters}.   We use the choice of $r_t$ during the evaporation to enforce $r_b<r_h'<r_t$ as we change $r_h'$.  A simple way to achieve this is the linear dependence
\begin{equation}
r_t=r_h+\alpha(r_h-r_h'),
\label{eq:lineardependence}
\end{equation}
with $\alpha>0$ (Fig.~\ref{fig:protocol1} ).\footnote{Note that as long as $r_t(r_h)=r_h$ and the derivative at this point is finite any form of the function $r_t(r_h')$ seems to give a non-singular evaporation protocol.} Fig.~\ref{fig:protocol1} shows the dependence of the real and inception horizon entropies $S_{\rm real} = 2\pi r_h/4G_N$ and $S_{\rm inception} = 2\pi r_h'/4G_N'$ as we change the protocol parameter $r_h'$. Note that the inception entropy changes non-linearly with $r_h'$ since $G_N'$ is changing also. Fig.~\ref{fig:protocol1} also shows the location of the Page transition where the inception and real entropies exchange dominance.   An  appealing feature of this protocol is that it turns out $r_b$ is imaginary below the Page transition so that brane trajectory \eqref{eq:genbranetrajectory} remains well-defined. Intriguingly, $r_b$ is zero at the transition and then becomes real.

In order for our protocol to cover the Page transition, which occurs when the inception and real black holes have equal entropy, 
the transition must happen before the horizons have the same size ($r_h'=r_h$). This means that we need the condition $G_N'<G_N$.
Plugging in the linear dependence of $r_t$ \eqref{eq:lineardependence} into the relationship \eqref{eq:braneparameters} and enforcing $G_N'<G_N$
gives the constraint $\hat c<\sqrt{\frac{\alpha}{1+\alpha}}$. So, in our protocol, the inception CFT has a bigger central charge than the real CFT $c'>c$. 


\subsection{Proof by replica trick}
\label{sec:replica}

\begin{figure}[!h]
    \centering
\begin{tabular}{c c}
 \includegraphics[height=4cm]{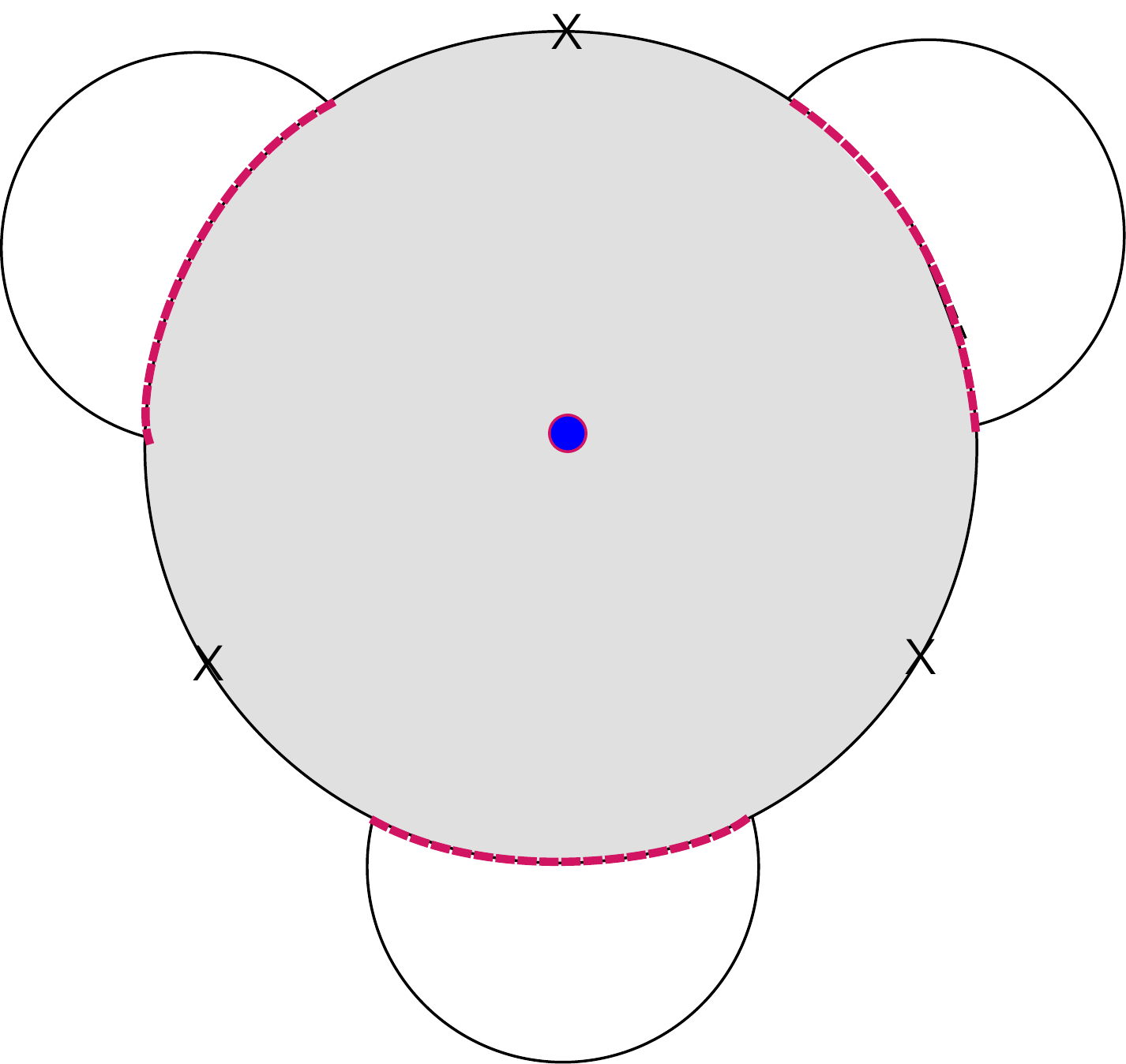}  & \includegraphics[height=4cm]{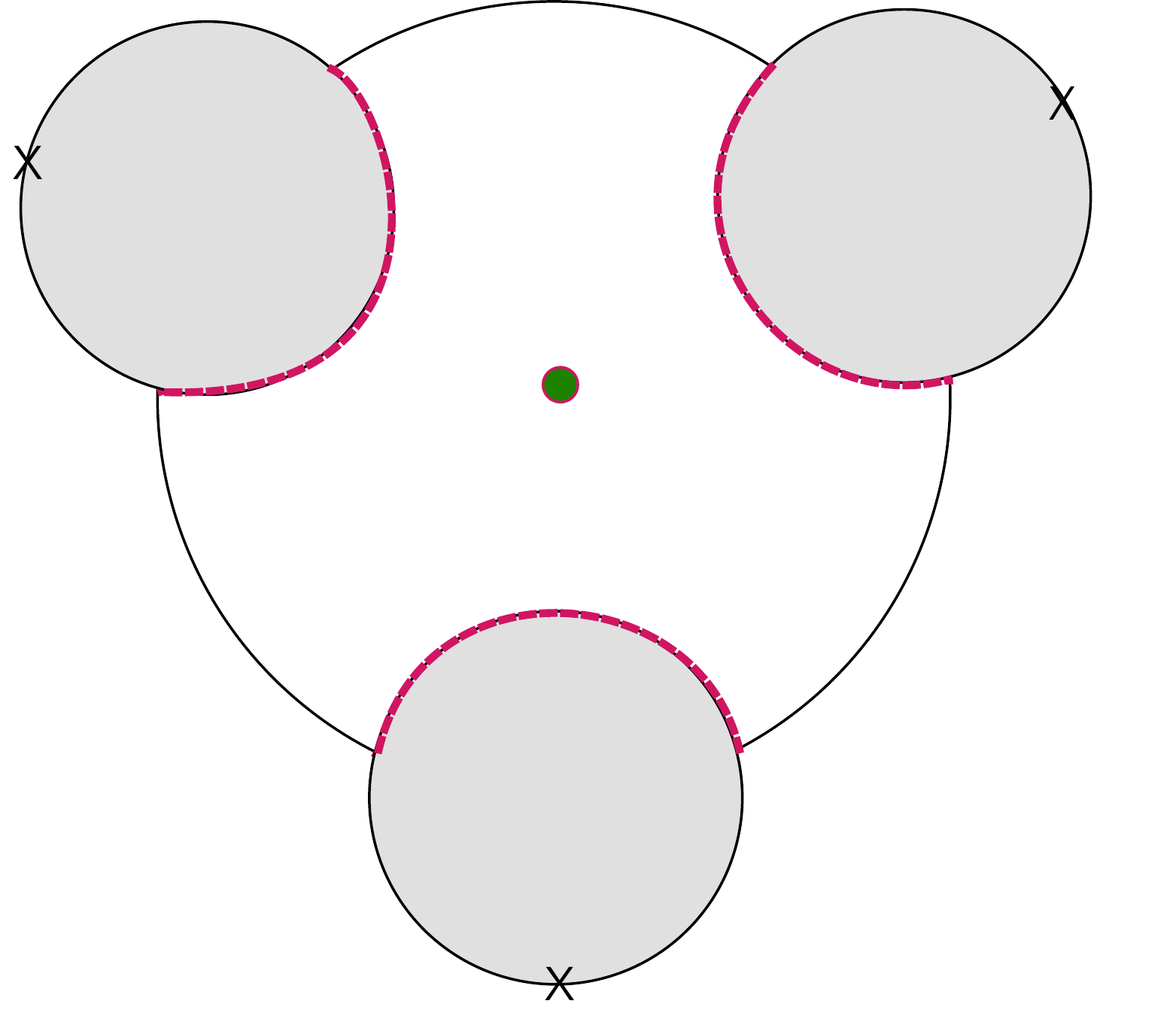} 
 \includegraphics[scale=.18]{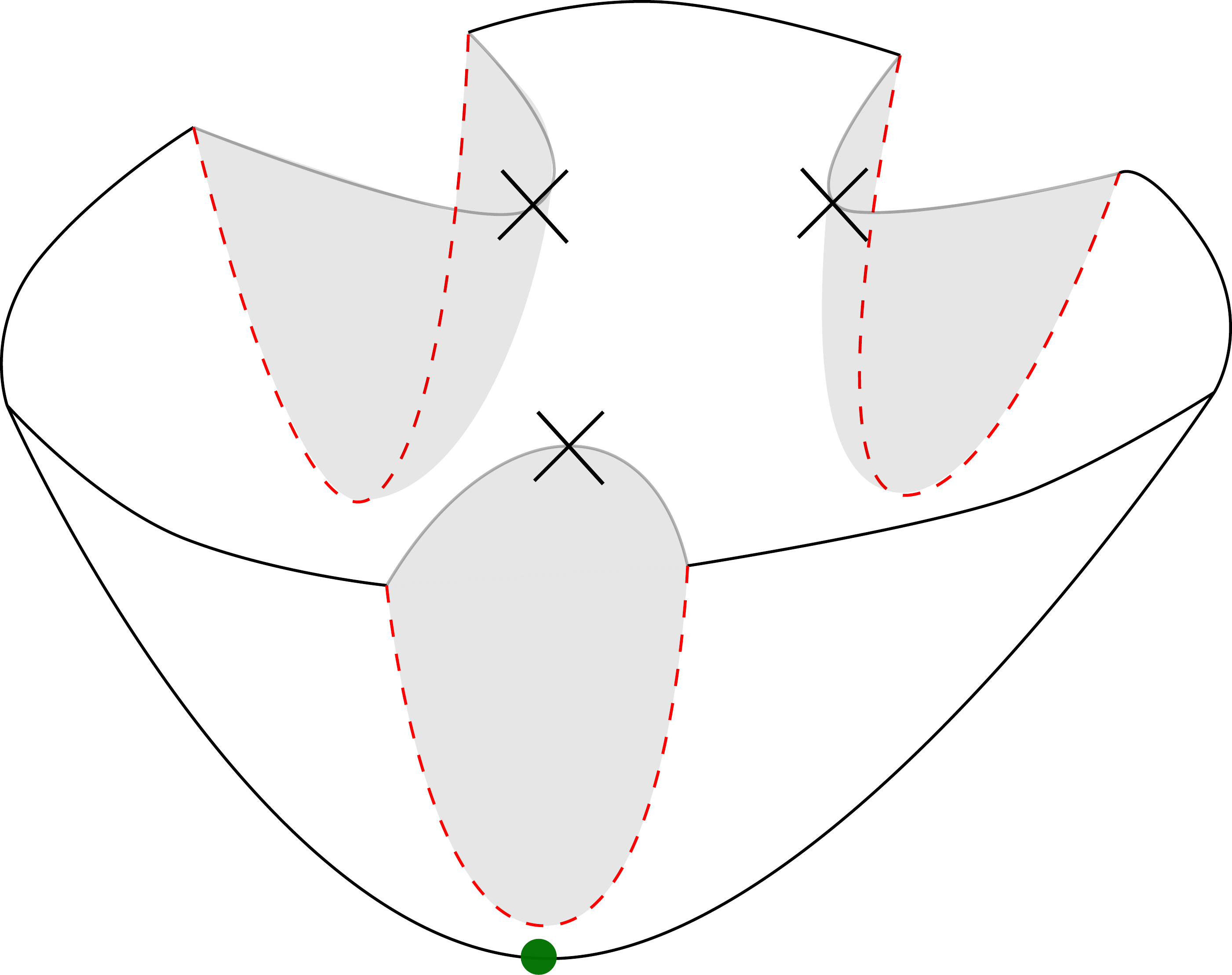}
\end{tabular}
\caption{\small{ 
({\bf Left} and {\bf Middle}) Two geometries which contribute to the calculation of the third R\'enyi entropy $\mathrm{Tr}\,\rho_{R}^3$ of the radiation in our setup. The white regions are the original black hole side of the spacetime, while the shaded regions are the inception side. The EOW branes are marked out by red dashed lines. 
Note that these are three-dimensional Euclidean spacetime geometries, we have simply suppressed the spatial circle, and deformed them into a two dimensional plane. The left geometry smoothly caps off at the RT surface (blue dot) in the inception black hole and dominates in the Hawking phase, while the middle geometry caps off smoothly at the RT surface (green dot) in the original black hole and dominates in the island phase. The crosses label points at which two copies of the asymptotic boundary are glued together in the replicated manifold. ({\bf Right}) A visualization of the two-dimensional surface in the middle panel, embedded in three dimensions instead of two, in the style of Fig.~\ref{fig:cigars}. }}
\label{fig:replica}
\end{figure}

We can give evidence for the extended RT formula that we proposed after inception by using the replica trick, following the ideas explained in \cite{Almheiri:2019qdq, Penington:2019kki}. In Fig.~\ref{fig:replica}, we have displayed two bulk geometries which contribute to the calculation of the third R\'enyi entropy $\mathrm{Tr}\,\rho_{R}^3$ of the radiation in our setup; note that they both have the same asymptotic boundary (corresponding to the original CFT and the radiation). Given the asymptotic boundary, the gravity calculation involves finding the various geometries which fill in this asymptotic boundary, with EOW branes appropriately separating the original black hole from the radiation side of the geometry. The asymptotic boundary to fill in is obtained the following way. We take the asymptotic boundary of the glued cigars of Fig.~\ref{fig:cigars} and cut open the arc corresponding to the radiation (inception) CFT along the time reflection symmetric slice, in order to define the density matrix. 
For the third R\'enyi, we take three copies of this, and cyclicly glue them together along the cut. There are two replica symmetric ways to fill in the resulting boundary. In the left geometry in Fig.~\ref{fig:replica}, the central gray disc denotes the cigar geometry (i.e. the Euclidean black hole) on the radiation side with asymptotic circle of length $3\beta_R$, with the three white regions denoting the individual cigar geometries on the original black hole side cut off by EOW branes (in red). Note that this is a three-dimensional Euclidean geometry (not a two-dimensional spatial slice). For this geometry, the $\mathbb{Z}_n$ symmetric point lies at the horizon of the radiation black hole, and thus from \cite{Lewkowycz:2013nqa}, the leading contribution to the entanglement entropy from this saddle is given by $A_{R}/4G_N'$. On the other hand, in the right geometry of Fig.~\ref{fig:replica}, the central white disc denotes the cigar geometry on the original black hole side with asymptotic circle of length $3\beta_{BH}$, with the three shaded regions denoting the individual cigar geometries on the inception side. In this case, the $\mathbb{Z}_n$ symmetric point lies at the horizon of the original black hole, and the corresponding contribution to the entanglement entropy is given by $A_{BH}/4G_N$. The true entanglement entropy is the minimum of these two:
\beq
S_{Rad.} = \mathrm{min}\,\left(\frac{A_R}{4G_N'}, \frac{A_{BH}}{4G_N}\right).
\eeq
Therefore, when $\frac{A_R}{4G_N'} > \frac{A_{BH}}{4G_N}$, we have a phase transition and the true RT surface is the horizon of the original black hole. This is precisely what we deduced previously from the generalized homology rule. Thus, our generalized homology rule follows from the Euclidean path integral for gravity, more or less the same way as shown in \cite{Almheiri:2019qdq, Penington:2019kki} for the island formula.

\section{Secret sharing in Hawking radiation }
\label{sec:eyelands}

Hawking radiation could have an intricate entanglement structure  with the black hole microstates \cite{Almheiri:2012rt,Mathur:2011wg,Shenker:2013yza, Brown:2019rox}, and also between subsystems of the radiation, e.g. between early and late time radiation that is spatially separated on a fixed late time surface.  Indeed, such correlations have long been suggested as a potential mechanism for information recovery from black holes (see  \cite{Almheiri:2012rt,Mathur:2011wg} and the review \cite{Balasubramanian:2011dm}). The correlations could also affect the way in which information is recovered from Hawking radiation -- for example, given a subset of the radiation of a given size we might be able to reconstruct all, some, or none of the observables in the black hole interior.

\begin{figure}
    \centering
    \includegraphics[width=0.8\textwidth]{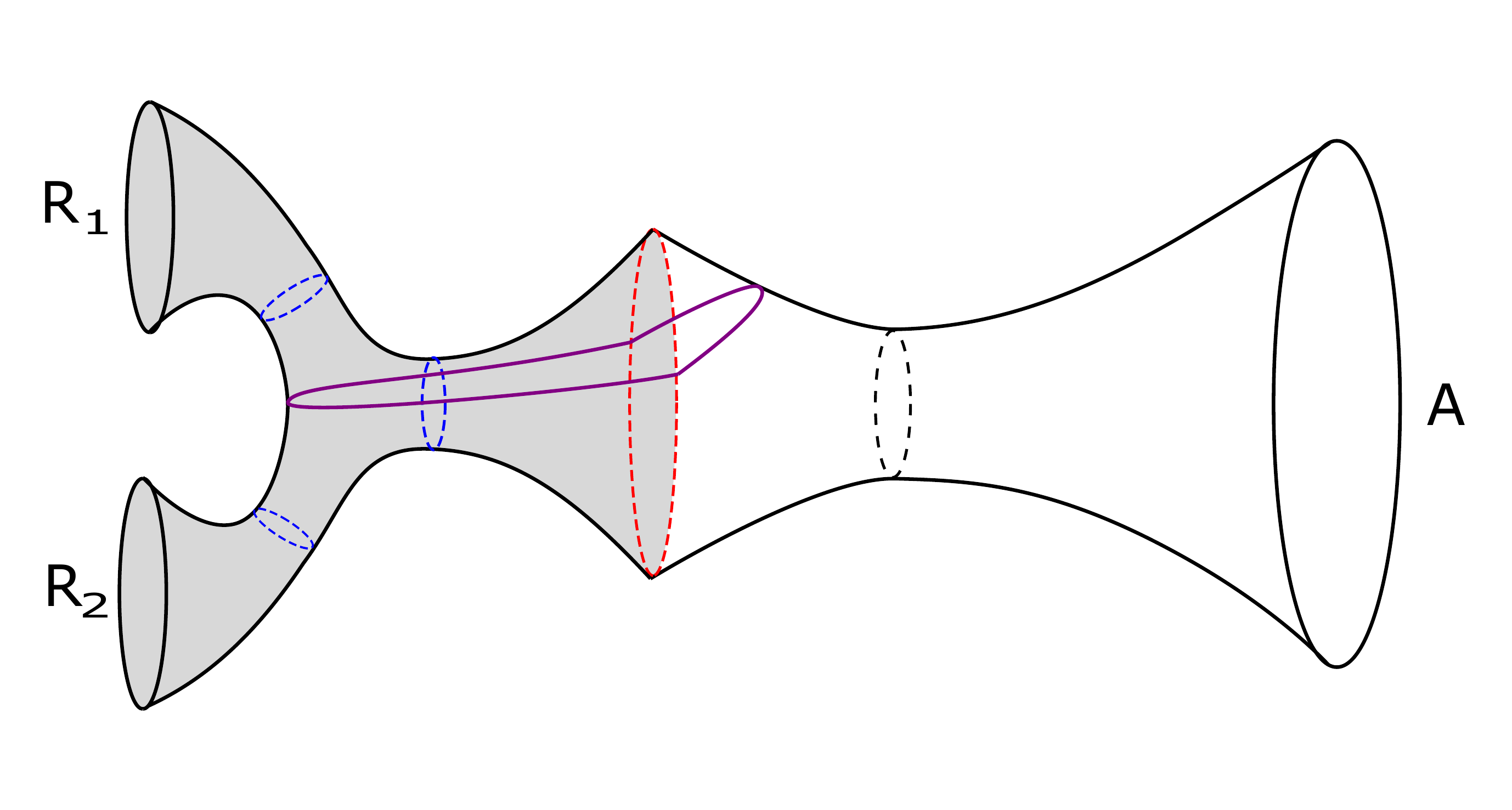}
    \caption{\small The purification of the inception disk by a three boundary wormhole with two radiation legs $R_1$ and $R_2$. The red dashed line is the EOW brane position, the rest of the dashed lines are causal horizons. There is a new RT surface drawn in purple that is homologous to $R_1$ and involves part of the total island that is between the EOW position and the horizon of the real black hole $A$.
    }
    \label{fig:octopus_1}
\end{figure}

To this end, imagine collecting  Hawking radiation in $n$ different detectors at asymptotic infinity. The radiation Hilbert space $\mathcal{H}_{R}$ thus factorizes into $n$ pieces: $\mathcal{H}_{R} = \mathcal{H}_{R_{1}} \otimes \cdots  \mathcal{H}_{R_{n}}$. The state of the combined black hole plus radiation can then be written as 
 \begin{equation}
     |\Psi \rangle = \sum_{i_{1} \cdots i_{n}} c_{i_{1} \cdots i_{n}} |i_{1} \rangle_{R_{1}} \otimes \cdots |i_{n} \rangle_{R_{n}} \otimes |\psi_{i_{1} \cdots i_{n}} \rangle_{B} 
     \label{eq:multibdystate}
 \end{equation}
where $|\psi_{i_{1} \cdots i_{n}} \rangle_{B}$ is the state of the original black hole. We seek to study the entanglement structure of \eqref{eq:multibdystate} using the inception geometry technique that we introduced above. The black hole microstate  $|\psi_{i_{1} \cdots i_{n}} \rangle_{B}$ can again be realized by  insertion of an EOW brane carrying a CFT behind the black hole horizon. As before, we consider the inception geometry -- the holographic dual of the EOW CFT state -- which ``fills in'' space behind the brane.   Because of the entanglement of the brane microstates with the radiation, we expect the inception geometry to contain a black hole, which we proceed to purify with an auxiliary system that can be identified with the Hawking radiation.  To model the partitioned state in (\ref{eq:multibdystate}), we first prepare $n$ CFT Hilbert spaces, and identify the $i^{\text{th}}$ radiation Hilbert space $\mathcal{H}_{R_{i}}$ with the $i^{\text{th}}$ CFT Hilbert space. 
We then take the purifying inception geometry to be a multiboundary wormhole connecting $n$ asymptotic AdS boundaries with these CFTs on them. The inception geometry is furthermore connected to the real black hole through the EOW brane (see Fig.~\ref{fig:octopus_1}). 
Multiboundary wormholes have been extensively studied, especially in AdS$_3$, for example in \cite{Krasnov:2003ye,Krasnov:2000zq,Skenderis:2009ju,Balasubramanian:2014hda}, and provide an interesting class of entanglement structures for the state \eqref{eq:multibdystate} which we can study using AdS/CFT.

We will first briefly review the construction of multiboundary wormhole geometries in AdS$_3$.  For the most part, we will focus on the simplest case, i.e. 
wormholes with three boundaries, and discuss the computation of entanglement entropy for one of the asymptotic boundaries (i.e., one of the radiation subsystems).  In this setting, we will demonstrate the existence of a new class of extremal surfaces, the ``infalling geodesics", which start from the inception side of the geometry, penetrate the EOW brane, and cross over to the interior of the original black hole (Fig.~\ref{fig:octopus_1}).  These new surfaces will lead to partial information recovery from subsets of the radiation. This partial recovery will be possible from islands of space that occupy part of the region between the EOW brane and the real black hole horizon.

\subsection{Covering space construction of multiboundary wormholes: review} 
\label{sec:covering}
\begin{figure}
    \centering
    \includegraphics[width=7.5cm]{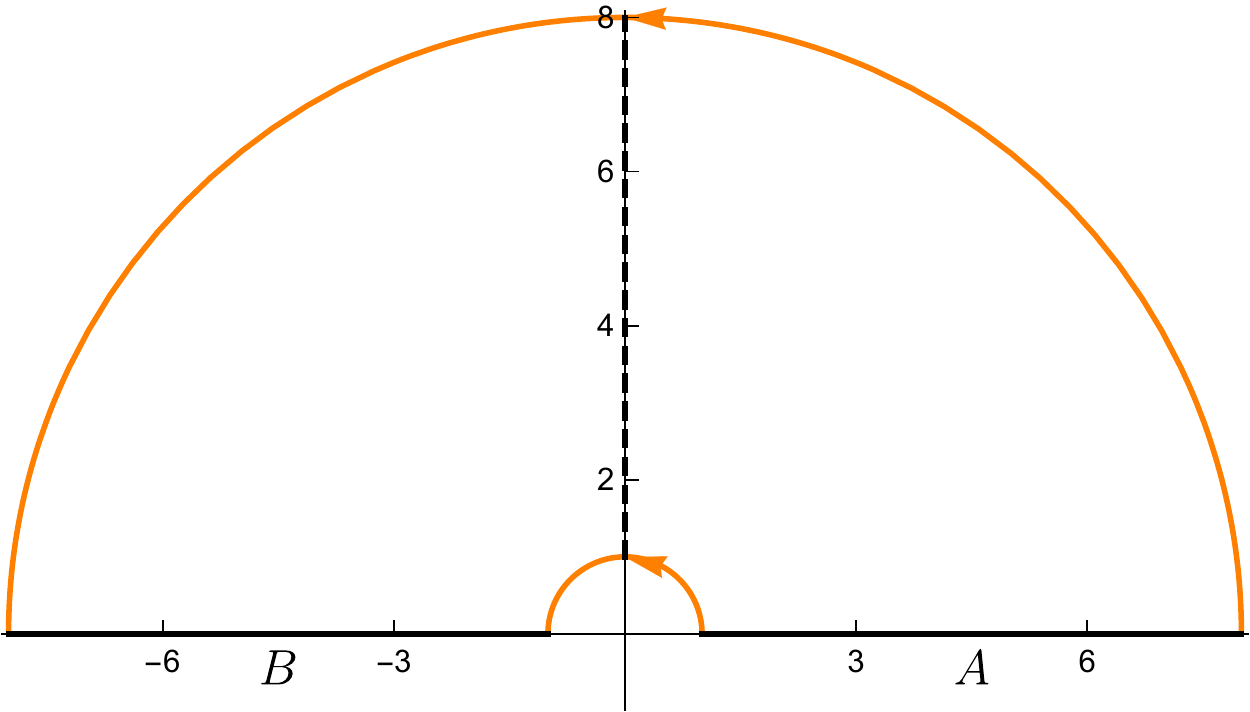}
    \caption{\small A covering space (upper half space $\mathbb{H}^2$) depiction of Fig.~\ref{fig:tfd}a, the time-reflection-symmetric slice of the eternal two-sided black hole.
    The orange geodesics are identified, and the black segments are asymptotic boundaries $A$ and $B$.
    The vertical dashed black line is the bifurcate horizon, the modulus of the real geometry.
    The region between the orange and black arcs is a fundamental region for $\Sigma_2$.
    }
    \label{fig:covering-tfd}
\end{figure}

Multiboundary wormhole geometries are vacuum solutions of Einstein's equations in three dimensions with a negative cosmological constant
They are constructed by quotienting the hyperbolic upper half plane $\mathbb{H}^2$ (which we will refer to as the covering space) by a discrete diagonal isometry subgroup $\Gamma \subset PSL(2,\mathbb{R})$ with hyperbolic generators.
The action of $\Gamma$ identifies pairs of boundary-anchored geodesics on $\mathbb{H}^2$, so $\mathbb{H}^2/\Gamma$ is a Riemann surface $\Sigma_m$ with some number of asymptotic boundaries $m$.\footnote{In principle, we can have a surface of arbitrary genus $g$ formed by attaching asymptotic regions to a closed surface, but we restrict to the case $g=0$.}
For example, in the case $m=2$, the Riemann surface is a cylinder, and the resulting geometry $\Sigma_2$ is the $t=0$ slice of the two-sided eternal AdS black hole which is dual to a thermofield double state in the boundary CFT (Fig.~\ref{fig:covering-tfd}).
The discrete group in this case is generated by just a single element, which acts on the upper half plane by a linear fractional transformation
\begin{equation}
    \gamma_1(z) = \mu^2 z ,
\end{equation}
where a larger $\mu \in \mathbb{R}$ generates a larger cylinder.
This transformation sends points on the smaller orange semicircle in Fig.~\ref{fig:covering-tfd} to points on the larger orange semicircle.
It can also be written in $SL(2,\mathbb{R})$ form as
\begin{equation}
    \gamma_1 = 
\left( 
\begin{array}{cc}
   \mu  & 0 \\
  0   & \frac{1}{\mu}
\end{array}
\right) .
\end{equation}
Notice that the black dashed segment in Fig.~\ref{fig:covering-tfd} is invariant under the action of $\gamma_1$.
This can be taken as the defining feature of the causal horizons in multiboundary wormhole geometries; they are always invariant under a combination of the generators.
This principle can be used to extract both their location and length as we explain in Appendix \ref{sec:partial-islands-appendix}.
Furthermore, this method of constructing wormholes extends to generate full Lorentzian geometries, and the Riemann surface $\Sigma_m$ can always be interpreted as the $t=0$ time-reflection-symmetric slice of a Lorentzian 2+1d geometry with metric\footnote{These coordinates cover only the the Wheeler-De Witt patch of the $t=0$ slice in Lorentzian, but they do cover the entire spacetime in Euclidean.}
\begin{equation}
    ds^2 = -dt^2 + \ell^2 \cos^2 \frac{t}{\ell} d\Sigma_m^2,
    \label{eq:multiboundarymetric}
\end{equation}
where $d\Sigma_m^2$ is the constant negative curvature metric with unit curvature radius on $\Sigma_m$ inherited from the covering space $\mathbb{H}^2$.
We will analyze the $t=0$ slice.  We consider an identification that produces a $t=0$ slice with $m$ asymptotic regions, i.e., an $m$-boundary wormhole. Since our geometries will always be time-reflection-symmetric, extremal surfaces that start at $t=0$ remain in this slice (see Appendix~\ref{app:identifications} for details of the $\Sigma_3$ case).

\begin{figure}
    \centering
    \begin{tabular}{c c}
     \includegraphics[width=7.5cm]{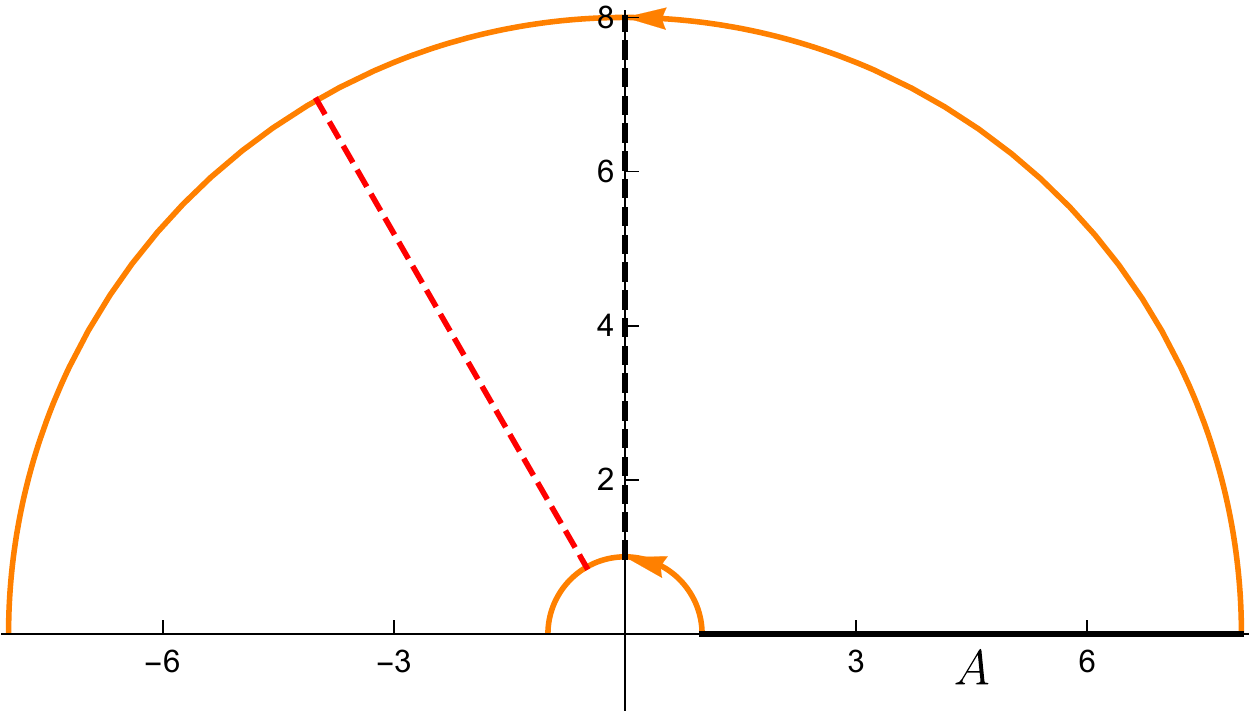}    & \includegraphics[width=7.5cm]{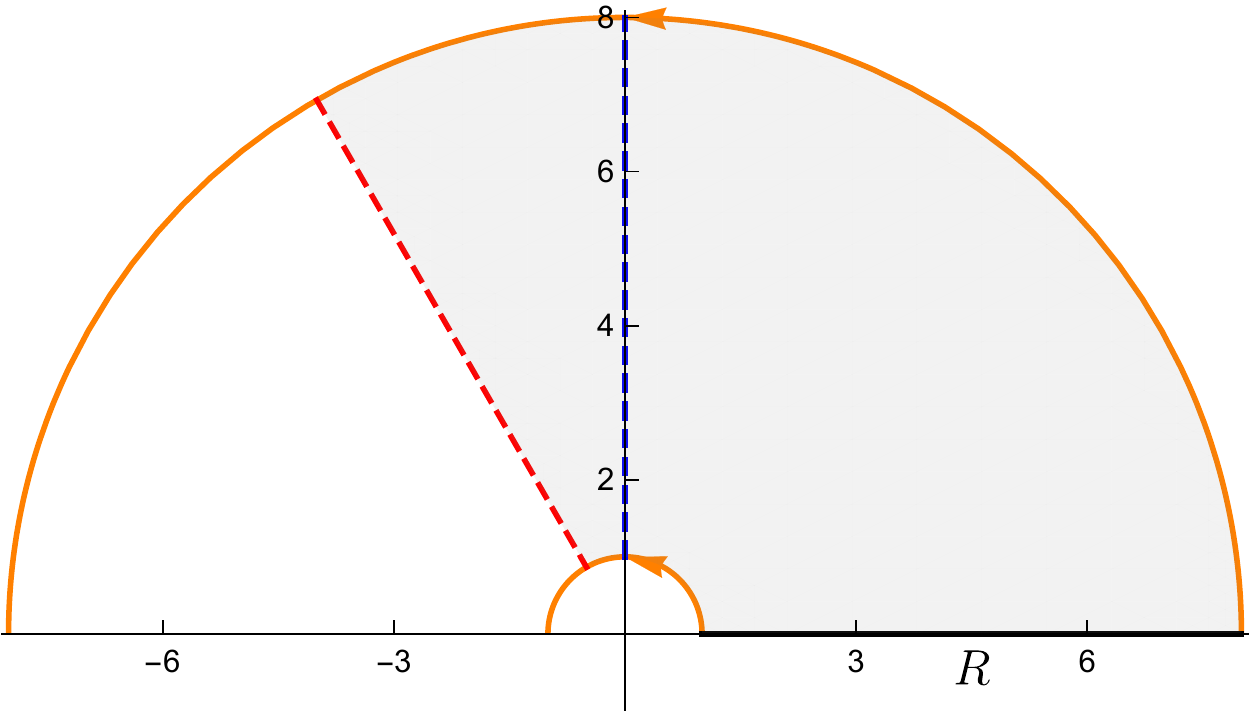}
    \end{tabular}\\
    \includegraphics[width=9cm]{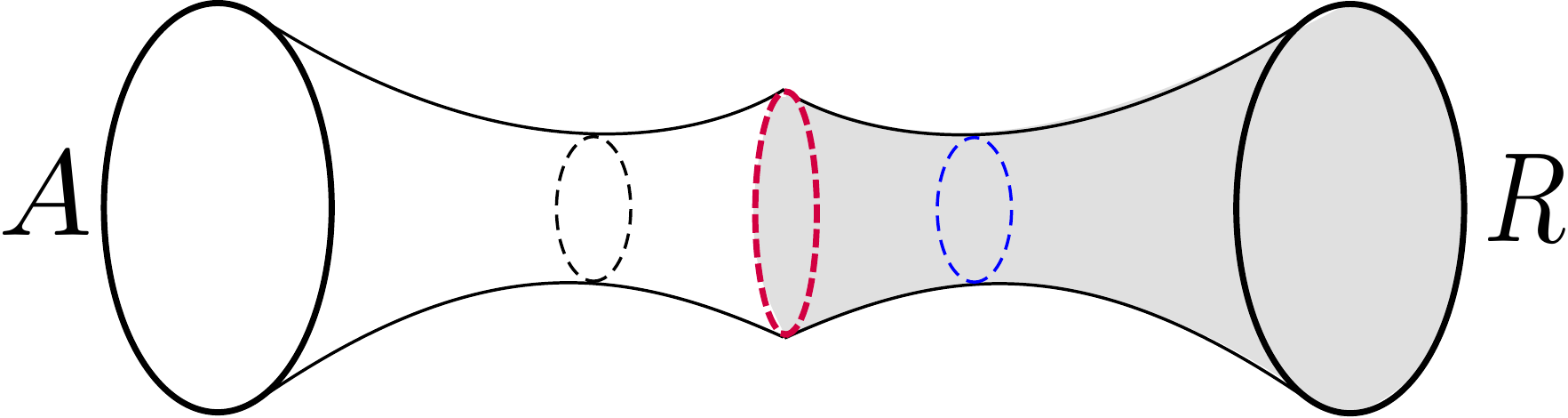}
    \caption{\small Top left: a covering space depiction of the ``real" region, Fig.~\ref{fig:tfd}b, the surface $\Sigma_2$ after introducing an EOW brane.
    The dashed red segment is the EOW brane, which cuts off one asymptotic region (region $B$ from Fig.~\ref{fig:covering-tfd}).
    The region bounded by the black line (asymptotic boundary $A$), orange lines, and dashed red line forms a fundamental region for the geometry.
    Top right: the geometrization of the brane CFT into the inception region (gray), another cut off copy of $\Sigma_2$.
    The length of the brane in the inception region must match its length in the real region (top left).
    The dashed red segments on the top left and top right are identified, causing the real region to form a long wormhole into the inception region, stretching from the real asymptotic region $A$ to the radiation asymptotic region $R$.
    The dashed blue line is the horizon of the inception black hole.
    Bottom: the glued geometry.
    }
    \label{fig:covering-tfd-eow}
\end{figure}

We can introduce an EOW brane in the geometry (\ref{eq:multiboundarymetric}), located on a circle in the $t=0$ slice $\Sigma_m$ (Fig.~\ref{fig:covering-tfd-eow}).
This brane sits in front of one of the causal horizons of the multiboundary wormhole, and effectively cuts off one of the asymptotic regions.
We choose this brane to intersect the $t=0$ slice at constant Schwarzschild $r$ coordinate, as in the brane trajectory solutions from Sec.~\ref{sec:quantincep}.
In the two-boundary covering space picture, constant Schwarzschild $r$ corresponds to a straight line from the origin that makes some angle with the horizontal axis \cite{Skenderis:2009ju}.
This is true for multiple boundary wormholes as well, provided that the brane only cuts off a single asymptotic boundary.
To compute the entanglement entropy of asymptotic regions (or subregions of these), we will employ our extended Ryu-Takayanagi prescription where extremal surfaces are geodesics homologous to the desired (sub)region relative to the brane.  Thus, the extremal surfaces are permitted to go through the brane, subject to a refraction condition imposed by the cosmological constants on either side.  The parts of the geodesics on the wormhole geometries $\Sigma_m$ can be easily understood in covering space, where they take the form of either semicircles or vertical lines in the upper half plane, both of which must end orthogonally on $\partial \mathbb{H}^2$.  Thus, the covering space picture will be our main tool in computing extremal surfaces.


\subsection{Secret sharing between two radiation subsystems}

We start by considering a black hole in AdS with entropy $S_{\text{BH}}$, and an  EOW brane behind the horizon.  We want to consider splitting the radiation into two subsystems as in (\ref{eq:multibdystate}), perhaps corresponding to early and late time radiation.  To model this situation in the inception geometry we need two asymptotic boundaries.  Thus we can take the inception geometry to be a three boundary wormhole, with two boundaries identified with the radiation system, and 
(Fig.~\ref{fig:octopus_1}).

\begin{figure}
    \centering
    \includegraphics{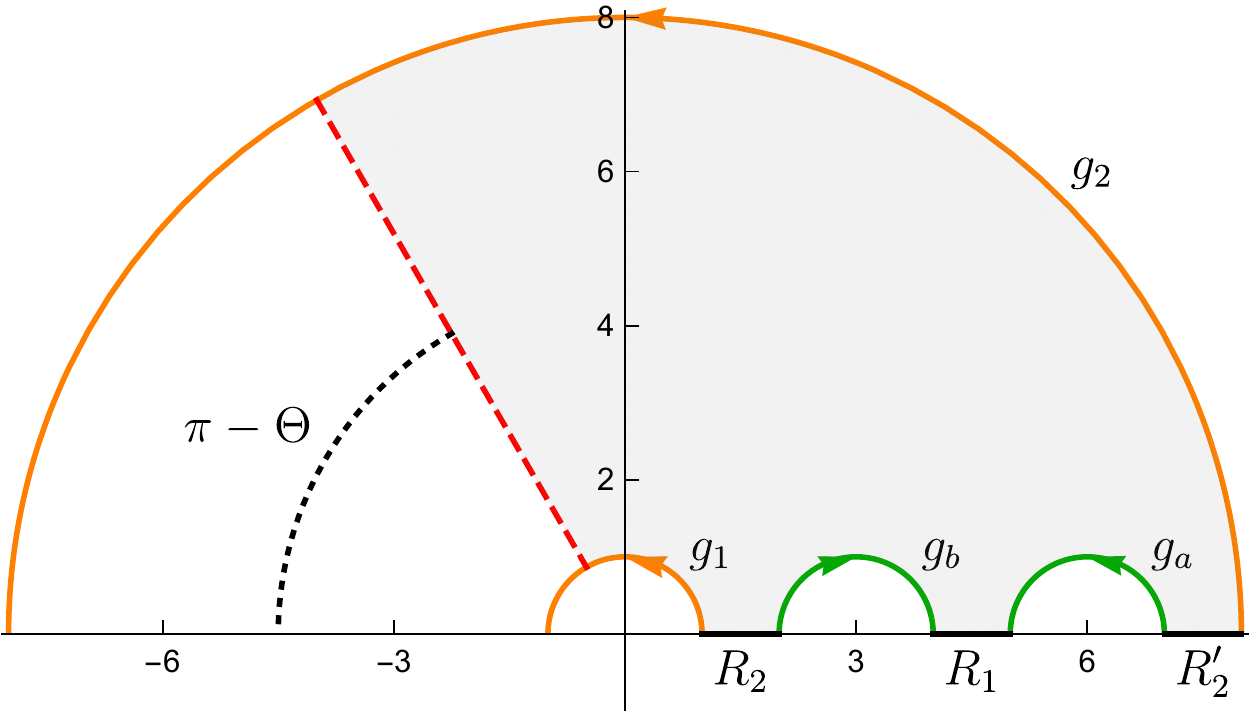}
    \caption{\small A covering space (upper half plane $\mathbb{H}^2$) depiction of the inception geometry with $D_2 = 8, D_1 = 1, X_a = 6, D_a = 1, X_b = 3, D_b = 1, \Theta = \frac{2\pi}{3}$.
    Note that this does not include the ``real" black hole region, which appears as a thermofield double with one asymptotic region cut off by the brane, as in the top left panel of Fig.~\ref{fig:covering-tfd-eow}.
    The shaded gray region is a fundamental region.
    The orange (resp. green) semicircles are identified with the orientations shown.
    The dashed red segment is the brane location as seen from the inception region; if extended, it would intersect the origin.
    The circles are labeled according to their parametric equations \eqref{eq:circle-data-1}-\eqref{eq:circle-data-4}.
    The radiation is split into two disjoint subregions, $R_1$ and $R_2$ ($R_2$ and $R_2'$ join to form a single asymptotic region after the identifications).
    These radiation regions are asymptotic regions (black segments) in the multiboundary wormhole geometry, where the gray shading approaches the boundary (horizontal axis).
    The angle (dotted black arc) formed by the negative horizontal axis and the brane is $\pi - \Theta$.
    Notice that the brane location cuts off what would have become a third asymptotic region in the standard multiboundary wormhole geometry $\Sigma_3$.
    }
    \label{fig:circle-data-1}
\end{figure}

The covering space depiction of this inception geometry is shown in Fig.~\ref{fig:circle-data-1}.
The discrete group $\Gamma$ which generates this three boundary wormhole (without the EOW brane) makes identifcations between the geodesics 
\begin{align}
g_1(\lambda) & = D_1 e^{i\lambda} ,\label{eq:circle-data-1}\\
g_2(\lambda) & = D_2 e^{i\lambda} ,\\
g_a(\lambda) & = X_a + D_a e^{i\lambda} , \\
g_b(\lambda) & = X_b - D_b e^{-i\lambda}  ,
\label{eq:circle-data-4}
\end{align}
where the curve parameter is $\lambda \in [0,\pi]$, and we take $D_2 > D_1$, $D_1 < X_b - D_b$, $D_2 > X_a + D_a$, and $X_b + D_b < X_a - D_a$ as an ansatz.
The particular identifications generated by $\Gamma$ are
\begin{equation}
g_1(\lambda) \sim g_2(\lambda), \hspace{.5cm} g_a(\lambda) \sim g_b(\lambda) ,
\end{equation}
and the generators of  $\Gamma \subset SL(2,\mathbb{R}) $ acting on the upper half space are (see appendix \ref{app:identifications})
\beq
\gamma_1
= 
\left( \begin{array}{cc}
   \sqrt{\frac{D_2}{D_1}}  & 0 \\
  0   & \sqrt{\frac{D_1}{D_2}}
\end{array}
\right) ,
\eeq
and
\beq
\label{eq:gamma2-1}
\gamma_2
= 
\left( \begin{array}{cc}
   \sqrt{D_a}  &  \frac{X_a}{\sqrt{D_a}} \\
0 & \frac{1}{\sqrt{D_a}}
\end{array}
\right)
\left( \begin{array}{cc}
  0  & -1 \\
 1 & 0
\end{array}
\right)
\left( \begin{array}{cc}
  \frac{1}{\sqrt{D_b}}   &  -\frac{X_b}{\sqrt{D_b}} \\
0 & \sqrt{D_b}
\end{array} 
\right) \, .
\eeq

\begin{figure}
    \centering
    \includegraphics{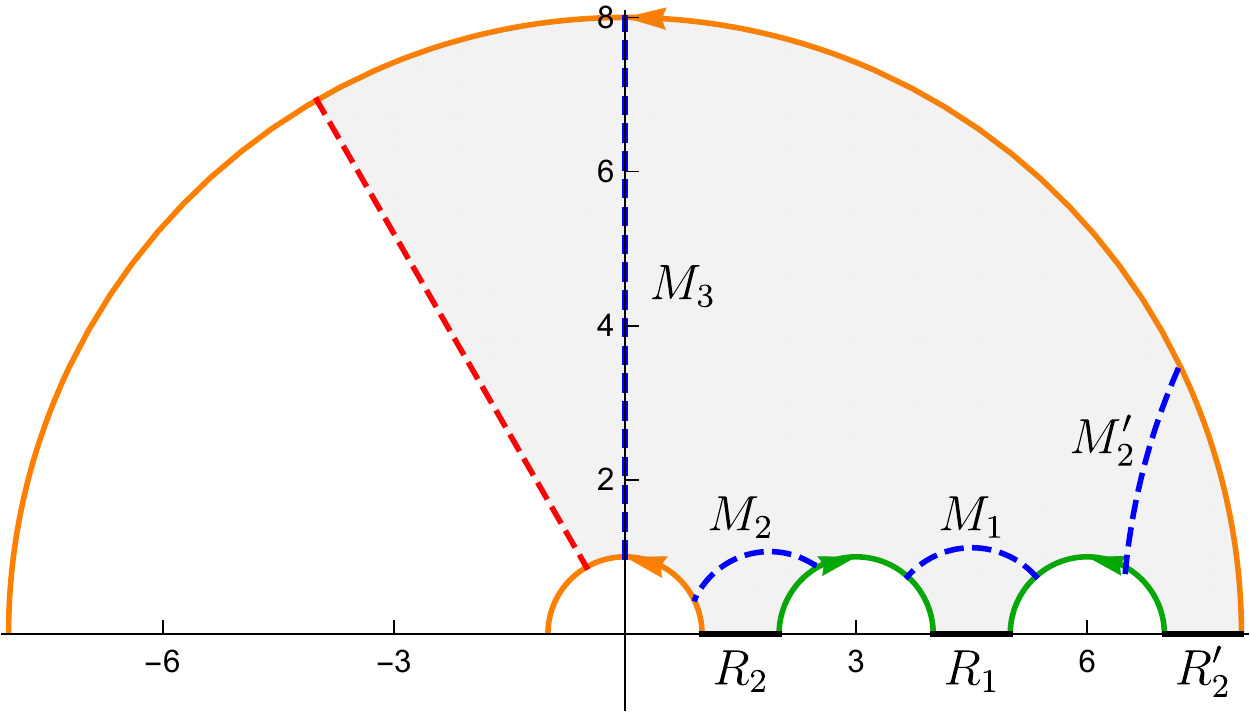}
    \caption{\small 
    Geodesic arcs $M_1$, $M_2$, and $M_3$ (dashed blue lines) in the fundamental region whose lengths ($m_1$, $m_2$, and $m_3$ respectively) are the moduli of the inception geometry.
    The arcs $M_2$ and $M_2'$ are joined smoothly under the identification.
    }
    \label{fig:octopus-moduli-1}
\end{figure}

\begin{figure}
    \centering
    \includegraphics{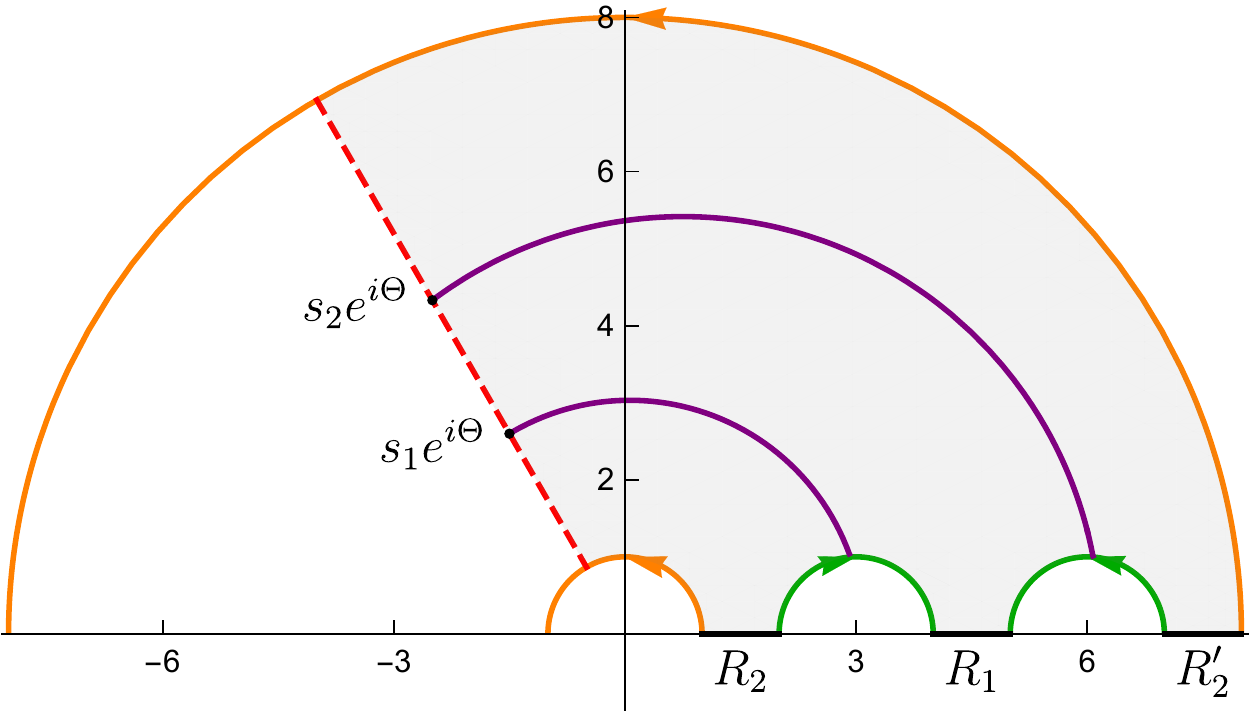}
    \caption{\small An infalling geodesic (purple) with brane endpoints $s_1 = 3$ and $s_2 = 5$ (black points on dashed red segment), and circle data unchanged from Fig.~\ref{fig:circle-data-1}.
    The larger purple arc is part of a large semicircle which is the image under $\gamma_2$ \eqref{eq:gamma2-1} of the smaller semicircle that includes the smaller purple arc.
    The two purple arcs are joined smoothly by the identification of the green semicircles.
    For fixed $s_1$ and $s_2$, the curve shown is the unique infalling geodesic which passes through the specified points.
    }
    \label{fig:infalling-1}
\end{figure}

The three boundary wormhole has three independent moduli (Fig.~\ref{fig:octopus-moduli-1}), which are lengths $m_1$, $m_2$, and $m_3$ of geodesic curves $M_1$, $M_2$, and $M_3$, homologous to their respective boundaries.
Above the Page transition, there is a new class of geodesics which which can pass through the surface where the EOW brane was located before inception -- namely the locus where the real and inception geometries are spliced.\footnote{These surfaces never dominate before the Page transition because of entanglement wedge nesting, i.e. because the entanglement wedge of region $R_1$ must be contained in the entanglement wedge of $R_1 \cup R_2$}.  We refer to these as ``infalling" geodesics, and an example is shown in Fig.~\ref{fig:infalling-1}.  Such infalling geodesics can start on the real side or the inception side, but we will be interested in the latter (as in Fig.~\ref{fig:octopus_1}) as we are computing the entanglement entropy of subsets of the radiation.
If the brane is at an angle $\Theta$ in the upper half plane (see Fig.~\ref{fig:circle-data-1}),\footnote{The brane sits at a constant Schwarzschild coordinate on the $t=0$ slice which translates to a line with fixed angle on the upper half plane. The relation of the angle to the Schwarzschild coordinate $r_t$ of the brane is $2\pi r_t=\frac{\ell' \log(D_2/D_1)}{\sin \Theta}$.} we can choose the two endpoints of the infalling geodesic to be $s_1 e^{i\Theta}$ and $s_2 e^{i\Theta}$ with $s_2 \geq s_1$.
The infalling geodesic is uniquely specified by the brane endpoints $s_1e^{i\Theta}$ and $s_2e^{i\Theta}$.
Let us call its length $L_I(s_1,s_2)$.

If we hope to use an infalling geodesic as part of an extremal surface, we must complete it in a way such that it is homologous to an asymptotic region.  The component that is extended from the inception side into the real side will take the form of a geodesic in a global BTZ spacetime which is cut off by the EOW brane (which sits at a constant Schwarzschild radial coordinate).
The infalling geodesic must extend continuously through the brane, so the portion in the real space should have brane endpoints which match the infalling endpoints $s_1e^{i\Theta}$ and $s_2e^{i\Theta}$ on the inception side.
Let us call the length of the part of the geodesic that is in the real space $L_b(s_1,s_2)$.  When we compute the entropy of one radiation subsystem,
the complete infalling geodesic, which lives partly in the inception geometry and partly in the real geometry, will compete for dominance with the causal horizon of the corresponding asymptotic region in the inception geometry.
The entropy associated with this part of the geodesic, $S_{\text{brane}}$, is analogous to a bulk entanglement contribution and it depends only on the fraction of the brane contained between the infalling endpoints.
We write the dependence on $s_1$ and $s_2$, but emphasize that the true dependence is only on the brane length and brane subregion length (between the infalling endpoints).
\begin{equation}
    S_{\text{brane}} = \frac{L_b(s_1,s_2)}{4G_N} .
\end{equation}

The entanglement entropy of one asymptotic radiation region (say $R_1$), then, is given by the following double minimization (using our extended Ryu-Takayangi prescription):\footnote{Note that the usual Ryu-Takayanagi formula deals with extremal surfaces.  However, in our case, the Newton constants in the real geometry and inception geometry are different.  Since these constants appear in the gravitational action, the philosophy of Lewkowycz and Maldacena \cite{Lewkowycz:2013nqa} instructs us to instead extremize the total entropy, which is proportional to length divided by the Newton constant in the appropriate region.  Properly speaking, we should refer to the surfaces in this paper as extremal/minimal entropy surfaces, but we will abuse terminology and continue to refer to them as extremal/minimal surfaces.}
\begin{equation}
    S_{R_1} = \min \left[ \underset{s_1,s_2}{\min} \left[ \frac{L_I(s_1,s_2)}{4G_N'} + S_{\text{brane}} \right] , \frac{m_1}{4G_N'} \right] .
    \label{eq:extendedRT}
\end{equation}
Notice that when the first term appearing in \eqref{eq:extendedRT} is minimal, we have infalling geodesic dominance in the entropy calculation.
This means that there is a piece of the minimal surface which lives in between the real black hole horizon and the EOW brane position, a geodesic in a cutoff global BTZ geometry.
The entanglement wedge of the region $R_1$, then, includes the region between this geodesic and the EOW brane position, which we call a ``partial island" because it is part of the total island between the real black hole horizon and the EOW brane position that was described in \cite{Almheiri:2019hni}.
We work out the explicit formulas for $L_I$ and $S_{\text{brane}}$ in Appendix \ref{sec:partial-islands-appendix}. With the entropy formula in hand, we  wish to choose an evaporation protocol and draw Page curves.
To simplify the situation, we choose all the moduli to be symmetric, i.e. 
\begin{equation}
    m_1 = m_2 = m_3 .
    \label{eq:symmetric}
\end{equation}
Notice that, independently of any relationships between the moduli, the inception black hole radius $r_h'$ is simply related to the $m_3$ modulus:
\begin{equation}
    m_3 = 2\pi r_h'.
\end{equation}
The inception black hole radius $r_h'$ will be our ``protocol parameter", which we tune from zero to $r_h$ while obeying the constraints of Sec.~\ref{sec:protocol}. 
By evaporation protocol we mean a procedure for varying the moduli of the inception geometry in such a way that the entanglement between the radiation and the black hole microstates increases, while maintaining a non-singular gluing of the inception geometry and the real geometry.
Of course, it is dynamics that determines how the real black hole evaporates. 

In the two boundary case, there was only one way to increase the entanglement, namely to increase the size of the inception black hole.
However, with three boundaries, we have three moduli and must decide how to vary them.
The modulus which controls the aforementioned entanglement is $m_3$, so in principle the only constraint on our protocol is to somehow increase the entropy associated with $m_3$.
Our choice, as outlined above, is to take all the moduli to be equal and then increase them, while fixing constants $\ell,G_N,\ell'/G_N'$.   During the protocol  $G_N'$ is allowed to change as described in Sec.~\ref{sec:protocol}.
Under this protocol, there will be an exchange of minimal surface between the causal horizon of one asymptotic radiation region (say $R_1$) and an infalling geodesic with components on both sides of the brane, as in Fig.~\ref{fig:infalling-1}.

Before proceeding, we emphasize that our results involving the dominance of infalling geodesics and partial islands are largely independent of the details of the gluing.
That is to say, we analyzed a wide variety of alternate protocols, many of which did not obey the particular gluing constraints that we required here, and in every case we found infalling geodesics and partial islands.We will also discuss other examples of robustness of partial islands in the next subsection.
Reassuringly, protocols which do not correspond to ``evaporation", i.e. protocols which do not increase the entanglement between the radiation and the black hole microstates, do not lead to partial islands.
(An example of such an invalid protocol is to fix $m_3/G_N'$ to be less than the real black hole entropy and then to increase $m_1$ and $m_2$; this only increases entanglement within the Hawking radiation and not between the radiation and the black hole microstates.)
This is as it should be, since under such protocols we do not cross the Page transition.

\begin{figure}
    \centering
    \includegraphics{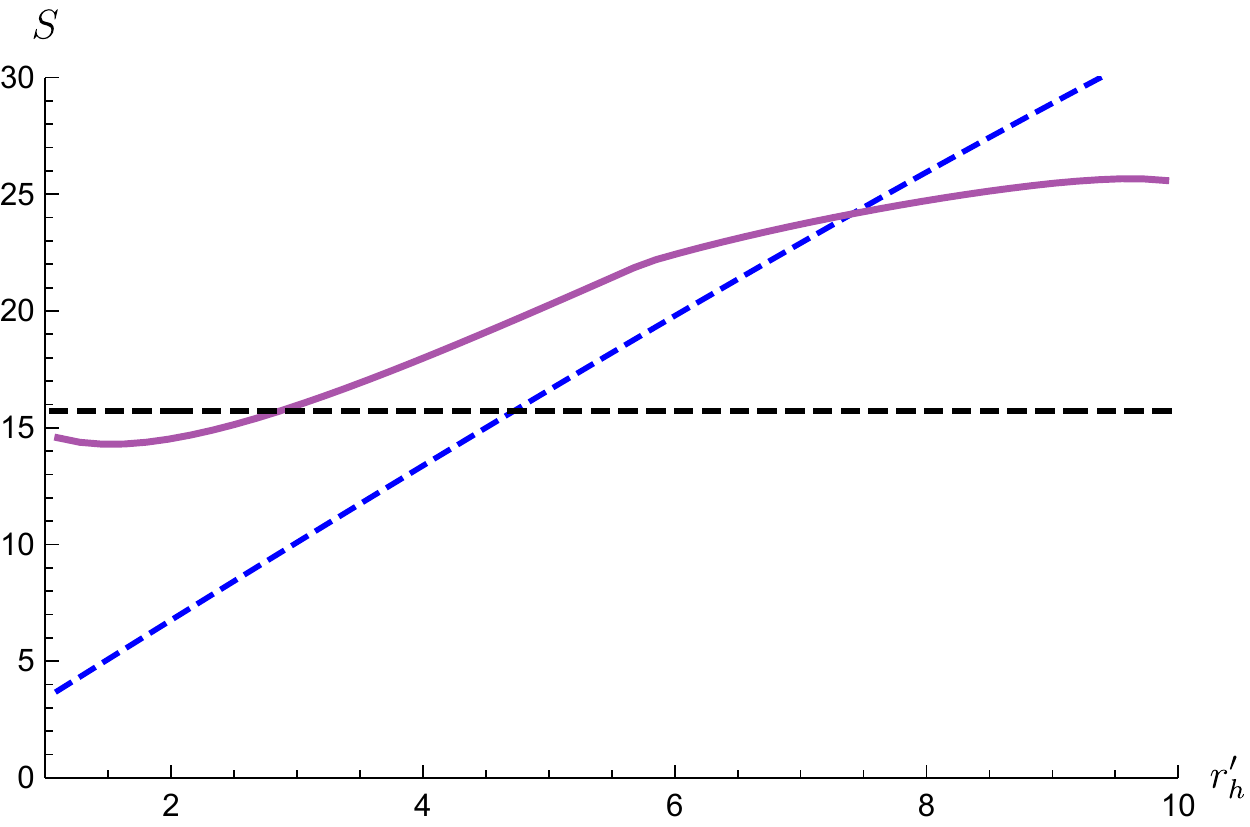}
    \caption{\small A plot of various entropies computed using our extended Ryu-Takayanagi formula as the evaporation protocol proceeds.  The entropies associated  to the purple, individual dashed blue, and dashed black curves in Fig.~\ref{fig:octopus_1} are shown here as the purple, dashed blue, and dashed black curves respectively.  The dashed black line is the entropy of the real black hole horizon with $r_h = 10$ and $G_N = 1$.  The minimum of the dashed blue line and dashed black line is the entropy of the radiation region $R_1 \cup R_2$.
    The entropy of region $R_1$ is the minimum of the dashed blue curve and purple curve.
    We have chosen $\ell = 1$ and $c/c' = 0.2$, which determines $G_N'$ and $\ell'$.  The dashed blue line grows approximately linearly as the inception modulus $r_h'$ is increased.  The solid purple line is the entropy of the minimal entropy infalling geodesic, including a possible contribution from the real black hole spacetime.  The Page transition occurs when the dashed blue line crosses the dashed black line. The partial island becomes reconstructible from a single radiation leg after the purple line crosses the dashed blue line.}
    \label{fig:radiation-entropy-1}
\end{figure}

\begin{figure}
    \centering
    \includegraphics[scale=1]{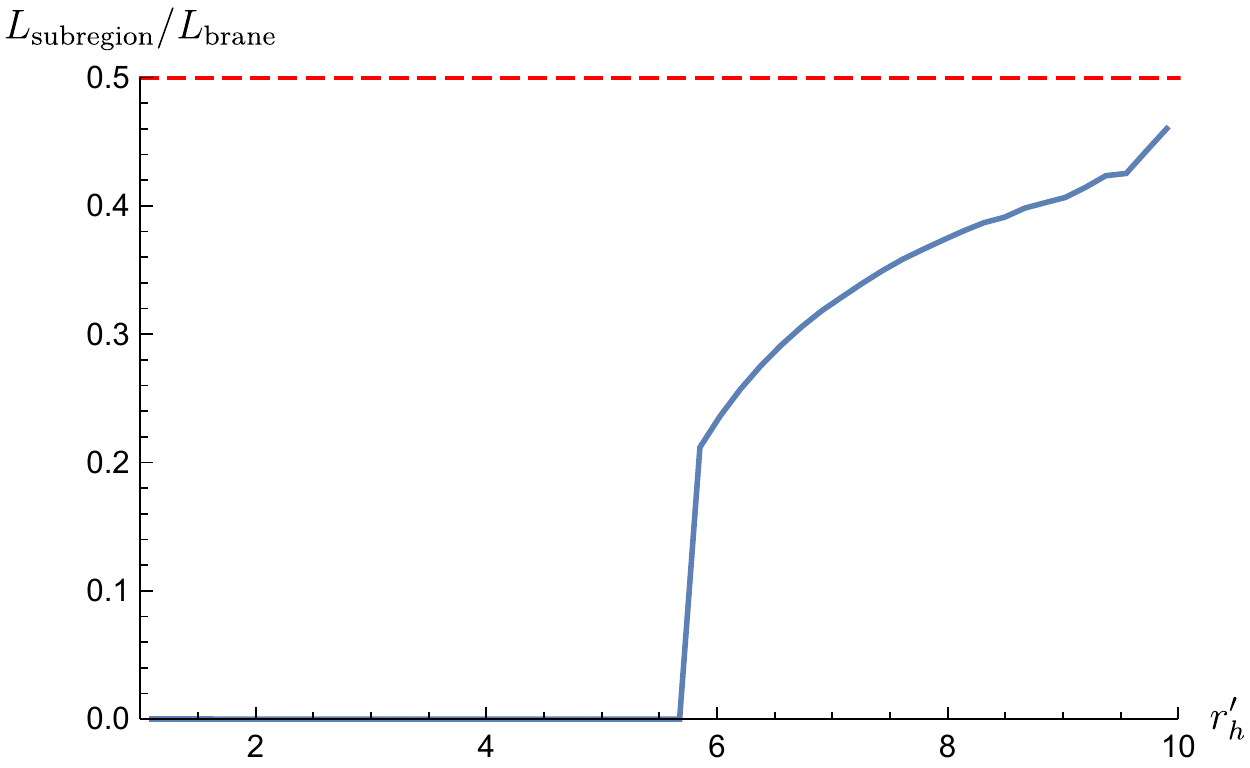}
    \caption{\small The growth of the brane fraction $L_\text{subregion} / L_{\text{brane}}$ captured by the minimal infalling geodesic plus brane entropy combination (with the same parameters as Fig.~\ref{fig:radiation-entropy-1}).  Notice that, comparing with Fig.~\ref{fig:radiation-entropy-1}, the partial island is already more than 30\% of the brane by the time the infalling geodesic becomes more minimal than the inception causal horizon (around $r_h' \approx 7$).
    The red dashed line at fraction $\frac{1}{2}$ is the maximum possible value of the blue curve for our choice of equal inception moduli, and the protocol ends at about 45\% brane fraction.
    So, at the end of the protocol ($r_h' \approx 10$), the union of the entanglement wedges of $R_1$ and $R_2$ capture about 90\% of the total brane length.
    But, the entanglement wedge of $R_1 \cup R_2$ captures 100\% of the total brane length immediately after the Page transition.
    }
    \label{fig:eyeland-growth-1}
\end{figure}

A plot of the resulting entropy  $S_{R_1}$ of radiation region $R_1$ is shown in Fig.~\ref{fig:radiation-entropy-1}. 
The brane subregion captured by the infalling geodesic, i.e., the portion of the brane (for example, in Fig.~\ref{fig:infalling-1}) traced out by $se^{i\Theta}$ for $s \in [s_1,s_2]$, is a certain fraction of the full brane length, $L_{\text{subregion}}/L_{\text{brane}}$.
This fraction grows with the amount of radiation in region $R_1$; its development is shown in Fig.~\ref{fig:eyeland-growth-1}.
\begin{figure}
    \centering
    \includegraphics[scale=1]{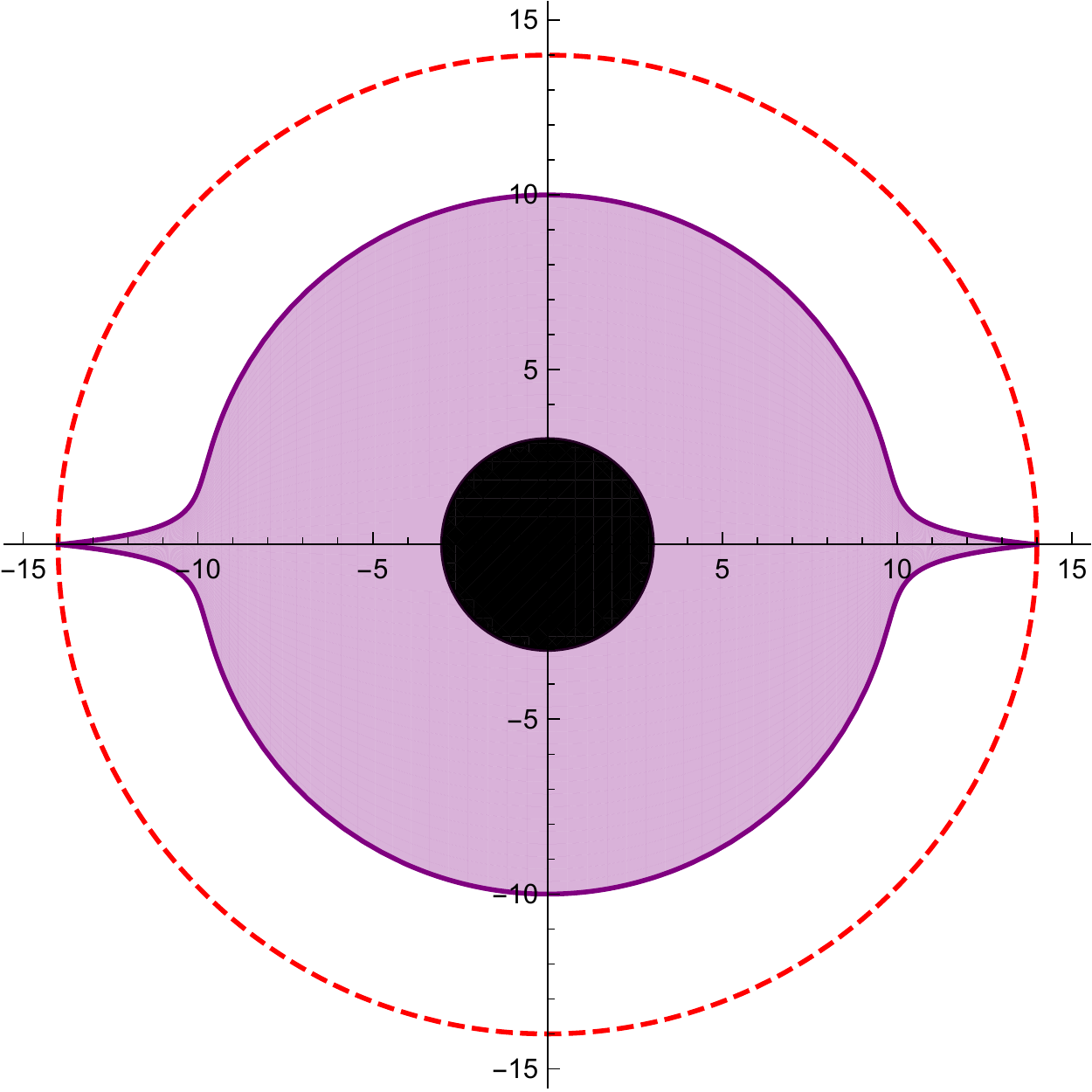}
    \caption{\small A schematic (not depicted using real parameter values or curves) of the geodesic behind the brane for both individual radiation legs if the brane fractions were to saturate at 50\%.
    The dashed red circle is the brane and the black disk is the real black hole.
    The darker purple curves are the pieces of spacelike geodesics which dip behind the brane and attach at the brane to the infalling geodesic coming from the radiation (inception) side of the geometry.
    The upper purple curve attaches to an infalling geodesic to become homologous to radiation leg $R_1$, and the lower purple curve plays a similar role for leg $R_2$.
    The lighter purple region in between the two purple curves cannot be reconstructed with access to only one radiation leg; even if two observers collecting radiation separately are communicating classically, they cannot reconstruct this ``eyeland".
    }
    \label{fig:eyeland-protocol-1}
\end{figure}
Notice from Fig.~\ref{fig:eyeland-growth-1} that the asymptotic value of the fraction of the brane captured by the infalling geodesic is almost $\frac{1}{2}$.
This is because we chose the moduli to be equal \eqref{eq:symmetric}.\footnote{If we had instead chosen (for example) $m_1 = 2m_2 = m_3$, the maximum fraction of the brane would be greater than $\frac{1}{2}$ for region $R_1$ and less than $\frac{1}{2}$ for region $R_2$.}
We can visualize the situation on the BTZ black hole geometry if this fraction were to actually saturate.   
Let us consider in this situation the entanglement of region $R_1$ and the entanglement of region $R_2$.
There will be two infalling geodesics, one associated with $R_1$ and another associated with $R_2$.
They will both have pieces that sit between the EOW brane position and the real black hole horizon, in a cutoff global BTZ geometry.
However, these components will not touch, and there is a leftover region (Fig.~\ref{fig:eyeland-protocol-1}), an ``eyeland", which cannot be reconstructed with access to only one radiation leg, or even by observers in both radiation legs who have only classical contact.  
Reconstructing the eyeland requires quantum mechanical access to both radiation regions simultaneously; by ``access" we simply mean the ability to act with operators and make measurements.

Note that if the split between the radiation regions was highly asymmetric, e.g. $R_1$ had a much larger causal horizon than $R_2$, there would not be an eyeland.  Instead, the entanglement wedge of $R_1$ would cover the entire region between the EOW brane position and the real black hole horizon, and the entanglement wedge of $R_2$ would end at its causal horizon.  This implies that the smaller region contains no information about the black hole interior and only encodes information about the rest of the Hawking radiation.  Interestingly, if we split the radiation into many very small parts, none of the individual entanglement wedges will contain the black hole interior.  However, the union of a sufficiently large number of the reservoirs will contain at least a portion of the interior.  This points to a multi-party character in the information about the black hole carried by the Hawking radiation.

In this way, the existence of the eyeland is similar to a quantum secret sharing scheme between the two radiation legs; an observer needs access to both legs, not just one (or even to one plus a friend in the other leg) to uncover a secret hidden in the eyeland.
At the Page transition, the secret sharing scheme is, in a sense, perfect.  This is because neither radiation leg can reconstruct \textit{anything} in the interior individually, but both combined can reconstruct the whole interior.  As the protocol continues after the Page transition, this secret sharing is weakened, since the individual radiation legs begin to reconstruct partial islands due to infalling geodesic dominance in the entropy calculation.  Na\"{i}vely, one might have expected the secret sharing scheme to eventually break completely; that is to say, the union of the entanglement wedges of $R_1$ and $R_2$ would cover the whole interior.
However, this is not what occurs. 
The eyeland represents a robust remnant of the initial secret sharing which is impossible to eliminate if the radiation is collected in the manner described.
This secret sharing scheme has a local nature; the geometrization of the radiation into the inception multiboundary wormhole has led to a geometric secret sharing scheme which has a built-in locality.
In other words, we can tell if the secret is obtainable just by specifying its spatial location in the interior.
This interpretation follows naturally from  our inception procedure, and the fact that we purify the real black hole with a multi boundary wormhole geometry. It would be interesting to understand to what extent the locality of information in this sense holds true in more realistic models of black hole evaporation.

\subsection{Other eyelands}\label{subsection:oeye}


Although we have been only discussing a particular example so far,  now we would like to argue that the emergence of partial islands as well as eyelands is quite universal phenomena. In order to see this, in this subsection we list
several other setups which  contain these regions.

\begin{figure}
    \centering
    \includegraphics[scale=.4]{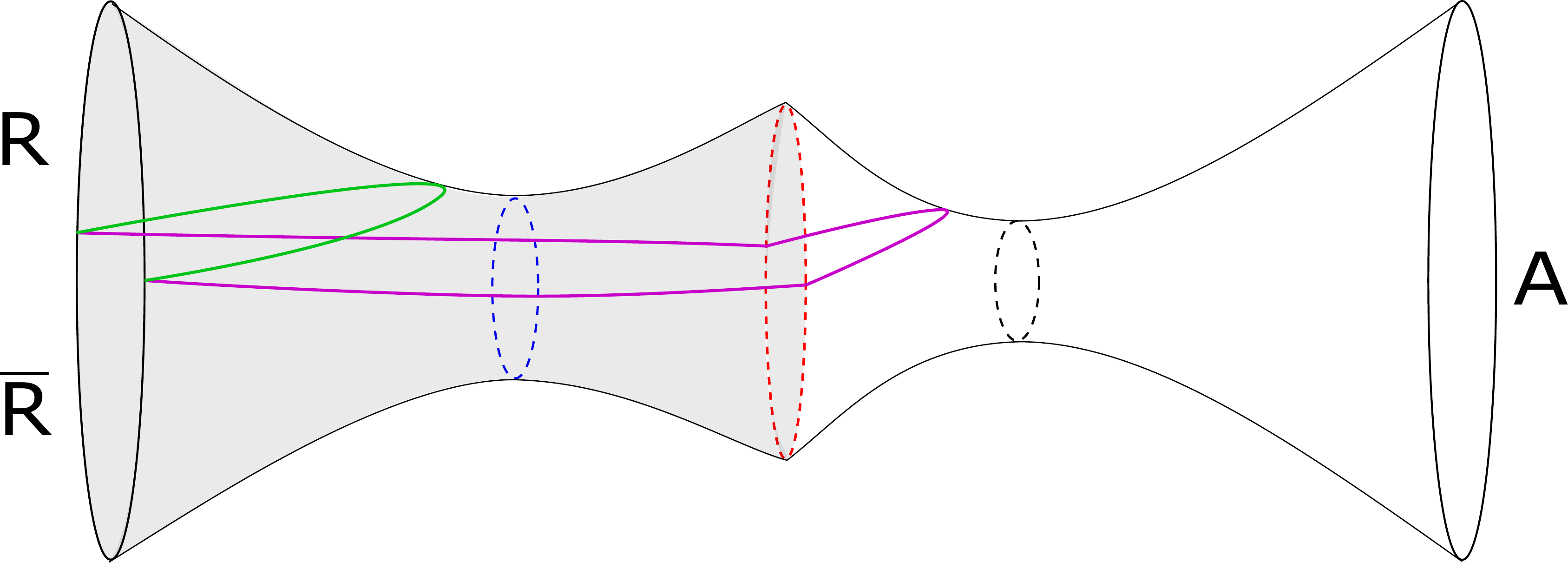}
    \caption{\small An infalling geodesic (purple) and partial island for a subregion $R$ of a single asymptotic radiation region $R \cup \bar{R}$.
    The green geodesic, which is contained entirely in the asymptotic BTZ patch of $R \cup \bar{R}$, has entropy $S_1$, and the purple infalling geodesic has entropy $S_2$.
    When the temperature of the inception black hole is large, the geodesic with entropy $S_2$ is the dominant RT surface.
    }
    \label{fig:subregion-infalling}
\end{figure}

For example, in the two boundary situation, we geometrize the radiation by a single boundary black hole and divide the asymptotic boundary of the radiation region into two parts, $R$ and $\bar{R}$.
Let $L_R$ be the size of region $R$. 
There are two candidate RT surfaces which are homologous to the region $R$ (Fig.~\ref{fig:subregion-infalling}).
One is the geodesic connecting the end points of $R$, which lies solely in the asymptotic radiation region, i.e. a global BTZ patch.
The contribution  of this geodesic to the entropy is $S_{1}  =\frac{c'}{3}\log \sinh \frac{L_{R}}{\beta'} \sim c'L_R/3\beta'$, where $\beta'$ is the temperature of the inception black hole, and $c'$ is the central charge of brane CFT. 
The second geodesic is naturally split into two pieces: the first piece starts from the boundary of the region $R$, falls through the horizon in the inception region, then ends on the brane.  
The second piece is on the real black hole side, and starts and ends on the brane in a cutoff global BTZ patch, just as $L_b$ did in the three boundary discussion.
The length of the first part gives the so-called boundary entropy $S_{\text{bdy}}$, which does not depend on $\beta'$ or $L_R$.  The length of the second piece can be approximated by $\frac{c'L_R}{3\beta}$, where 
$c$ is the central charge of the real CFT, and $\beta$ is the temperature of the real black hole. 
So, the total length of the second candidate is $S_2 \approx 4G_N' S_{\text{bdy}} + \frac{cL_R}{3\beta}$.
Thus we conclude that when the temperature of radiation is low, i.e. we have $ \frac{c'}{\beta'} \ll \frac{c}{\beta}$, $S_1$ dominates, but in the opposite limit $S_2$ becomes dominant.
\begin{figure}[!h]
    \centering
    \includegraphics[height=6cm]{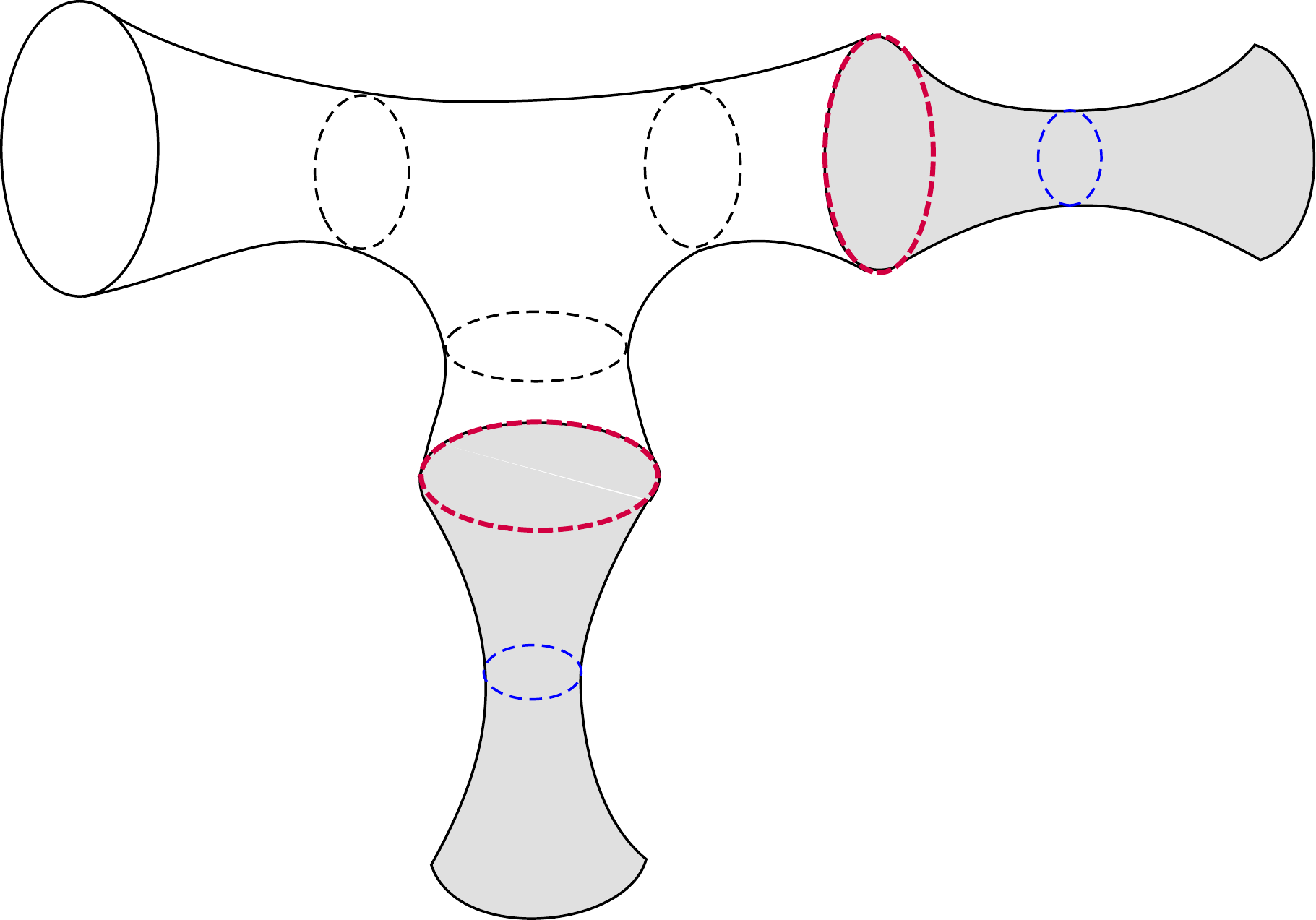}
    \caption{\small{A situation where the original black hole (unshaded) has a pair of pants region behind the horizon with two separate EOW branes (red), each entangled with separate radiation systems (shaded regions). }}
    \label{fig:partisles}
\end{figure}

Another multiboundary situation with partial islands is sketched in Fig.~\ref{fig:partisles}. Here the original black hole has a pair of pants region behind its horizon with two separate EOW branes. These branes are further entangled with independent radiation systems. Assume for simplicity that all the extremal surfaces on the original black hole side are of the same length $L_1$, and those on the radiation side are of the same length $L_2$. Then, as we crank up the entanglement between the radiation and the branes, i.e., when $L_2 \gg L_1$ (and assuming that we are in a regime where the in-falling geodesics are not dominant), the entire pair of pants interior goes over to the entanglement wedge of the radiation systems. But if we only had access to one radiation leg, then the entanglement wedge of that radiation subsystem would not include the interior pair of pants region, although it would have a partial island, bounded by the extremal surface adjacent to the corresponding EOW brane.

\begin{figure}[!h]
    \centering
    \includegraphics[height=5cm]{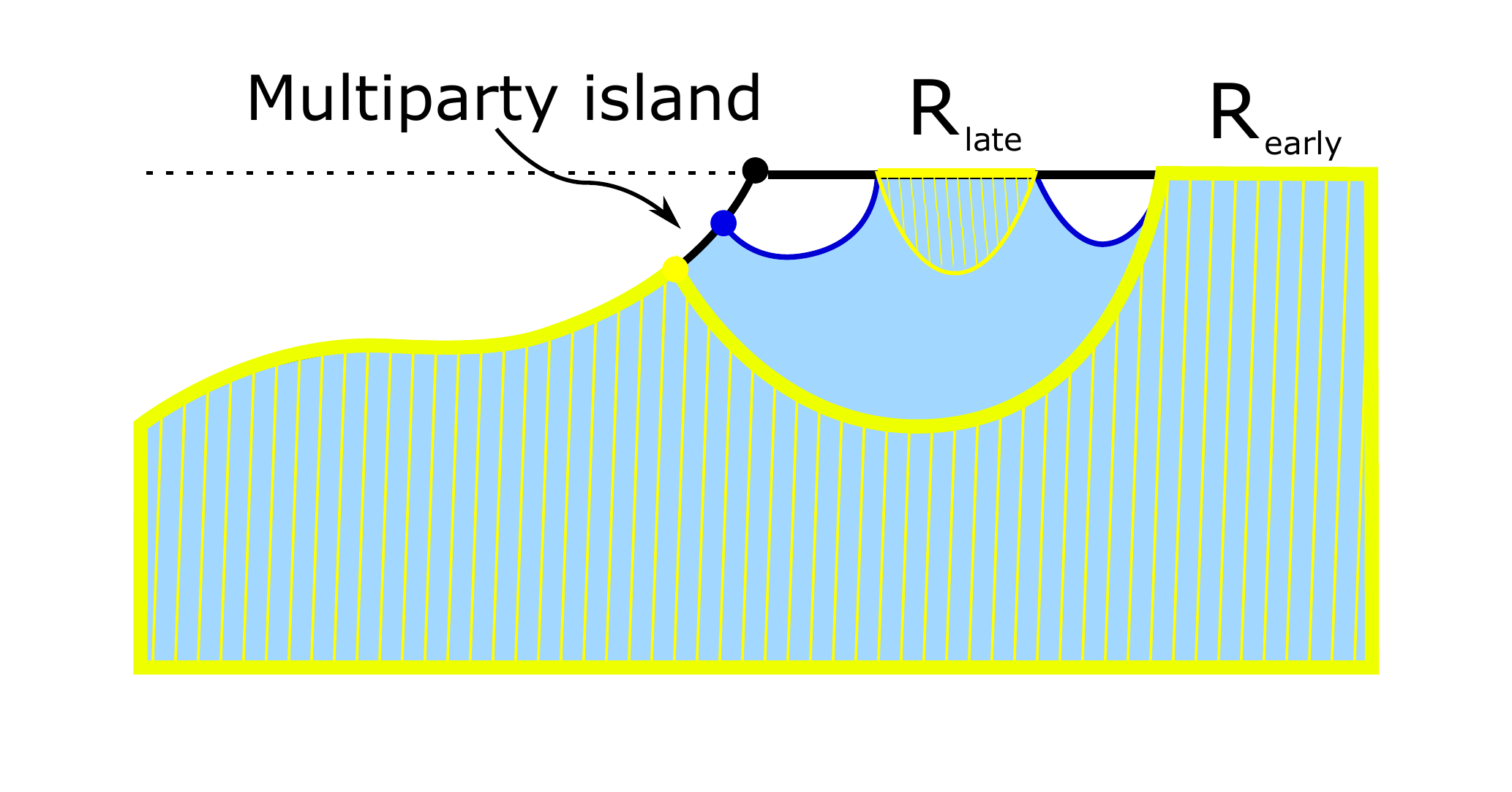}
    \caption{\small{We can divide the radiation into early and late parts in the model of \cite{Almheiri:2019hni}. The entanglement wedge of the union (blue shading) has the structure of a multiboundary wormhole, albeit in one higher dimension than where the black hole lives. Having access to only the early radiation gives a partial island (where the yellow hatched region meets the Planck brane), the remaining multiparty island is only accessible by having control over $R_{\rm early}$ and $R_{\rm late}$ simultaneously. }}
    \label{fig:otherinception}
\end{figure}

Partial islands emerge also in the doubly holographic model of \cite{Almheiri:2019hni}. We refrain from recapping the details of the construction in \cite{Almheiri:2019hni}, but sketch how a partial island comes about in their setup on Fig.~\ref{fig:otherinception}. In this case the multiparty island is not a true ``eyeland", since it is difficult to split the radiation in a way that both parts have non-trivial partial islands in their respective entanglement wedges.

\section{Discussion}\label{sec:discussion}

In this paper, we studied information recovery from three-dimensional asymptotically AdS black holes evaporating into an external reservoir.  We first modeled  black hole microstates in terms of states on an End-of-the-World brane that are entangled with the emitted radiation.  Tracing out the radiation leaves a thermal state on the brane.  We then took the brane theory to be a 2d CFT with a geometric dual, and thus replaced the brane by a new black hole geometry spliced in behind the original horizon, a process that we called ``inception".  We considered an example of such  splicing, which satisfies a modified version of the Israel junction conditions.  The inception black hole can in turn be purified by replacing it with a long wormhole leading to one or more new asymptotic regions.  These regions, which purify the black hole microstate, can be regarded equivalently as a  representation of the emitted radiation, and thus realize the ER=EPR scenario.

In this context, we proposed that the entropy of subsets of the radiation could be computed by applying a novel version of the Ryu-Takayanagi prescription to segments of the new asymptotic regions of the inception geometry.  In this novel  prescription, the extremal surfaces that geometrize entanglement entropy are allowed to pass through the locus where the inception and real geometries are spliced together.  The same prescription can be applied from the real black hole side.  If an RT surface passes between the inception and real geometries, and is homologous to a radiation reservoir, this implies that the entanglement wedge of the radiation contains some piece of the real black hole spacetime.   Conversely, if an RT surface passes between the real and inception geometries and is homologous to a real asymptotic region, this implies that the entanglement wedge of a part of the CFT dual to the original black hole contains a piece of the radiation.  We argued that our extended Ryu-Takayanagi is correct via the replica trick and an exchange of dominance between saddle points in Euclidean gravity.

Our construction has several consequences: (1) We recover the Page curve for AdS black holes, namely the entropy of the radiation saturates at the coarse-grained entropy of the black hole, (2) Access to a portion of the radiation will allow us to reconstruct a part of the region behind the original black hole horizon (a partial island), (3) The union of such partial islands, the union of regions that can be reconstructed from different subsets of the radiation, cannot cover the entire black hole interior, and (4) The entire region behind the black hole horizon can only be reconstructed by observers who can perform simultaneous quantum measurements on all subsets of the radiation.  In this sense, Hawking radiation implements a quantum secret sharing scheme across all the radiation quanta.

In our radiation protocol we tuned the gravitational parameters of the inception theory, $G_N'$ and $\ell'$, as we increased the entanglement of the black hole microstates with the radiation, while keeping the ratio $\ell'/G_N'$, and hence the central charge of the EOW brane CFT, fixed. In our model, the inception geometry is dual to some effective field theory describing black hole microstates  entangled with the radiation reservoir.  In our evaporation protocol, we had to move the splicing locus of the inception  and real geometries as we changed the amount of entanglement between the microstates and the radiation.  We can interpret this as a change in the UV cutoff of the effective theory of the microstates.  Indeed, because we keep $c' \sim \ell'/G_N'$  fixed, we can regard the sequence of effective theories arising in the evaporation protocol as a sequence of irrelevant deformations of the same CFT.   As we showed, the change in UV cutoff is equivalent to the change in $G_N'$ and $\ell'$, so we can perhaps view the running values of these parameters as renormalized quantities in the effective theory.

In fact, it is possible to define alternative evaporation protocols that lead to the same results, including recovery of the Page transition and secret sharing in Hawking radiation. For example, we could imagine modifying our conditions for splicing the real and inception geometries by introducing a shell that has some localized stress tensor along the gluing surface, or we could glue in black holes that have non-trivial topology behind the horizon. By introducing such extra freedom, we expect that we could also fix $\ell'$ and $G_N'$ to constant values during the evaporation process.  In such scenarios, we expect on general grounds that the infalling geodesics and the secret sharing results will persist, but it would be nice to check this explicitly.

It is not clear that Hawking radiation can be naturally purified in the geometric manner that we implemented in our model.  However, it is clear there exists an equivalence class of bases for the radiation Hilbert spaces in which our secret sharing scheme is manifest.
That is to say, we need not collect radiation which is geometrized in precisely the way we have described, but rather, we can allow for the action of any product of local unitary operators on each radiation Hilbert space.
Such unitaries may destroy the geometric character of our purification within the entanglement wedge of the radiation subsets on which they act, but they will preserve the local secret sharing scheme, since such schemes by definition cannot be broken by local unitaries.   In the bulk, the dual statement to this is that the geometric ``eyeland'' cannot be destroyed by local unitaries, since it is outside of the entanglement wedges of the respective individual radiation subsystems. In this sense, the shared secret remains geometric in this equivalence class of purifications.
Interestingly, the horizon itself is the most secret of secrets -- whenever there is an eyeland, the horizon is part of it, and in particular, its smoothness is protected.
It would be interesting to understand if the actual Hawking radiation from an evaporating black hole produces an entanglement structure similar to the one we have modeled.   If it is the case that black hole evaporation generally involves a chaotic dynamics that is well-captured by holographic models, it is possible that picture that we have presented will hold in general.

The authors of \cite{Akers:2019nfi} also used muiltiboundary wormholes and the classic Ryu-Takayanagi formula to study the evolution of quantum entanglement during black hole evaporation. 
A new  ingredient in our construction is the EOW brane, which we dualized to get the inception geometry.
In \cite{Akers:2019nfi}, because there is no EOW brane, the only candidate RT surfaces were unions of causal horizons. 
However, in our model there are additional candidates, namely the infalling geodesics, which exist because we have an explicit model of the microstate theory.
The infalling geodesics lead to the formation of ``partial islands" and ``eyelands" that make secret sharing in the radiation manifest.
We believe these constructs are essential to correctly model a realistic information recovery process in more that two dimensions.

In this work, we reduced the problem of information recovery to the problem of entanglement wedge reconstruction in holography. It would be interesting to study further details of this construction, e.g. by using the Petz map. For example, in our model, modular flow of an operator acting on the radiation Hilbert space is related to the geometric flow around the RT surface. In order to study such a geometric flow, we will need the full Lorentzian geometry rather than just its time slice. In this sort of analysis, the details of the Lorentzian gluing (which we presented in the main text) will become important, whereas the infalling geodesic and eyeland analysis was largely insensitive to these details.

More generally, it would be interesting to construct the
full Lorenzian geometry for the evaporating black hole. 
For the purposes of this paper, it was enough to focus on a time independent entangled state dual to a time-reflection-symmetric geometry and to change the parameters of the associated state by hand.
One particular obstruction to a more realistic construction is the fact that the masses of the inception and real sides of the geometry are different and change in time, and therefore there will be an energy current from one side to the other. Thus the total geometry is time dependent.  A possible strategy to tackle this problem is to look at quantities at late time.  We speculate that in the late time regime, the geometry again becomes a static black hole, as the system reaches equilibrium.  This should simplify the relevant calculations.

Adding EOW branes to multiboundary wormholes presents an interesting model for many situations.
For example, since we can have different cosmological constants on either side of the brane, it would be interesting to study cosmologically motivated models like \cite{Cooper:2018cmb} or the so-called bag-of-gold spacetimes (see \cite{Fu:2019oyc} and references therein).
Furthermore, we can extend these wormhole-brane geometries indefinitely using iterated holographic inception, since our version of inception creates a spacetime of the same dimension rather than one greater dimension as in \cite{Almheiri:2019hni}.
Such extended geometries seem interesting in their own right.
Also, though we did not do so in this work, it is possible to have a multiboundary wormhole on the real side of the geometry, which would model the evaporation of collections of  entangled black holes.
These scenarios all deserve further study, as we move closer to more realistic evaporation processes.

\subsection*{Acknowledgments}
We thank Ben Craps, Matt DeCross, Mikhail Khramtsov, Cathy Li, Tokiro Numasawa, and Simon Ross for useful conversations.
VB, AK, and TU were supported in part by the Simons Foundation through the It From Qubit Collaboration (Grant No. 38559). VB and TU were also supported by the DOE grants FG02-05ER-41367 and QuantISED grant DE-SC0020360. 
AK was supported in part by DOE grant DE-SC0013528.
VB, AK, OP, GS, and TU thank the Simons Foundation for hospitality during the annual It from Qubit collaboration meeting, where this work was initiated.

\appendix

\section{Useful formulas for identifications on the upper half plane}\label{app:identifications}

Identifications are given by modding out with Fuchsian groups, see \cite{Skenderis:2009ju}. For two concentric circles, we just need the group generated by a scaling $z\mapsto \mu^2 z$, where $\mu^2=D_2/D_1$ and $D_i$ are the radii of the concentric circles. The corresponding generator is
\beq
\gamma_1
= 
\left( \begin{array}{cc}
   \mu & 0 \\
  0   & \frac{1}{\mu}
\end{array}
\right).
\eeq
When we have more than two boundaries, the group is non-abelian and there are other generators identifying pairs of circles that are not concentric. 
When the circles are concentric, we identify without reversing the orientations of the semicircles, while when the circles are non-intersecting, we identify with reversing the orientations. This is important since, for the small circles (e.g. the green semicircles in Fig.~\ref{fig:circle-data-1}) we also need to map the outside of one circle into the inside of the other, otherwise we do not have a well defined fundamental region for the quotient. We can achieve this by doing a combined translation/scaling by an upper triangular element to bring the circle to the unit circle centered at the origin, then doing a modular transformation $z \to -1/z$ of the upper half space coordinate which leaves the circle invariant while reversing the outside and inside, and then mapping it to the target circle by another translation/scaling.  The upshot is that the generator for identifying a circle at $X_a$ and radius $D_a$ with another one at $X_b$ and radius $D_b$ is
\beq
\label{eq:gamma2}
\gamma_2
= 
\left( \begin{array}{cc}
   \sqrt{D_a}  &  \frac{X_a}{\sqrt{D_a}} \\
0 & \frac{1}{\sqrt{D_a}}
\end{array}
\right)
\left( \begin{array}{cc}
  0  & 1 \\
-1 & 0
\end{array}
\right)
\left( \begin{array}{cc}
  \frac{1}{\sqrt{D_b}}   &  -\frac{X_b}{\sqrt{D_b}} \\
0 & \sqrt{D_b}
\end{array}
\right)
\eeq
We have $\text{Tr}\gamma_2 = \frac{X_b-X_a}{\sqrt{D_a D_b}}$, therefore the hyperbolic condition $|\text{Tr}\gamma_2|>2$ is satisfied when $(X_a-X_b)^2>4 D_a D_b$. 
The elements $\gamma_1$ and $\gamma_2$ generate a Fuchsian group $\Gamma$ which creates the multiboundary wormhole $t=0$ slice $\Sigma_3$ via $\Sigma_3 = \mathbb{H}_2/\Gamma$.
The curves in covering space which are the causal horizons can be computed by understanding which geodesics are left invariant under specific combinations of the generators.
For example, there is a geodesic that is left invariant by the transformation $\gamma_2$ which corresponds to the causal horizon of radiation region $R_1$. It is given by $(X-X_*)^2+Y^2=D_*^2$ where
\beq
X_*=\frac{X_a+X_b}{2}, \quad \quad D_*=\frac{1}{2}\sqrt{(X_a-X_b)^2-4 D_a D_b}.
\eeq
We will return to the other causal horizons in the next subsection.

\section{Partial islands for the three-boundary inception geometry}\label{sec:partial-islands-appendix}

We want to obtain an explicit Page curve of the radiation when the radiation itself has been split into multiple factors. We begin with a few geometric preliminaries in BTZ and the multiboundary wormhole covering space, and then turn to a computation of the entanglement entropy of the radiation.

\subsection{BTZ subregion entropy at finite cutoff}
We will first need some understanding of geodesics  in the real black geometry that start and end on the EOW brane.  The BTZ black hole in global coordinates is given by
\begin{equation}
ds^2 = -\frac{r^2-r_h^2}{\ell^2} dt^2 + \frac{\ell^2 dr^2}{r^2-r_h^2} + r^2d\phi^2 ,
\end{equation}
and the spacelike geodesic equations are
\begin{align}
1 & = -\frac{r^2-r_h^2}{\ell^2} \dot{t}^2 + \frac{\ell^2 \dot{r}^2}{r^2-r_h^2} + r^2 \dot{\phi}^2 \\
E & = \frac{r^2 - r_h^2}{\ell^2} \dot{t} \\
L & = r^2 \dot{\phi} .
\end{align}
The geodesics that live in the $t=0$ slice obey $E = 0$ since they must remain on an equal time slice;  we restrict to this case now.
We can rewrite the first equation using the angular momentum $L$ as
\begin{equation}
\ell^2 \dot{r}^2 r^2 = (r^2-L^2)(r^2-r_h^2) ,
\end{equation}
and changing variables to $u \equiv r^2$ simplifies this to
\begin{equation}
\frac{1}{4}\ell^2 \dot{u}^2 = (u-L^2)(u-r_h^2) .
\end{equation}
The solution subject to $u'(0) = 0$ (which implies $u(0) = L^2$ is the point of closest approach to the horizon) is
\begin{equation}
u(s) = \frac{1}{2} \left( (L^2-r_h^2) \cosh(2s/\ell) + L^2 + r_h^2 \right) .
\end{equation}
The angular equation can now be integrated and choosing $\phi(0) = 0$ yields
\begin{equation}
\phi(s) = \frac{\ell}{r_h} \text{arctanh} \left[ \frac{r_h}{L} \tanh (s/\ell) \right] .
\end{equation}
The above geodesic in the real geometry with endpoints at the EOW brane location attaches to the infalling geodesic coming from the inception side. 

Recall that 
the  EOW brane locus is at finite Schwarzschild radial coordinate $r_{\text{brane}}$ (this quantity is equivalent to $r_t$, the turning point radius, in the Euclidean brane trajectories discussed in Sec.~\ref{sec:quantincep}).  This quantity is determined by the tension (or equivalently, the proper length $L_{\text{brane}}$).
The relationship is
\begin{equation}
L_{\text{brane}} = 2\pi r_{\text{brane}} .
\end{equation}
In the main text, in some regimes of parameters, the entropy of the radiation was computed by the infalling geodesic in the inception side completed on the real side by the geodesic segment described about.  The contribution to the entropy from this segment, $S_{\text{brane}}$, can thus be regarded as a arising from the EOW brane.
The geodesic segment hits the brane at arclength parameters $\pm s_{\text{brane}}$, where this quantity is
\begin{equation}
s_{\text{brane}} = \frac{\ell}{2} \text{arccosh} \left[ \frac{2r_{\text{brane}}^2 - L^2 - r_h^2 }{L^2-r_h^2} \right] .
\label{eq:brane-arclength}
\end{equation}
The angular extent on the brane is
\begin{equation}
\Delta \phi_{\text{brane}} \equiv \phi(s_{\text{brane}}) - \phi(-s_{\text{brane}}) = \frac{2\ell}{r_h} \text{arctanh} \left[ \frac{r_h}{L} \sqrt{\frac{r_{\text{brane}}^2-L^2}{r_{\text{brane}}^2-r_h^2} } \right].
\end{equation}
We invert this to solve for $L$ and find\footnote{At this point, we must restrict the range of $\Delta\phi_{\text{brane}}$ to be within $[0,2\pi]$.  Notice that the formula for $L^2$ is not periodic in $\Delta\phi_{\text{brane}}$; if we take values outside this fundamental range, we will obtain non-minimal geodesics which cross themselves and wind around the black hole horizon.}
\begin{equation}
L^2(\Delta \phi_{\text{brane}}) = \frac{r_{\text{brane}}^2 r_h^2}{r_{\text{brane}}^2 \tanh \left[ \frac{r_h \Delta\phi_{\text{brane}}}{2\ell} \right]^2 + r_h^2 \text{sech} \left[ \frac{r_h \Delta\phi_{\text{brane}}}{2\ell} \right]^2} .
\label{eq:geodesic-angular-momentum}
\end{equation}

To use the RT formula in our setup, we must deal with the standard competition between two homologous geodesics for a given brane subregion of length $L_\text{subregion}$.
This subregion length is determined in terms of the angular extent by
\begin{equation}
L_\text{subregion} = r_{\text{brane}} \Delta\phi_{\text{brane}} .
\label{eq:brane-subregion-length}
\end{equation}
The first candidate minimal geodesic is the conventionally homologous one which is smoothly deformable to the desired brane region; this geodesic dominates for small enough angular extent.
The other candidate is a disconnected geodesic which takes over for large angular extent.
This geodesic has two components: one which wraps the black hole horizon and another which winds around it.
It has angular extent $2\pi - \Delta\phi_{\text{brane}}$.
Let us write
\begin{equation}
s_{\text{brane}}(L_\text{subregion})
\end{equation}
to represent the brane intersection arclength parameter for a brane subregion of size $L_s$, where the $L^2$ appearing in \eqref{eq:brane-arclength} is given by \eqref{eq:geodesic-angular-momentum} after obtaining $\Delta\phi_{\text{brane}}$ from \eqref{eq:brane-subregion-length}.
The entropy associated to a brane subregion of length $L_\text{subregion}$ is then
\begin{equation}
S_{\text{brane}} (L_\text{subregion}) = \frac{1}{4G_N}\text{min} \biggl[ 2s_{\text{brane}}(L_\text{subregion}) ,\ 2s_{\text{brane}}(2\pi r_{\text{brane}} -  L_\text{subregion}) + 2\pi r_h \biggr] .
\label{eq:brane-entropy}
\end{equation}
In the language of Sec.~\ref{sec:eyelands}, we have used the formula
\begin{equation}
    L_b(s_1,s_2) = \text{min} \biggl[ 2s_{\text{brane}}(L_\text{subregion}) ,\ 2s_{\text{brane}}(2\pi r_{\text{brane}} -  L_\text{subregion}) + 2\pi r_h \biggr] .
\end{equation}
We will show how the subregion length $L_{\text{subregion}}$ relates to the infalling endpoints $s_1$ and $s_2$ in the next subsection.

\subsection{Moduli and geodesics in covering space}

Our next step is to understand the relationship between the inception multiboundary wormhole moduli and the covering space depiction of the multiboundary wormhole.
We also need to understand geodesics in the covering space, and in particular the appearance of a new class of geodesics which has endpoints on the brane.
In principle, the metric of covering space includes the AdS radius (or, since we are modeling the inception geometry, the inception AdS radius $\ell'$), but we set $\ell' = 1$ since this overall constant can be reinstated by simply multiplying our results by $\ell'$.

The covering space setup consists of several semicircles in the upper half plane which are identified in a particular way.
The equations of these semicircles are
\begin{align}
g_1(\lambda) & = D_1 e^{i\lambda} ,\\
g_2(\lambda) & = D_2 e^{i\lambda} ,\\
g_a(\lambda) & = X_a + D_a e^{i\lambda} , \\
g_b(\lambda) & = X_b - D_b e^{-i\lambda}  ,
\label{eq:circle-data}
\end{align}
where the parameter is $\lambda \in [0,\pi]$ and we take $D_2 > D_1$, $D_1 < X_b - D_b$, $D_2 > X_a + D_a$, and $X_b + D_b < X_a - D_a$ as an ansatz.
The identifications are
\begin{equation}
g_1(\lambda) \sim g_2(\lambda), \hspace{.5cm} g_a(\lambda) \sim g_b(\lambda) ,
\end{equation}
and these are generated by the $SL(2,\mathbb{R})$ elements discussed in Appendix \ref{app:identifications}.
Our chosen ansatz for the wormhole moduli allows us to draw our covering space as in Fig.~\ref{fig:circle-data-1}.

There are three moduli ($m_1$, $m_2$, and $m_3$) associated with the inception geometry, which are lengths of geodesics $M_1$, $M_2$, and $M_3$ visualized in Fig.~\ref{fig:octopus-moduli-1}.
The first modulus $m_3$ is computed by the length of the vertical geodesic 
\begin{equation} 
M_3(\lambda) = i \lambda ,
\end{equation}
between $g_1$ and $g_2$. 
To compute the length $m_3$, then, we must integrate over the range $\lambda \in [D_1,D_2]$.
This length is
\begin{equation}
m_3 = \int_{D_1}^{D_2} \frac{d\lambda}{\lambda} = \ln \frac{D_2}{D_1} .
\end{equation}
For the second two moduli $m_1$ and $m_2$, it is more difficult to obtain analytic expressions.
There is a fixed geodesic under the identification $g_a \sim g_b$ which was written previously.
It is
\begin{equation}
M_1(\lambda) = \frac{X_a+X_b}{2} + \frac{1}{2} e^{i\lambda} \sqrt{(X_a-X_b)^2-4D_aD_b}  .
\end{equation}
The length of this geodesic between $g_a$ and $g_b$ computes the length $m_1$.
There is a second geodesic which connects $g_a$ to the image of $g_b$ under the scaling identification which generates $g_1 \sim g_2$; the relevant $SL(2,\mathbb{R})$ element is $z \to \frac{D_2}{D_1} z$, so the fixed geodesic is just the above with $g_b \to g_a$ and $g_a \to \frac{D_2}{D_1} g_b$.
Let us define $\zeta \equiv \frac{D_2}{D_1}$.
\begin{equation}
M_2'(\lambda) = \frac{\zeta X_b + X_a}{2} + \frac{1}{2} e^{i\lambda} \sqrt{(\zeta X_b - X_a)^2 - 4\zeta D_b D_a} .
\end{equation}
The length of this geodesic between $\zeta g_b$ and $g_a$ computes the length $m_2$.
To obtain the remaining component of this arc within the fundamental region, we simply act with $z \to \frac{z}{\zeta}$ and have
\begin{equation}
    M_2(\lambda) = \frac{X_b + \zeta^{-1} X_a}{2} + \frac{1}{2} e^{i\lambda} \sqrt{(X_b-\zeta^{-1}X_a)^2-4\zeta^{-1}D_bD_a} .
\end{equation}
To get integral expressions for these moduli, we need the intersection points.
For two intersecting circles with center distance $\Delta X$ and radii $D$ (left) and $D'$ (right), the horizontal distance to the intersection point from the left center is
\begin{equation}
\frac{\Delta X^2 + D^2 - {D'}^2}{2\Delta X} .
\end{equation}
This formula can be used to compute sines of the intersection point angles $\theta_1$ and $\theta_2$.
The length of a semicircle portion is given by $\int_{\theta_1}^{\theta_2} d\theta/\sin\theta$, so the moduli are
\begin{align}
m_1 & = \log \tan \left[ \frac{\pi}{2} - \frac{1}{2} \arccos \frac{X_a-X_b - \frac{2D_b(D_a+D_b)}{X_a-X_b}}{\sqrt{(X_a-X_b)^2-4D_aD_b}} \right] - \log \tan \left[ \frac{1}{2} \arccos \frac{X_a - X_b - \frac{2D_a(D_a+D_b)}{X_a-X_b}}{\sqrt{(X_a-X_b)^2-4D_aD_b}} \right] ,\\
m_2 & = \log \tan \left[ \frac{\pi}{2} - \frac{1}{2} \arccos \frac{\zeta X_b - X_a - \frac{2D_a(\zeta D_b + D_a)}{\zeta X_b - X_a}}{\sqrt{(\zeta X_b-X_a)^2 - 4\zeta D_b D_a}} \right] - \log \tan \left[ \frac{1}{2} \arccos \frac{\zeta X_b - X_a - \frac{2 \zeta D_b(\zeta D_b + D_a)}{\zeta X_b - X_a}}{\sqrt{(\zeta X_b-X_a)^2-4\zeta D_b D_a}} \right] .
\end{align}

In the presence of the brane, there is a new class of geodesics which can end on the brane.
We refer to such geodesics as ``infalling", and an example is shown in Fig.~\ref{fig:infalling-1}.
If the brane is at an angle $\Theta$, we can choose two endpoints $s_1 e^{i\Theta}$ and $s_2 e^{i\Theta}$ for the infalling geodesic with $s_2 \geq s_1$.
We are constraining the ratio $s_2 / s_1 \leq \zeta$.
To find the infalling geodesic which intersects these two points, it is useful to map the second intersection $s_2 e^{i\Theta}$ to the interior of $g_b$ using the map (see appendix \ref{app:identifications})
\begin{equation}
z \to X_b + \frac{D_aD_b}{X_a - z} .
\end{equation}
Then, we must simply find a geodesic which intersects both $s_1e^{i\Theta}$ and $X_b + \frac{D_aD_b}{X_a - s_2e^{i\Theta}}$.
Since any geodesic is a generalized semicircle with a center on the horizontal axis, these two points determine the infalling geodesic completely.
Its center is located at
\begin{equation}
\footnotesize
    X_I = \frac{D_a^2 D_b^2+2 s_2 \cos\Theta \left(s_1^2
   X_a-X_b (D_a D_b+X_a X_b)\right)+2 D_a
   D_b X_a X_b-(s_1-X_b) (s_1+X_b)
   \left(s_2^2+X_a^2\right)}{2 \left(-\cos\Theta \left(D_a
   D_b s_2+s_1 \left(s_2^2+X_a^2\right)+2 s_2
   X_a X_b\right)+D_a D_b X_a+s_1 s_2
   X_a \cos 2\Theta+s_1 s_2 X_a+X_b
   \left(s_2^2+X_a^2\right)\right)} .
\end{equation}
Its radius is given by the distance between the center and intersection point
\begin{equation}
D_I^2 = (X_I - s_1 \cos \Theta )^2 + s_1^2 \sin^2\Theta .
\end{equation}
To compute the length, we again need the angles of the two intersection points.
The sines (cosines) of these are given by ratios of the vertical (horizontal) distance of the intersection point from the center to the radius.
The length of the infalling geodesic is then
\begin{equation}
L_I(s_1,s_2) = \log \tan \left[ \frac{\pi}{2} - \frac{1}{2} \arcsin \frac{s_1 \sin \Theta}{D_I} \right] - \log \tan \left[ \frac{1}{2} \arcsin \frac{s_2 D_a D_b \sin \Theta}{ D_I (s_2^2-2s_2X_a\cos\Theta+X_a^2)} \right] .
\label{eq:infalling-length}
\end{equation}
Note that the subregion length captured by this infalling geodesic can be computed from the covering space as
\begin{equation}
    L_{\text{subregion}} = \frac{1}{\sin\Theta} \log \frac{s_2}{s_1} ,
\end{equation}
and we must of course reinstate $\ell'$ by multiplying it out front when we use any of these lengths in a more general context.

\subsection{An eyeland from inception}

We are interested in the Page curve of one component of the radiation system in a setup where the inception geometry has two legs $R_1$ and $R_2$.
To make the calculation easier, we now restrict to a ``symmetric setup", where all three radiation moduli are equal
\begin{equation}
    m_1 = m_2 = m_3 .
\end{equation}
To enforce this condition, we need only require that the $SL(2,\mathbb{R})$ elements which leave the moduli curves invariant be similar to each other as group elements, since in that case there exists an isometry swapping them.
A simple way of achieving this is to equate the eigenvalues of their matrix representations.
The $SL(2,\mathbb{R})$ elements which leave $m_{3,1,2}$ invariant are $\gamma_1$, $\gamma_2$, and $\gamma_1 \circ \gamma_2^{-1}$, respectively.
Equating eigenvalues yields the following two relations among the circle data parameters, and we define $\zeta \equiv \mu^2$:
\begin{equation}
    X_a = \mu X_b , \hspace{.5cm} X_b = \frac{\sqrt{D_aD_b} (1+\mu^2)}{\mu(\mu-1)} .
\end{equation}
We fix the rest of the circle data so that $\mu$ is the only free parameter.
\begin{align}
    D_1 & = 1/\mu ,\\
    D_2 & = \mu ,\\
    D_a & = 2\left( \frac{\mu-1}{2\mu} \right) , \\
    D_b & = \frac{1}{2} \left( \frac{\mu-1}{2\mu} \right) .
\end{align}
Note that the prefactors of 2 in $D_a$ and $\frac{1}{2}$ in $D_b$ are not so important; they simply need to multiply to 1, and we have chosen them so that all of our curves lie within the fundamental region shown in Fig.~\ref{fig:octopus-moduli-1}.
Under these conditions, all moduli have the simple expression
\begin{equation}
    m_{1,2,3} = 2 \log \mu .
\end{equation}
Our evaporation protocol, then, will be to begin at $\mu=1$ and increase $\mu$ steadily.
This is equivalent to beginning at $r_h' = 0$ and increasing $r_h'$.
Under this protocol, there will be an exchange of minimal surface between the causal horizon of one asymptotic radiation region and an infalling geodesic with a component behind the brane locus, as in Fig.~\ref{fig:infalling-1}.
From this point on, we have all the formulas necessary to numerically produce Figs.~\ref{fig:radiation-entropy-1}, \ref{fig:eyeland-growth-1}, and \ref{fig:eyeland-protocol-1}.

\section{Long wormholes from multiboundary black holes with EOW branes}\label{sec:AppendixC}
\begin{figure}[!h]
    \centering
    \includegraphics[height=5cm]{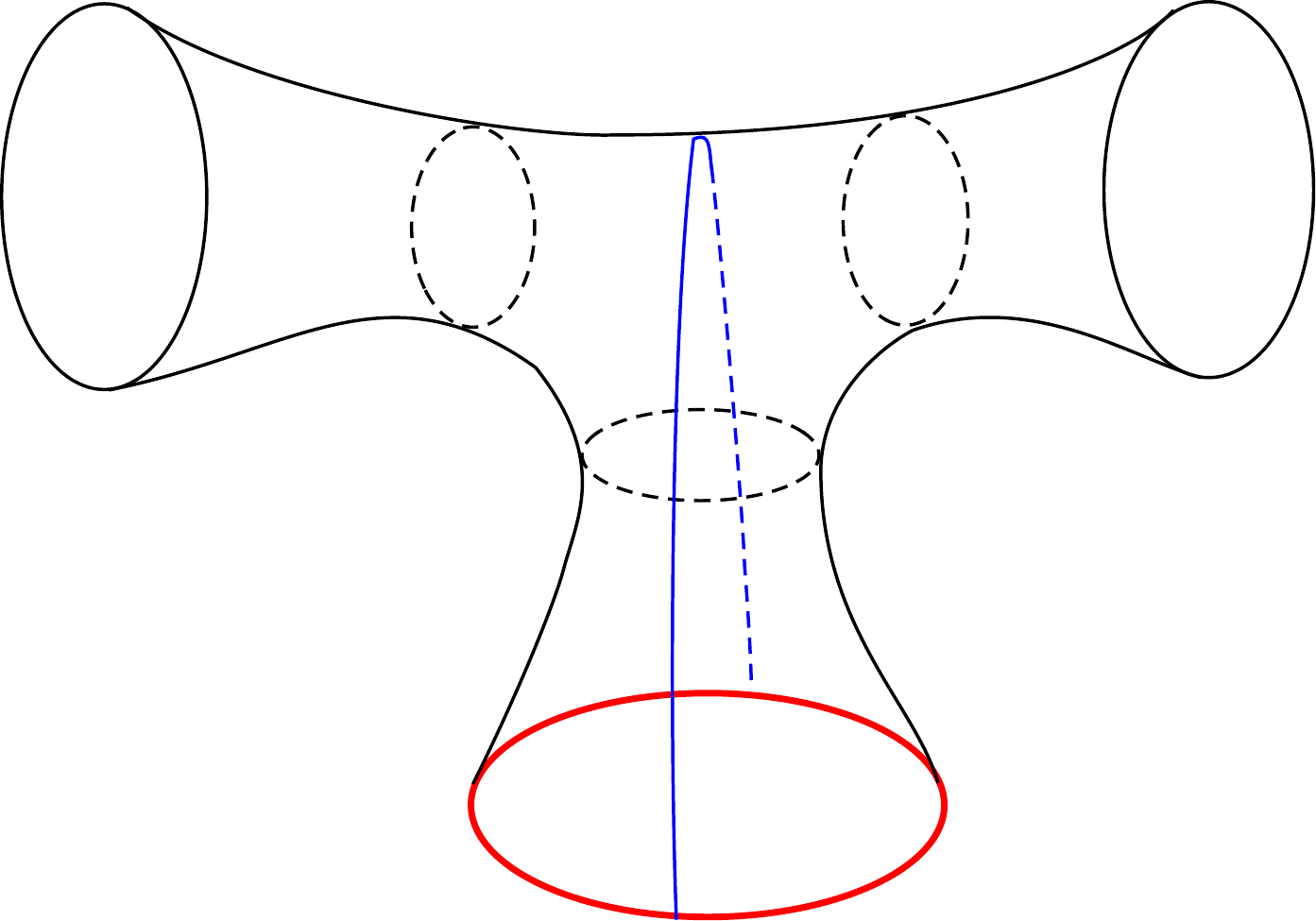}
    \caption{\small{A two-boundary long wormhole obtained from a three-boundary wormhole with an EOW brane (red). The blue surface is the extremal surface which measures the maximal cross section of the wormhole. }}
    \label{fig:3BWormhole}
\end{figure}

Multiboundary wormholes with EOW branes on various legs provide interesting examples of  long wormholes, with and without inception. In Fig.~\ref{fig:3BWormhole} we have shown the time-reflection symmetric slice of a three-boundary wormhole with an EOW brane on one of the legs. The resulting geometry is a two-boundary long wormhole. There are several interesting extremal surfaces in this geometry -- the smaller area of the two enclosing the two asymptotic boundalaries controls the entropy of entanglement between the two boundary subsystems. There is also another extremal surface which controls the maximal cross section area of the wormhole -- this was recently conjectured to be related to the restricted complexity of the boundary state \cite{Brown:2019rox}. 


\subsection{Euclidean geometry}
We can put EOW branes in any legs of a multi-boundary wormhole. Let us restrict to the case when the brane has constant tension such as in \eqref{eq:branetrajectory}. Such time dependent brane trajectories can be constructed the following way. We seek codimension one timelike surfaces with constant extrinsic curvature. It is easiest to start with the WDW representation of the wormholes
\beq
\label{eq:WDW}
ds^2=d{\hat \tau}^2 +  \cosh^2 \frac{\hat \tau}{\ell} d\Sigma^2,
\eeq
where $\hat \tau$ is Euclidean time and $d\Sigma^2$ is the metric on the $\hat \tau=0$ slice. We can take this Riemann surface to be the upper half plane with identifications, as we do in Sec.~\ref{sec:eyelands}. We can pick Poincar\'e coordinates $(x,y)$ on this surface and then it is useful to introduce polar coordinates $x=R \cos \Theta$, $y=R \sin \Theta$ so that the brane sits on constant $\Theta$ on the $\hat \tau=0$ slice. In these coordinates, the metric on the Riemann surface is
\beq
\label{eq:WDWslice}
d\Sigma^2 =\ell^2 \frac{dR^2+R^2 d\Theta^2}{R^2 \sin^2 \Theta}.
\eeq
It is natural to expect then the brane trajectory to be $\Theta=\Theta(\hat \tau)$, for some function of only $\hat \tau$. It is difficult to solve the resulting differential equation for $\Theta(\hat \tau)$, but we can instead just pull the solution \eqref{eq:branetrajectory} to this coordinate system, since everything is a patch of AdS. The relation of \eqref{eq:WDW}-\eqref{eq:WDWslice} to the BTZ coordinates \eqref{eq:btz} is
\beq
\tau= \frac{\ell^2}{r_h} \arctan \left( \tan \Theta \tanh \frac{\hat \tau}{\ell} \right), \quad r=r_h \frac{\cosh \frac{\hat \tau}{\ell}}{\sin \Theta}, \quad \varphi = \frac{\ell}{r_h} \log R.
\eeq
Using this, the surface \eqref{eq:branetrajectory} becomes the implicit surface
\beq
\frac{\cosh \frac{\hat \tau}{\ell}}{\sin \Theta} - \sqrt{\frac{1+\ell^2 T^2 \tan^2 \Theta \tanh^2 \frac{\hat \tau}{\ell}}{1-\ell^2 T^2}}=0
\eeq
The coordinate $R$ does not appear, therefore the brane indeed sits at $\Theta=\Theta(\hat \tau)$. The solution consistent with finite $\Theta$ at $\hat \tau=0$ is
\beq
\Theta(\hat \tau)=\arctan \left(\sqrt{\frac{(1-\ell^2 T^2)(1+\cosh \frac{2 \hat \tau}{\ell})}{2 \ell^2 T^2}} \right)
\eeq
This is then also consistent with the identifications on the time slice, when the problem is set up such as on Fig.~\ref{fig:octopus-moduli-1}, i.e. there are two concentric identified circles. We can move this setup around by $SL(2)$ Mobius transformations and obtain branes sitting in non-concentric throats.\footnote{ In other words, in a multi boundary setup, we can always find a Mobius transformation that brings any throat concentric. We need to transform the above curve with the inverse of that transformation.} 
\subsection{Lorentzian geometry}

We can continue $\hat \tau$ in \eqref{eq:WDW} to Lorentzian, in which case this coordinate system covers the WDW patch of the $\hat \tau=0$ slice. The brane eventually leaves this patch, but it is possible to track its motion by constructing overlapping coordinate patches that cover the entire Lorentzian geometry, as follows \cite{Skenderis:2009ju}.
 For simplicity, we will focus here on the case of a three-boundary wormhole with a simple pair-of pants topology behind the horizons, but our discussion can be extended to other more complicated geometries as well. Let us first recall from \cite{Skenderis:2009ju} that the Lorentzian multiboundary wormhole geometry can be constructed explicitly by gluing together various patches of the Lorentzian BTZ black hole. 
\begin{figure}[!h]
\centering
\begin{tabular}{c c c}
\includegraphics[height=5cm]{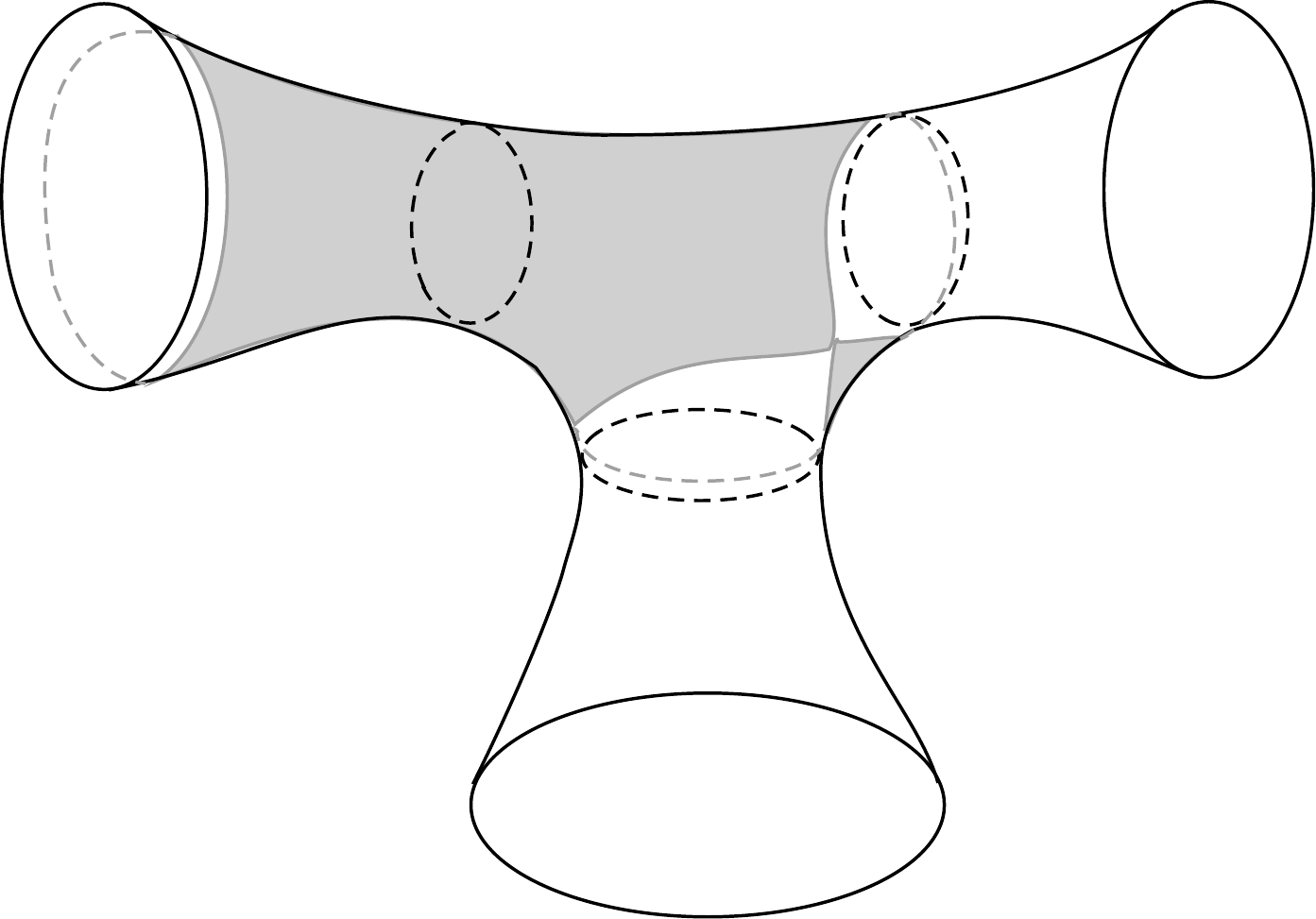} & \hspace{0.5cm} & \includegraphics[height=5cm]{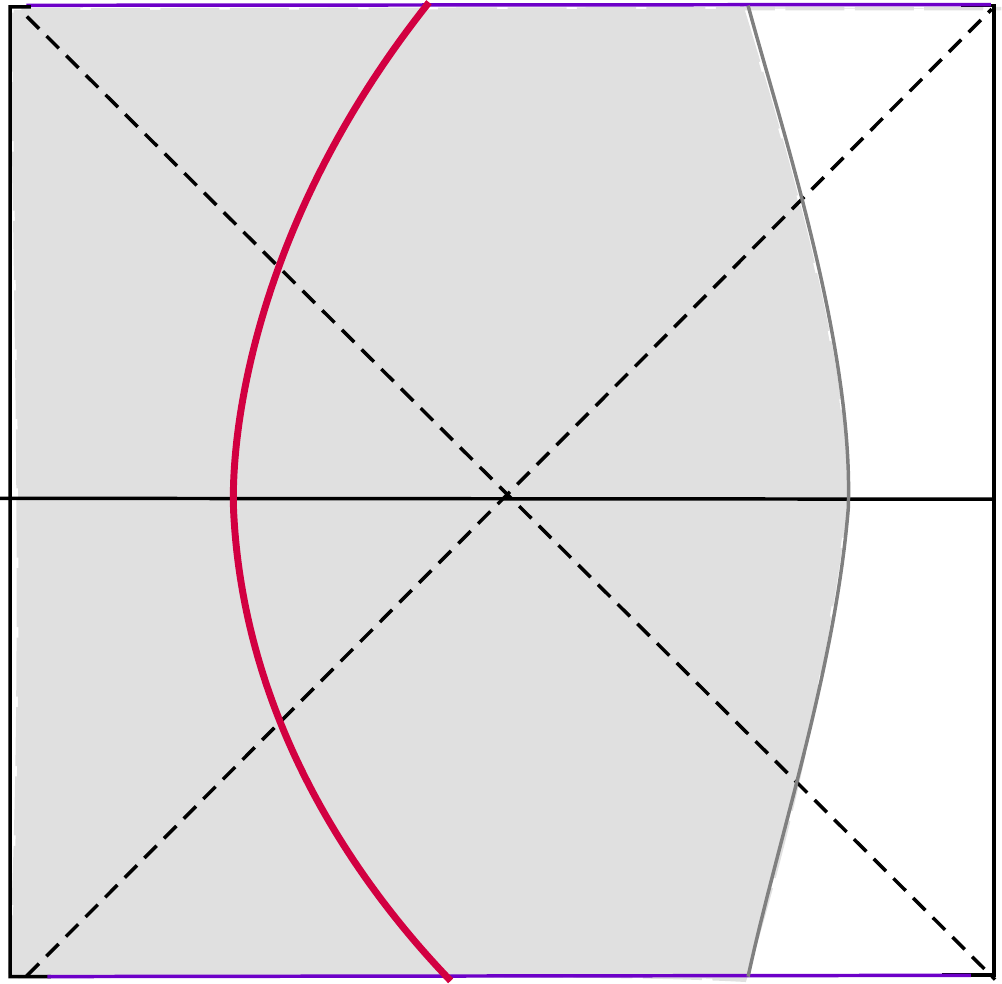}
\end{tabular}
\caption{\small{(Left) The shaded region denotes the part of the $t=0$ Cauchy slice covered by a patch of BTZ coordinates. (Right) The Lorentzian patch of the BTZ spacetime correspoding to one asymptotic region. The EOW brane is shown in red.}}
\label{fig:3BWormholeCoord}
\end{figure}
For the three-boundary wormhole with a simple pair of pants topology behind the horizons, we can cover the entire geometry with three patches of BTZ spacetime. One such patch is shown in the right panel of Fig.~\ref{fig:3BWormholeCoord}. The patch extends all the way out to one of the asymptotic boundaries; let us call it $B_1$. On the $t=0$ slice, this coordinate patch is topologically a cylinder, and extends beyond the horizon of $B_1$ to a point where the circle direction of the cylinder starts to self-intersect, as shown in the left panel of the Fig.~\ref{fig:3BWormholeCoord}. In the Lorentzian geometry this bounding circle of the coordinate patch; in Schwarzschild coordinates adapted to the right side of the BTZ black hole, is given by
\beq
r^2 = \frac{r_h^2 \cosh^2(r_h t/\ell^2)}{\cosh^2(r_h t/\ell^2)-c^2},
\eeq
where $0 < c^2< 1$ is a constant determined by the moduli (i.e., the geodesic lengths of the three extremal surfaces) of the Riemann surface at $t=0$. For instance, it can be computed from the proper length of the circle boundary of the coordinate patch when it self-intersects. At any rate, the trajectory of the EOW brane (shown in red in the figure) can be described in one BTZ coordinate patch, as already done in \cite{Cooper:2018cmb}:
\beq
\cosh\left(\frac{r_ht}{\ell^2}\right)\sqrt{\frac{r^2}{r_h^2}-1} = \frac{T}{\sqrt{1-T^2}},
\eeq
where $0<T<1$ is the tension. Of course, the three coordinate patches corresponding to the three asymptotic regions overlap in the interior in a non-trivial way; the transition functions can be found in \cite{Skenderis:2009ju}. Using these transition functions, we can describe the trajectory of the EOW brane in terms of the coordinates corresponding to any other coordinate patch, should the brane ever enter the overlapping region. Similarly, if the brane exits the original patch, then we can describe its trajectory in the other patches using these transition functions.

\bibliographystyle{JHEP}
\bibliography{eyelands}

\end{document}